\documentclass[9pt,twoside]{pnas-clean}

\templatetype{pnasresearcharticle} 

\doi{}
\setboolean{displaywatermark}{false} 
\usepackage{times}
\usepackage{amsmath}
\usepackage{graphicx}
\usepackage{array,multirow}

\newcommand{\jcm}[1]{}
\newcommand{\pp}{\ }
\newcommand{\boundary}{\mathrm{boundary}}
\newcommand{\pop}{\mathrm{pop}}
\newcommand{\con}{\mathrm{conflicted}}

\newcommand{\area}{\mathrm{area}}
\newcommand{\mino}{\mathrm{AA}}

\renewcommand{\d}{\partial}
\newcommand{\J}{J}
\newcommand{\JJ}{J}
\newcommand{\dist}{\mathcal{R}}

\newcommand{\pr}{\mathcal{P}}
\renewcommand{\Pr}{\pr_{\beta}}

\title{Quantifying Gerrymandering in North Carolina}

\author[a,b]{Gregory Herschlag}
\author[a,c,d]{Han Sung Kang} 
\author[a,d]{Justin Luo}
\author[f]{Christy Vaughn Graves}
\author[e]{Sachet Bangia}
\author[a]{Robert Ravier}
\author[a,g,1]{Jonathan C. Mattingly}
\affil[a]{Department of Mathematics, Duke University, Durham NC 27708}
\affil[b]{Department of Biomedical Engineering, Duke University,
  Durham NC 27708}
\affil[c]{Department of Computer Science, Duke University, Durham NC 27708}
\affil[d]{Department of Electrical and Computer Engineering, Duke University, Durham NC 27708}
\affil[e]{Department of Economics, Duke University, Durham NC 27708}
\affil[f]{Program in Applied and Computational Mathematics, Princeton University, Princeton NJ}
\affil[g]{Department of Statistical Science, Duke University, Durham NC 27708}

\leadauthor{Herschlag} 

\authorcontributions{J.M. and G.H. co-wrote the text. J.M. came up
  with the original idea to sample redistricting plans from a probability
  distribution and originally explored this idea with C.V.G.
  S.B. initiated the original visualization technique with the
  box-plots. J.M. conceptualized the two new indices. H.S. wrote the majority of the code used to run these simulations; H.S. and G.H. jointly debugged the code and tested it for errors.  J.L. was responsible for gathering VTD data and map visualization with ArcGIS.  R.R., G.H., and J.M. mentored H.S. and J.L. over the summer of 2016. G.H. was responsible for running simulations, data analysis, and all visualizations not done in Arc-GIS.}
\authordeclaration{None of the authors have any conflicts of interest to declare.}
\correspondingauthor{\textsuperscript{1}To whom correspondence should be addressed. E-mail: \url{jonm@math.duke.edu}}


\begin{abstract}
Using an ensemble of redistricting plans, we evaluate whether a given 
political districting faithfully represents the geo-political landscape.  
Redistricting plans are sampled by a Monte Carlo algorithm from a 
probability distribution that adheres to realistic and non-partisan criteria.
Using the sampled redistricting plans and historical voting data, we produce an 
ensemble of elections that reveal geo-political structure within the state.
We showcase our methods on the two most recent districtings of NC for 
the U.S. House of Representatives, as well as a plan drawn by a bipartisan 
redistricting panel. We find the two state enacted plans are highly atypical outliers whereas the
bipartisan plan accurately represents the ensemble both in partisan
outcome and in the fine scale structure of district-level results.
\end{abstract}

\dates{January 9, 2018}
\date{January 9, 2018}
\doi{}

\begin{document} 

\verticaladjustment{-2pt}

\maketitle
\thispagestyle{firststyle}
\ifthenelse{\boolean{shortarticle}}{\ifthenelse{\boolean{singlecolumn}}{\abscontentformatted}{\abscontent}}{}

\dropcap{I}n the 2012 NC congressional election, over half the total
votes went to Democratic candidates, yet only four of the thirteen congressional representatives were Democrats.  
Furthermore,  the most Democratic district had 29.63\% margin of victory, 
whereas the most Republican district had a 13.11\% margin of victory.
These results may be due to political gerrymandering or, alternatively,
be natural outcomes of NC's geo-political structure as
determined by the spatial distribution of partisan votes.

To probe the geo-political structure and its effect on election
outcomes we (i) sample from the space of congressional
redistricting plans that
adheres to non-partisan redistricting criteria;
 (ii) we simulate an election with
each of our sampled redistricting plans using the actual partisan votes
cast by North Carolinians in the 2012 and 2016 congressional
elections; and (iii) we aggregate election results to construct the
distributions of partisan vote balance on each district and of
the congressional delegation's partisan composition.  Districts that
do not respect typical election results are considered
gerrymandered.  When a districting is gerrymandered, the congressional delegation's partisan
composition may be not representative of what is typical.

Having probed the impact of the geo-political structure, we analyze
three specific districting plans: the two most recent districting plans of NC for the U.S. House of Representatives and a plan proposed by
a bipartisan panel of retired NC judges.  
By situating the
election outcomes of these three districting plans in our sampled ensemble,
we determine whether the three districting plans contain unlikely partisan
favoritism and thwart the underlying geo-political structure, as
expressed by the people's votes, by shifting each district's partisan
vote balance significantly away from what is typical.

\section*{Methods}
\label{sec:methods}
To sample from the space of congressional redistricting plans, we construct a family of probability distributions that are concentrated on
plans adhering to non-partisan design criteria from proposed legislation.
The non-partisan design criteria ensures that
\begin{enumerate}
\item the state population is evenly divided between the thirteen congressional districts,
\item the districts are connected and compact,
\item splitting counties is minimized, and
\item African-American voters are sufficiently
  concentrated in two districts to
  affect the winner.
\end{enumerate}
The first three criteria come from House Bill 92 (HB92) of the NC
General Assembly, which passed the House during the 2015 legislative
session.  HB92 also states that a districting should comply with the
Voting Rights Act (VRA); the fourth criteria ensures the VRA is
satisfied and is based on a redistricting plan proposed by the legislature
along with recent court rulings.  HB92 proposed establishing a bipartisan
redistricting commission guided solely by these principles (see the Appendix for the precise criteria).

There is no consensus probability distribution to select compliant redistricting plans.  For example, there is no criteria that determines when a plan contains districts that are not compact enough; it is also unclear if a distribution of plans should more heavily weight more compliant plans, or whether it should equally weight all compliant plans.  In short, there is no `correct' choice for this distribution.  We select a particular distribution for our main results and then demonstrate that these results are remarkably stable when the distribution is changed (see Section~C of the appendix).

Probability distributions are sampled with a standard Markov Chain Monte Carlo algorithm, which produces about $24,\!000$
random redistricting plans (we test the fidelity and robustness of our sampling in Section~B of the appendix) .  For each generated redistricting, we re-tally the actual historic votes
from a variety of electoral races, including the 2012 and 2016 US congressional elections, 
producing ensembles of election outcomes. When re-tallying the votes, 
we make the assumption that people vote for parties rather than people. 
We use this ensemble of election outcomes to quantify how representative a particular
districting is by observing its place in this collection; we also use
the ensemble to quantify gerrymandering by identifying districts
which have an atypical partisan concentration of voters.

We analyze and critique the NC U.S. Congressional
districting plans used in the 2012 and 2016 elections, as well as the
districting developed by a bipartisan group of retired judges as part
of the ``Beyond Gerrymandering'' project spearheaded by Thomas Ross
and the Duke POLIS Center \cite{beyondGerry}. We refer to these
districting plans of interest as NC2012, NC2016, and Judges respectively
(see Figure~8 for the district maps).

Using a related methodology, we assess to what degree three
districting plans of interest (NC2012, NC2016, and Judges) are
engineered. This is done by seeing how close each districting's
properties are to the collection of nearby redistricting plans.  Small
changes to district boundaries should not have a significant effect on
the character of election results.

\section*{Results}
\label{sec:main-results}
Using our ensemble of over $24,\!000$ redistricting plans, we tabulate the observed
probability distribution of the congressional delegation's partisan
composition for the 2012 and 2016 vote counts.  We then situate the
NC2012, NC2016 and Judges districting plans on this probability
distribution (see Figure~1).  The partisan composition of the NC2012
and NC2016 districting plans occur in less than 1\% of our generated
redistricting plans for both sets of election data, and is heavily biased
toward the Republicans.  The partisan composition of Judges
districting occurs in 39\% and 28\% of our generated redistricting plans
for the 2012 and 2016 votes respectively, providing the second most
likely outcomes both times.

\begin{figure}
\centering
       \includegraphics[width=1\linewidth]{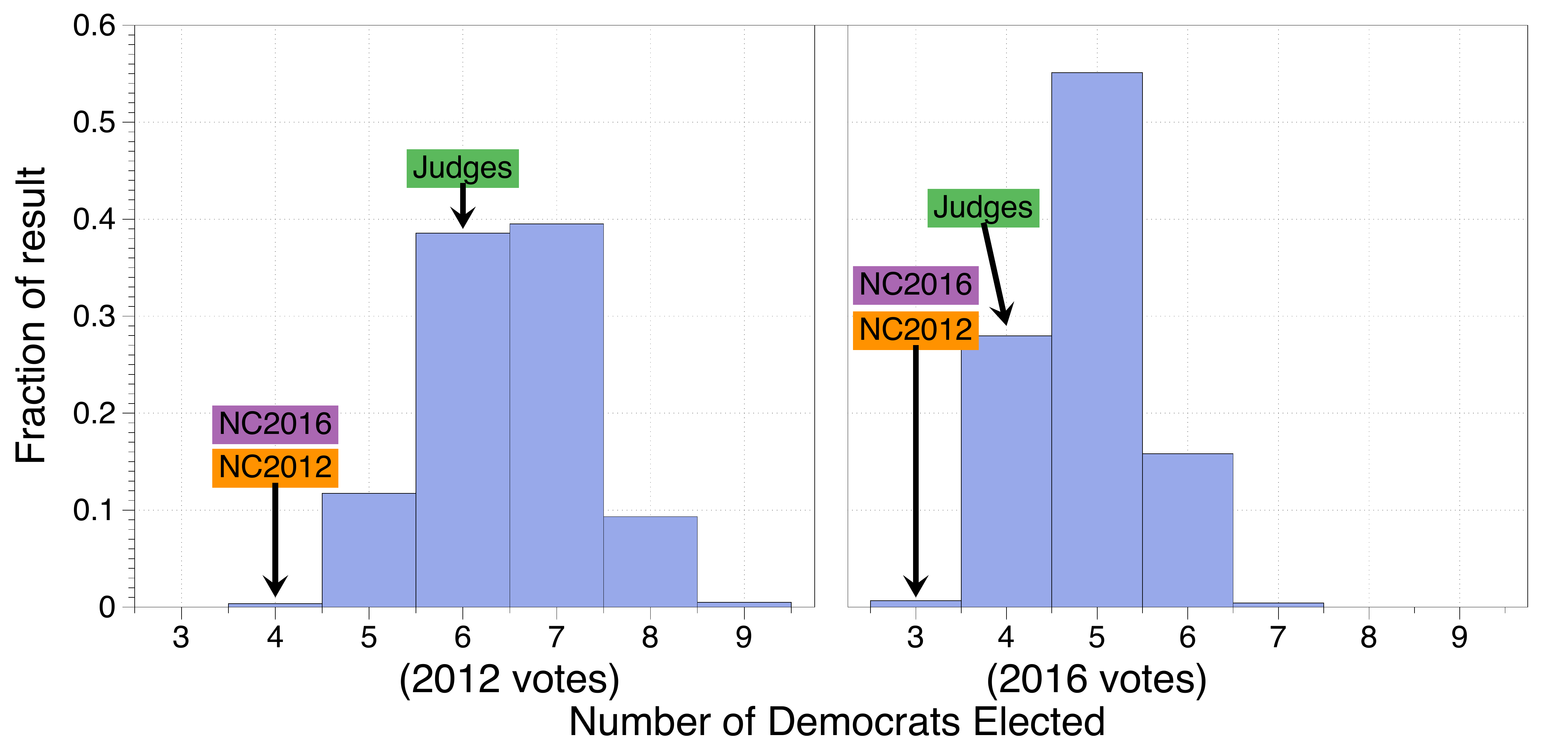}
\caption{Probability of a given  number of Democratics elected among the
     13 congressional seats using votes from the 2012 election (left) and 2016 election (right).}
\label{fig1}
\end{figure}

\begin{figure}
\centering
       \includegraphics[width=0.8\linewidth]{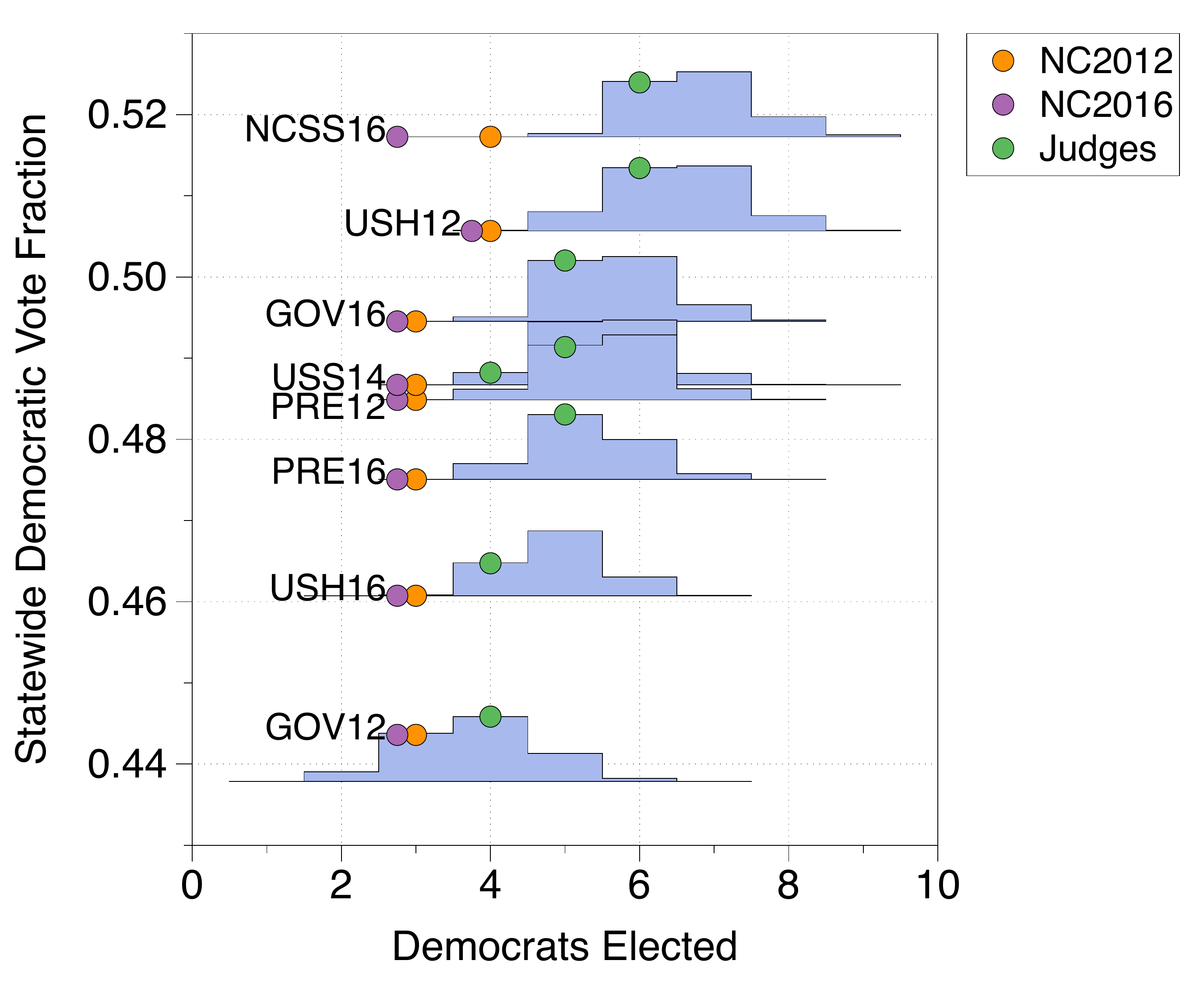}
\caption{Probability of a given number of Democratic wins among the
     13 congressional seats using vote counts from a variety of elections.  The y-axis shows the statewide democratic vote fraction.  Elections shown are the 2012 and 2016 presidential races (PRE12, PRE16), the 2016 North Carolina secretary of state race (NCSS16), the 2012 and 2016 gubernatorial races (GOV12, GOV16), the 2014 and 2016 US senatorial races (USS14), and the 2012 and 2016 US congressional races (USH12, USH16; also shown in Figure~\ref{fig1}).}
\label{fig:elsenior}
\end{figure}

By keeping the vote counts fixed and changing district boundaries, we have ignored any impact
on incumbency.  To test the effects of incumbency, we repeat the above analysis over a range of historic elections that include senatorial, presidential, gubernatorial races occurring between 2012 and 2016.  
We plot the histograms as a function of the statewide Democratic vote fraction in Figure~\ref{fig:elsenior}. 
The NC2012 and NC2016 districting plans robustly elect three Democratic candidates over nearly all sets of examined historical voting data, and these results nearly always occur in less than 1\% of the ensemble of redistricting plans when examined on the same set of election data.  In contrast, the Judges plan gradually shifts from electing four to six Democrats as the statewide Democratic vote fraction changes from 43.7\% to 51.6\% of the vote; when situated within the ensemble of redistricting plans, the results are nearly always one of the two most expected outcomes.

Although the above results are compelling, the partisan balance in election outcomes is not the only signature of gerrymandering and gives little detail of the structures that produce the atypical results. 
To further probe the geo-political structure, we order the thirteen
congressional districts in any given redistricting from the lowest to
highest percentage of Democratic votes in each district to construct
an ordered thirteen dimensional vector.  For each index, we construct a
marginal distribution.  We summarize the thirteen distributions in a
classical box-plot in Figure~3. On these box-plots, we overlay the
percentage of the Democratic vote for the ordered districts in the
NC2012, NC2016, and Judges districting plans.

     \begin{figure}
       \centering
       \includegraphics[width=1\linewidth]{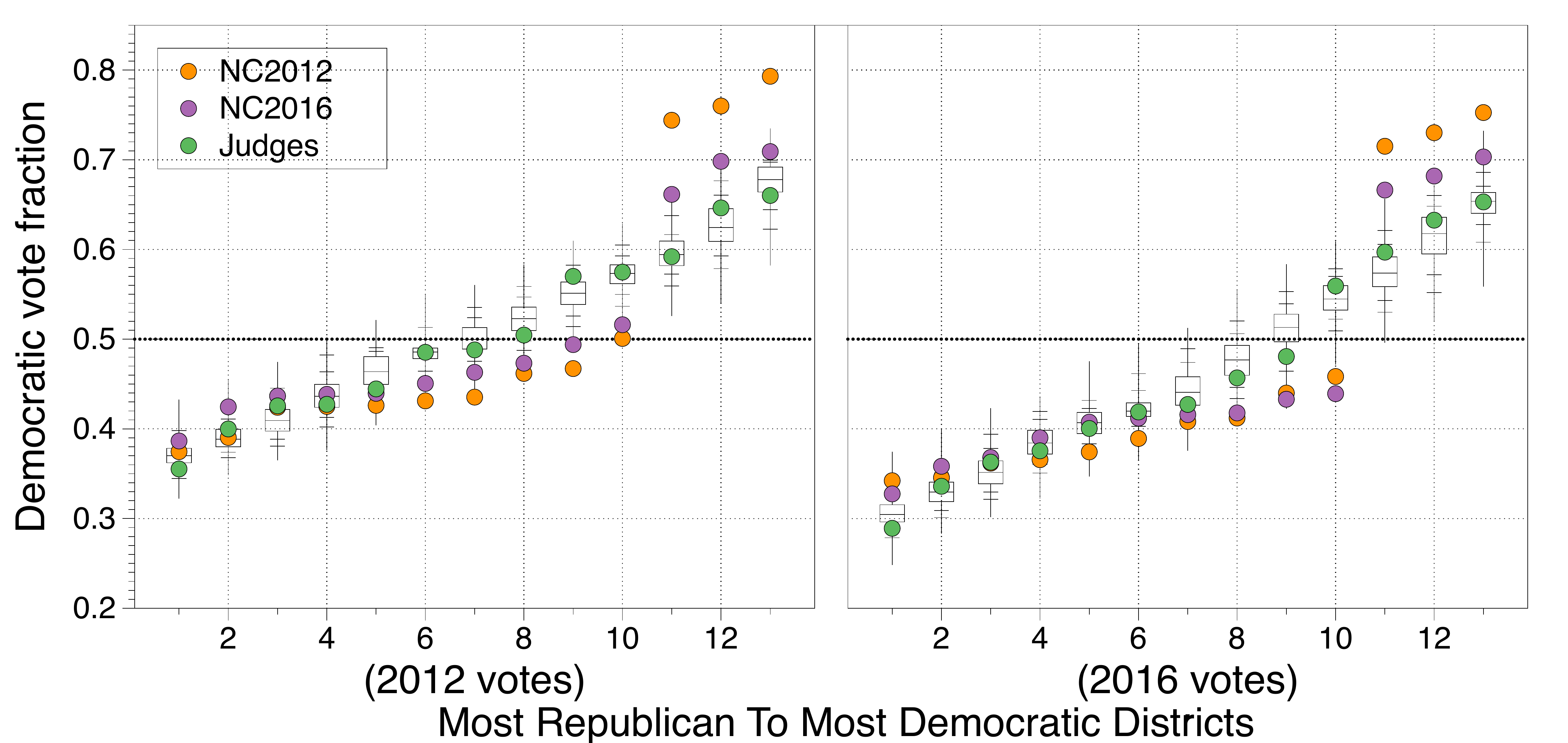}
       \caption{Box-plot summary of districts ordered from most
         Republican to most Democratic, for the voting data from 2012
         (left) and 2016 (right).  We compare our statistical results
         with the three redistricting plans of interest.}
       \label{fig3}
     \end{figure}

The structure of the box plot reveals interesting features in the
three examined districting plans.  The Judge's districts gradually
increase, roughly linearly, from the most Republican district (labeled
1) to the most Democratic (labeled 13); this behavior is identical to
the behavior of the marginal distributions. Furthermore, most of the
voting percentages from the Judges districts fall inside the boxes on
the box-plot which mark the central 50\% of the distribution.  The
NC2012 and NC2016 districting plans have a different structure: Both plans
jump in partisan voter percentage between the tenth and eleventh most
Republican districts (labeled 10 and 11, respectively). In the NC2012
districting, the fifth through tenth most Republican districts have
more Republicans than predicted by the ensemble (labeled 5-10). In the
NC2016 districting, the sixth through the tenth most Republican
districts have more Republicans than predicted by the ensemble.  In
both the NC2012 and NC2016 districting plans, votes are removed from the
central districts and added to the three most Democratic districts
(labeled 11-13) -- this is strong evidence that the middle districts
have been cracked to reduce the Democrats' influence whereas the most
Democratic districts have been packed with more Democrats than is
expected.

To quantify this observation, there are nearly no eleventh districts within the ensemble that have more Democratic votes than the eleventh district in the NC2012 and NC2016 data.  Similarly there are nearly no tenth districts within the ensemble that have fewer Democratic voters than the  NC2012 and NC206 redistricting plans.  The only exceptions to this trend are in the NC2016 plan under the 2012 votes: There are 15 tenth districts in the ensemble (0.06\%) with fewer Democratic voters than the tenth district of the NC2016 plan, and there are 25 eleventh districts in the ensemble  (0.1\%) with more Democratic voters than the eleventh district of the NC2016 plan.

Packing and cracking potentially reduce a party's political power.
When considering the 2012 votes, the box-plot demonstrates that the
7th-9th districts
are typically above the 50\% line, meaning
that we expect these districts to elect a Democratic representative; when
comparing these districts to the NC2012 and NC2016 districting plans, we
see  that the 7th-9th districts fall below this line, meaning that these districts
elected a Republican.  When considering the 2016 votes, the NC2012 and
NC2016 districting plans lead to a similar change in the delegation's
partisan composition, which can be seen by examining the ninth and
tenth most Republican districts (labeled 9-10).

In addition to the number of elected officials for each power, Figure~\ref{fig3} demonstrates add partisan safety -- districts 6-10 are all more robustly republican than they would otherwise be.  The effect is that elections within these districts may become more robustly Republican.  This effect maybe seen in both the NC2012 and NC2016 plans for districts five through seven under the 2012 vote counts, and districts seven through eight under the 2016 vote counts.

Under a standard uniform partisan swing hypothesis, a statewide shift in the votes results in the box-plots shifting globally up or down in the direction of the swing (e.g. \cite{King91}).  Hence, under this assumption, the jump in partisan vote fraction in the NC2012 and NC2016 plans results in a wide range of statewide partisan outcomes that produce an identical partisan composition of the Congressional delegation. This effect is absent from the typical ensemble plans as well as the Judges's plan.

The jump in Democratic vote fractions of the NC2012 and NC2016 plans relative to the fairly linear, gradually increasing  vote fractions of the typical ensemble and Judges's plan provides a signature of gerrymandering. This structure further reveals the districts which have had votes from one party removed and dispersed to other districts. The cracking and packing dilutes the Democratic party's political power. 
In the next section, we further quantify this signature by contextualizing the plans of interest within the ensemble of redistricting plans when analyzed with summary metrics.

\subsection*{Summary Metrics}

Although the above visualizations provide a clear picture of the signature of gerrymandering, there is a long history of employing summary metrics that seek to encapsulate the above structures with a single number:
Such metrics include electoral responsiveness, partisan bias, the efficiency gap, mean-median difference, declination and more (see, for example \cite{Gelman94,McGhee2015,Wang16,ChoLiu16,Warrington17}.  Typically these metrics are contextualized with historical data across past elections, districting plans, and states.  However it is unclear how meaningful the measures are, as they fail to consider the geo-political makeup of a region (for example, see \cite{ChenRodden13,ChoEG17}).   For example, in \cite{ChenRodden13}, the geo-political makeup of a state may cause it to be have a naturally partisan biased.

Based on these observations, we propose two novel metrics that contextualize a redistricting plan within the underlying space of possible plans:
The two indices are
\begin{itemize}
\item {\it Representativeness Index:} Measures how typical the
  resulting balance of power obtained by a given districting plan is by
  contextualizing the number of elected representatives with the ensemble of redistricting plans.
\item {\it Gerrymandering Index:} Quantifies how typical the observed
  level of packing and cracking is for a given redistricting by
  measuring how the individual districts deviate from the expected
  percentage of partisan votes.
\end{itemize}
All of the above mentioned summary statistics are based on the ordered districting vote percentages shown as the dots in Figure~\ref{fig3}, however the two novel statistics also utilized marginal distributional data (shown as the box plots in Figure~\ref{fig3}).

In the present work, we consider two of the established statistics -- partisan bias and the efficiency gap.  We also consider the two new proposed statistics.  We provide detailed descriptions of all utilized indices below.  Each summary statistic is computed for each redistricting plan in our ensemble and for each districting plan of interest (NC2012, NC2016, and Judges).  

     \begin{figure}
       \centering

\includegraphics[width=\linewidth]{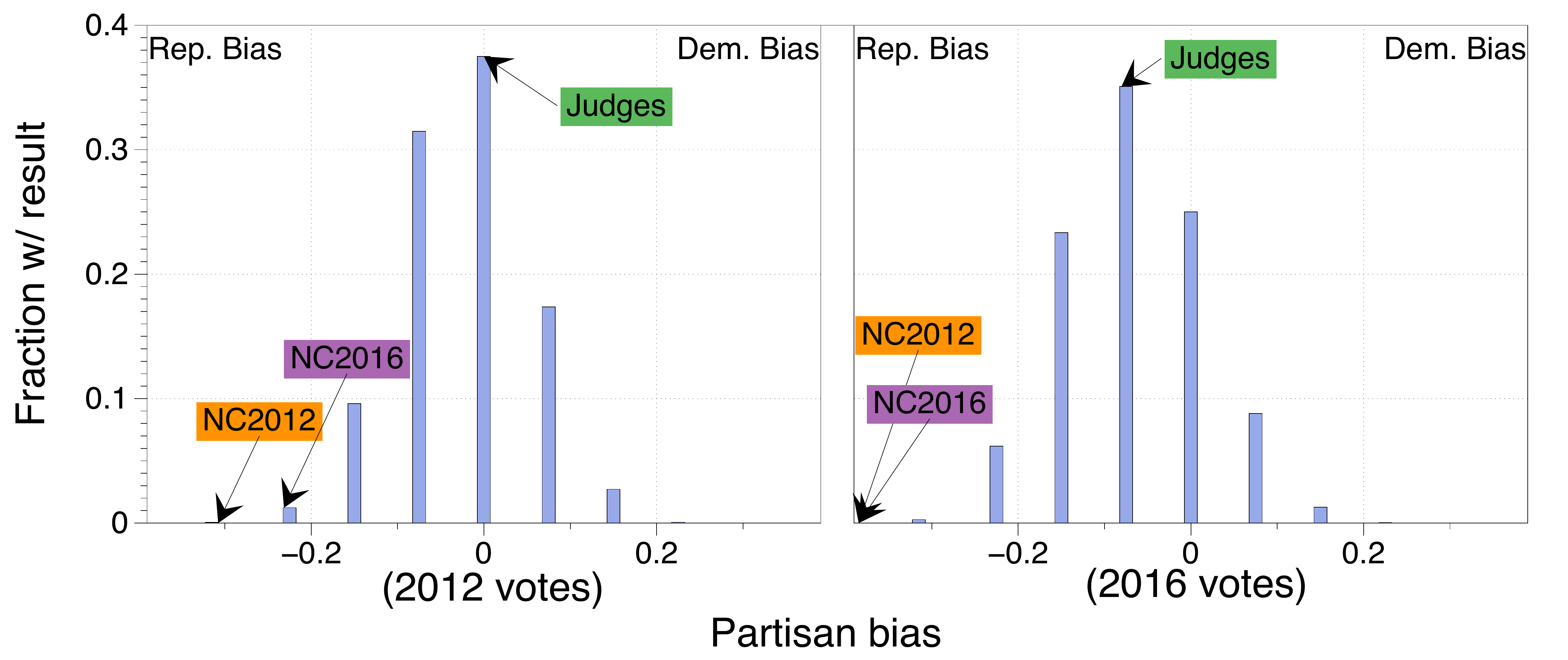}
\includegraphics[width=\linewidth]{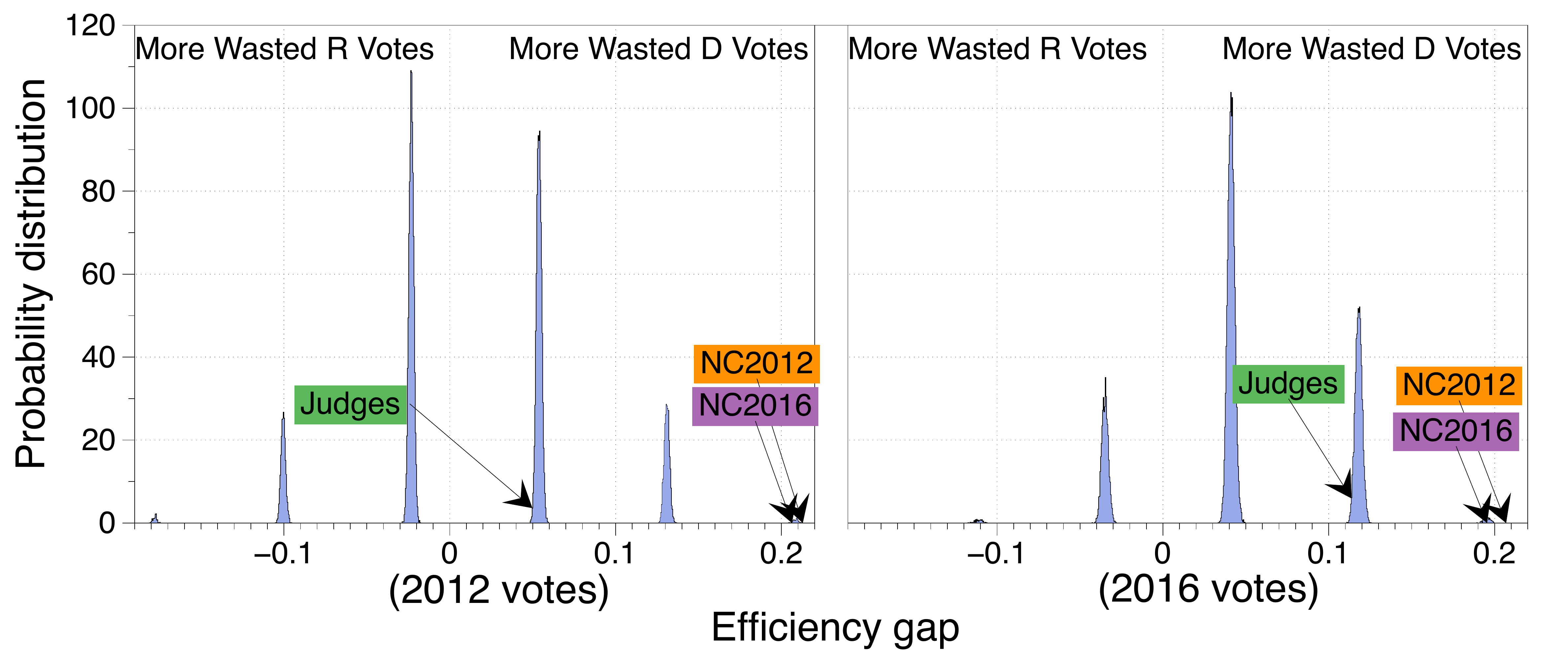}

       \caption{Absolute values of partisan bias (top) and the efficiency gap (bottom)
         for the three districts of interest based on the
         voting data from 2012 (left) and 2016 (right). }
       \label{figsymstats}
     \end{figure}

We contextualize the partisan bias and the efficiency gap, using the 2012 and 2016 congressional voting data, with histograms in Figure~\ref{figsymstats}.  In both statistics and under both vote counts, the NC2012 and NC2016 districting plans are extreme outliers; they show extreme bias toward the Republican party and waste an atypical number of Democratic party.  Under the 2012 voting data, only 14 of the generated redistricting plans (0.053\% of the plans) are as, or more, biased than the NC2012 districting plan, and only 334 redistricting plans (1.3\%)
are as, or more, biased than the NC2016.  For the 2016 voting data, no plan in the ensemble is as, or more, biased than either the NC2012 or NC2016 districting plans.  Under both sets of voting data, the NC2012 districting plan has a higher efficiency gap than any plan in the ensemble.  The efficiency gap for the NC2016 map is lower than 84 of the redistricting plans in the ensemble ($0.34\%$) for the 2012 voting data, and 86 of the redistricting plans in the ensemble ($0.35\%$) for the 2016 voting data.

In stark contrast, the Judges districting plan has less partisan biased than 15,327 redistricting plans (62.51\%) for the 2012 votes and 6130 redistricting plans (or $25.0\%$) for the 2016 votes.  The partisan bias of the Judges plan is also the most probable bias within the ensemble for both vote counts.
The Judges districting plan is also more efficient than 14,778 redistricting plans (60.02\%) and 6912 redistricting plans ($28.2\%$) under the 2012 and 2016 votes, respectively.

         \begin{figure}
       \centering
\includegraphics[width=\linewidth]{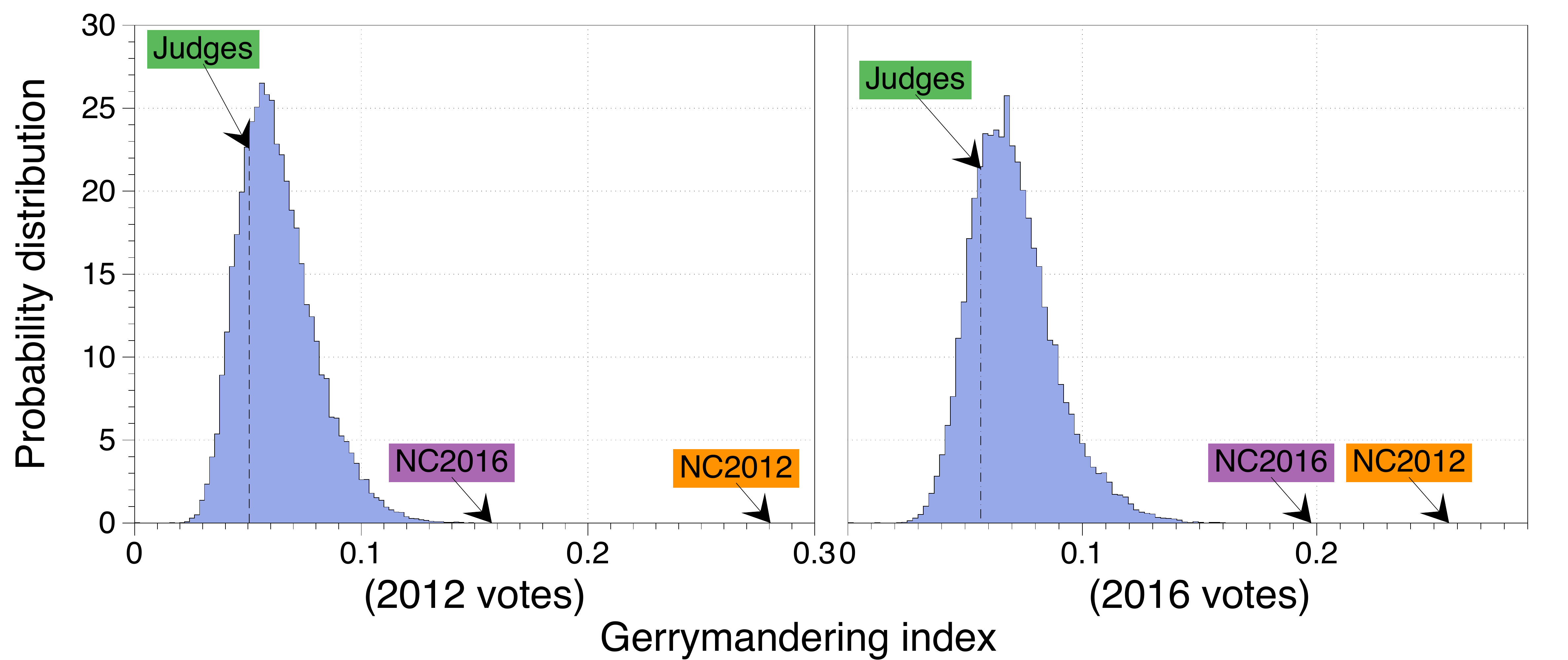}
\includegraphics[width=\linewidth]{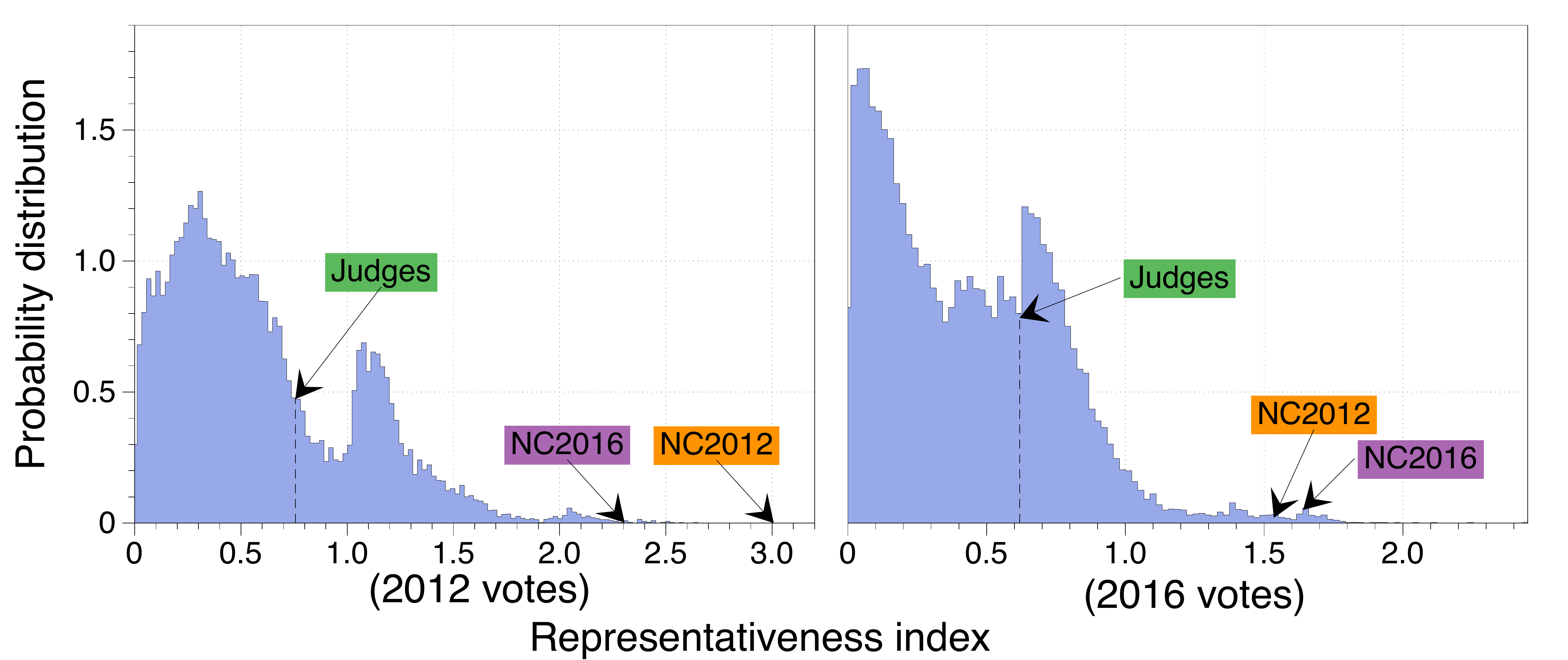}

       \caption{Gerrymandering (top) and Representativeness (bottom)
         Indices for the three districts of interest based on the
         voting data from 2012 (left) and 2016 (right). }
       \label{figourstats}
     \end{figure}

We also contextualize the Gerrymandering Index and Representativeness index with histograms in Figure~\ref{figourstats}.
None of the generated redistricting plans constructed have a partisan bias Gerrymandering Index
bigger than NC2012, regardless of the voting data used. Similarly, none of the redistricting plans have a Representativeness
Index greater than NC2012 when the 2012 votes are used and only 172
of the redistricting plans (0.7\%) have a greater Representativeness Index when the 2016 votes
are used. Again, none of the redistricting plans have a Gerrymandering
Index larger than NC2016 under both the 2012 and 2016 votes. Only 34
redistricting plans (0.14\%) and 105 redistricting plans (or 0.43\%) have a
Representativeness Index greater than NC2016 under the 2012 and 2016
votes, respectively.

Again, in stark contrast, the Judges districting plan has a lower Gerrymandering index than 18,670 redistricting plans (76.15\%) and 18,891
redistricting plans (77.05\%) under the 2012 and 2016 votes, respectively.  Similarly, the Judges districting plan has a lower Representativeness index than 7,250
redistricting plans (29.57\%) and 7,625 redistricting plans (31.1\%) under the 2012 and
2016 votes, respectively.

All four metrics over both election years indicate that the Judges plan is very typical. 
The Judges plan shows low partisan bias, has a reasonable efficiency gap, has a comparatively low level of gerrymandering, 
and is reasonably representative. In contrast, the NC2012 and NC2016 plans
have strong partisan bias toward the republicans, have a large efficiency gap, are unrepresentative, and are highly gerrymandered.

\paragraph*{Local engineering} 
If relatively small changes in a redistricting dramatically change the
partisan vote balance in each district then it raises questions how
representative the results generated by the redistricting are, and
suggests the redistricting was selected or engineered.\footnote{As the initial working paper reporting these results of this paper \cite{Bangia17} was being
 completed an the work in \cite{Chikina_Frieze_Pegden_2017} appeared
 which provides an interesting set of ideas to assess if samples being
 drawn are typical or outliers
 exactly in our context. We hope to explore these ideas in the near future.}

\begin{figure}
       \centering
       \includegraphics[width=\linewidth]{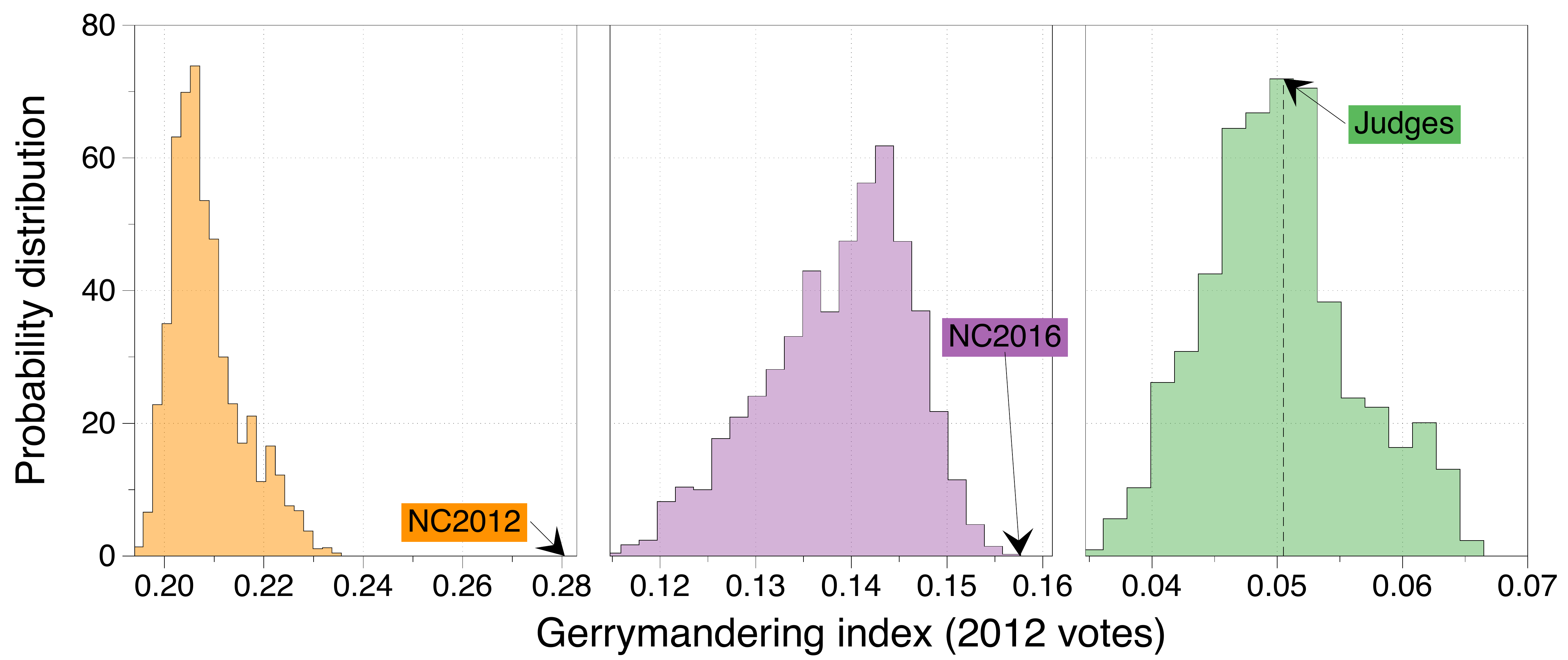}
       
       \caption{Gerrymandering Index based on random samples drawn
         from nearby the three redistricting plans of interest: NC2012
         (left), NC2016 (center), and Judges (right). Only for the
         Judges are the other points in the neighborhood similar to
         the redistricting of interest. All plots use the 2012 votes.}
       \label{fig4}
     \end{figure}

We explore the degree to which the NC2012,
NC2016 and Judges redistricting plans are locally typical by examining the
distribution of Gerrymandering Indices of an ensemble approximately 2,500 nearby
districting plans, (see Section~S5 for a precise
description of `nearby').  We select the Gerrymandering Index, because it is the only considered index that is independent of the number of elected officials, and we do not expect this number to vary significantly when examining nearby districting plans.  We display our results in
Figure~\ref{fig4}, and find that the NC2012 and NC2016 redistricting plans have
Gerrymandering Indices which are significantly larger than their
respective ensembles of nearby redistricting plans, while the Judges plan
has a Gerrymandering Index in the middle of the ensemble
produced by its nearby redistricting plans. In other words, small changes
to district boundaries make the NC2012 and NC2016 redistricting plans less
partisan but do not change the characteristics of the Judges
redistricting. This suggests that NC2012 and NC2016 plans, in contrast
to the Judge's plan, were precisely engineered to achieve a partisan
goal.

\section*{The distribution on redistricting plans}
Biased on the criteria for a reasonable redistricting outlined above, we define a family of probability distributions on the 
space of redistricting plans by first defining a score function on any 
given redistricting. This score function will return lower score to 
those redistricting which better adhere to the outlined design 
criteria. We will then use the score function to define a probability 
measure on the space of redistricting plans. 

\subsubsection*{Possible methods for generating the  ensemble of 
  redistricting plans} Before detailing the method we use we discuss a 
number of alternative methods and situate them relative to the one we 
have employed. Ideas to 
generate redistricting plans with 
computational algorithms have been being developed since the 1960's 
\cite{nagel1965simplified,thoreson1967computers,gearhart1969legislative}. 
There are three main classifications of redistricting algorithms: 
constructive randomized algorithms 
\cite{cirincione2000assessing,ChenRodden13,Chen15}, moving boundary MCMC algorithms 
\cite{macmillan2001redistricting,MattinglyVaughn2014,QuantifyingGerrymandering,Wu15,fifield2015}, 
and optimization algorithms \cite{mehrotra1998optimization,Liu16}.
Constructive randomized algorithms begin each new redistricting with an
initial random seed and either grow a fixed number of districts or combine small districts until the desired number of districts is achieved.   Moving boundary MCMC
algorithms find new redistricting plans by altering existing district
boundaries to sample from a specified distribution on redistricting plans.  In 
\cite{fifield2015}, the authors demonstrate that MCMC 
algorithms may be better at sampling the redistricting space than 
constructive algorithms.  MCMC algorithms can 
theoretically sample the space of redistricting plans with the correct probability 
distribution,\footnote{One can rigorously prove that the Markov Chain given by this
  algorithm converges to the desired distribution if run long
  enough. One only needs to establish that the Markov Chain transition
  matrix is irreducible and aperiodic. Since one can evolve from any
  connection redistricting to another through steps of the chain, it is
  irreducible. Aperiodicity follows as there exist redistricting plans which are
  connect to itself through a loop consisting of two steps and a loop
  consisting of three steps. Since
2 and 3 are prime and hence have greatest common divisor 1, the chain is
aperiodic. See the Perron-Frobenius Theorem for more details.
} whereas constructive algorithms may construct many similar 
redistricting plans while not generating other types of equally valid 
redistricting plans, leading to a skewed distribution. In particular,
one is unsure what distribution on redistricting the algorithm is producing. 

Recently an evolutionary algorithm has been proposed 
which begins with a constructive method, but then uses mixing to find 
either elite or `good enough' redistricting plans \cite{Liu16}; using the 
collection of `good enough' redistricting plans, the authors make statistical 
predictions on the fitness of districting plans, however as with the
constructive algorithms it is unclear how evolutionary algorithms 
compare with Monte Carlo models in terms of sampling the space 
properly and from what distribution they draw.  One advantage of the
MCMC approach over all of the other options is that it 
samples from a explicitly specified and constructed probability 
distribution on redistricting plans. Hence, this biases and preferences are
explicit and open to critique. 
  We remark that all of the above works have considered minimizing population deviation and compactness; a few of the works have considered minimizing county splitting; none of these works have included Voting Rights Act requirements, whereas our work does. For this work, we use a MCMC algorithm to sample the space of redistricting plans that is similar to those presented in \cite{MattinglyVaughn2014,QuantifyingGerrymandering,fifield2015}.  

Using a MCMC method requires one to define a family of
probability distribution on the space of redistricting plans.  We define
this family of distributions first by developing a score function that evaluates
the overall ``goodness'' of a districting plan, and then use the score
function to define a probability distribution.  The family of distributions focuses around compliant redistricting plans. 

\subsection*{Defining the score function}
To define the score function, we introduce several mathematical formalisms, the first of which is represents the state of North Carolina as a graph $G$ with 
edges $E$ and vertices $V$. Each vertex represents a Voting 
Tabulation District (VTD); an edge between two vertices exists if 
the two VTDs share boundaries with non-zero length.  In general VTDs may be split into census blocks, however (i) the utilized redistricting criteria requires this splitting to minimized and (ii) we demonstrate that splitting VTDs to achieve zero population deviation in a district has nearly no effect on our results (see Section~S4.A).

Defining the graph this way allows us to formally define a redistricting plan: Assuming each VTD belongs to a single district, a redistricting plan is defined as a function from the vertices, $V$, to one of the possible districts, which are represented by sequential integers -- there are thirteen congressional districts in North Carolina, so we define a redistricting plan as a function $\xi: V \rightarrow \{1,2,\dots,13\}$. 
The redistricting plan function $\xi$ is interpreted as follows: If a VTD is represented by a vertex $v \in V,$ then $\xi(v)=i$ means that the VTD in question belongs to district $i$; similarly, for any $i \in \{1,2,\dots, 13\}$ and plan $\xi$, the 
$i$-th district, denoted $D_i(\xi)$, is given by the set $\{v \in 
V: \xi(v)=i\}$. 
 We restrict the space of considered redistricting plans $\xi$ such 
that each district $D_i(\xi)$ is a single connected component; this restriction, along with our edge criteria, ensures that \S120-4.52(f) of HB92 is always satisfied.  We 
denote the collection of all redistricting plans with connected 
districts by $\mathcal{R}$. 

A plan $\xi$ is rated with our score function denoted $J$.  $J$ maps each redistricting $\xi \in \mathcal{R}$ to a nonnegative number.  Lower scores signify redistricting plans that more closely adhere to the criteria of HB92.  To define the score function $J$, we employ several sub-functions that measure how well a given redistricting satisfies the individual principles outlined in HB92.  We will denote these sub-functions as  $J_{p}, J_{I}, J_{c},$ and $J_{m}$: the \emph{population score}
$J_{p}(\xi)$ measures how well the redistricting $\xi$ partitions 
the population of North Carolina into 13 equal population groups; the 
\emph{isoperimetric score} $J_{I}(\xi)$
measures how compact the districts are; the \textit{county score} $J_{c}(\xi)$ measures the number of counties split between multiple districts; lastly, the \textit{minority score} $J_{m}(\xi)$ measures 
the extent to which a districting plan adheres to the VRA.  Once the
sub-functions are specified, our score function $J$ is defined as a
weighted sum of $J_{p}, J_{I}, J_{c},$ and $J_{m}$; since all of the sub-score functions are not on the same scale, we use a weighted combination to balance the influence of each criteria. Specifically, we define 
\begin{align}
  J(\xi) = w_{p} J_{p}(\xi)+w_{I} J_{I}(\xi)+ w_{c} J_{c}(\xi) + w_{m} J_{m}(\xi), 
  \label{eqn:score}
\end{align}
where $w_{p}$, $w_{I}$, $w_{c}$,  and $w_{m}$ are a collection of positive weights. 

To describe the individual sub-functions, data is associated to our 
graph $G$ which allows the recovery of relevant features on each VTD. 
The positive functions $\pop(v)$, $\area(v)$, and $\mino(v)$, defined on a vertex 
$v \in V$, represent (respectively) the total population, geographic area, and 
minority population of the VTD associated with $v$; the symbol $\mino$ represents the minority population because African-Americans are the only minority in North Carolina with a large enough population to gain representation under the VRA. 
The functions $\pop(v)$, $\area(v)$, and $\mino(v)$ are extended to a collection of vertices $B \subset V$ by 
{\medmuskip=-1mu
\thinmuskip=-1mu
\thickmuskip=-1mu
\begin{align}\label{eq:PopArea}
  \pop(B) = \sum_{v \in B} \pop(v), \ \area(B) =
  \sum_{v \in B} \area(v),\  \mino(B) =
  \sum_{v \in B} \mino(v)\,. 
\end{align}}

The boundary of a district $D_i(\xi)$, denoted $\d D_i(\xi)$, is the subset of the edges $E$ that connect vertices 
inside of $D_i(\xi)$ to vertices outside of $D_i(\xi)$. 
According to this definition, VTDs that border another state or the ocean will not have an edge that signifies this fact.  To incorporate state boundary information, we add the vertex $o$ to $V$, 
which represents the ``outside'' and connect it with an edge to each 
vertex representing a VTD which shares perimeter with the boundary of the state. We 
assume that any redistricting $\xi$ always satisfies 
$\xi(v)=0$ if and only if $v=o$; 
since $\xi$ always satisfies $\xi(o)=0$, and $o \not \in 
D_i(\xi)$ for $i \geq 1$, it does not matter that we have not defined 
$\pop(o)$, $\area(o)$ or $\mino(o)$, as $o$ is never included in the districts. 

Given an edge $e \in E$
which connects two vertices $v, \tilde v \in V$, 
$\boundary(e)$ will represent the length of common border of the VTDs 
associated with the vertex $v$ and $\tilde v$.  As before, 
the definition is extended to the boundary of a set of edges 
$B\subset E$ by 
\begin{align}
\label{eq:Boundary}
    \boundary(B) = \sum_{e \in B} \boundary(e)\,. 
\end{align}

With these preliminaries out of the way, we define the score sub-functions used to assess the goodness of a redistricting. 

\subsubsection*{The population score function}
The population score, which measures how evenly populated the districts are, is defined by 
\begin{align*}
  \JJ_{p}(\xi) = \sqrt{\sum_{i=1}^{13}
  \Big(\frac{\pop(D_i(\xi))}{\pop_{\text{Ideal}}} - 1\Big)^2}, 
  \quad \pop_{\text{Ideal}}=\frac{N_{pop}}{13}, 
\end{align*}
where $N_{pop}$ is the total population of North Carolina, 
$\pop(D_i(\xi))$ is the population of the district $D_i(\xi)$ as 
defined in equation \eqref{eq:PopArea}, and $\pop_{\text{Ideal}}$ is the population 
that each district should have according to the `one person one vote' 
standard: $\pop_{\text{Ideal}}$ is equal to 
one-thirteenth of the total state population. 

\subsubsection*{The Isoparametric score function}
The Isoperimetric score, which measures the overall compactness of a 
redistricting, is defined by 
\begin{align*}
  \JJ_{I}(\xi)= \sum_{i=1}^{13}\frac{\big[\boundary(\d D_i(\xi))\big]^2}{\area(D_i(\xi))}\,. 
\end{align*}
It is is the ratio of the 
square perimeter to the total area of each district. The Isoparametric score is 
minimized for a circle, which is the most compact shape. 

This compactness measure is one of two measures often used in the legal 
literature, where its reciprocal is proportional to as \textit{the Polsby-Popper score} or \textit{the perimeter 
  score} \cite{Pildes_Niemi_1993,Practice_Hebert_2010}. 
The second measure, 
usually referred to as \textit{the dispersion score}, is more 
sensitive to overly elongated districts, although the perimeter score 
also penalizes them. We select the \textit{the perimeter score} because it penalizes undulating boundaries, whereas the the dispersion score does not.  There are over 20 measures to evaluate compactness (see, for example, \cite{Niemi90}).  We chose the \textit{the Polsby-Popper score} both because of its historical significance and because it is consistent with the utilized compactness criteria.  To validate the robustness of this choice, we have also examined a dispersion score and find qualitatively similar results (see Section~S3.D).

\subsubsection*{The county score function}
The county score function measures how many, and to what degree, counties are split between districts.  If two VTDs belong to different districts but the same county, the county is called a split county. The score function is defined as 
\begin{align*}
  J_c(\xi)=& \{\# \textrm{ counties split between 2 
  districts}\}\cdot W_2(\xi) \\&+  M_C  \cdot \{\# \textrm{ counties split between 
  $\geq$ 3 
  districts}\}\cdot W_3(\xi), 
\end{align*}
where $M_C$ is a large constant that heavily penalizes three county splits, and $W_2$ and $W_3$ are weight factors that smooth abrupt transitions between split/non-split counties and are defined by 
\begin{align*}
  W_2(\xi) &= \sum_{\substack{\text{counties} \\ \text{split between}\\\text {2 
  districts}}} \Big(\parbox{16em}{Fraction of county VTDs in 2nd largest\\
  intersection of a district with the county }\Big)^{\frac12},\\
  W_3(\xi) &= \sum_{\substack{\text{counties} \\ \text{split 
  between}\\\text { $\geq$ 3 
  districts}}} \Big(\parbox{18em}{Fraction of county VTDs not in 1st 
  or 2nd \\largest 
  intersection of a district with the county }\Big)^{\frac12}.\\
\end{align*}

\subsubsection*{The Voting Rights Act or minority score function}
The VRA mandates that minorities have the ability to elect a number of representatives; the number is determined by the fraction of the population comprised of the minority.  In North Carolina, the only minority large enough to warrant consideration under the VRA is the African American population. 
African-American voters make up approximately 20\% of 
the eligible voters in North Carolina; since 0.2 is between 
$\frac{2}{13}$ and $\frac{3}{13}$, the current judicial interpretation 
of the VRA stipulates that at least two districts should have 
enough African-American voters so that this demographic may 
elect a candidate of their choice. 

African-American voters should not be overly represented in a district either.  The NC2012 districting 
plan was ruled unconstitutional because two districts, each containing 
over 50\% African-Americans, were ruled to have been packed too 
heavily with African-Americans, diluting their influence in other 
districts. The NC2016 districting was accepted based on 
racial considerations of the VRA and contained districts that held 
44.48\% African-Americans, and 36.20\% African-Americans.  The amount 
of deviation constitutionally allowed from these numbers is unclear. 

Based on these considerations, we chose a VRA score function which 
awards lower scores to redistricting plans which had one district close to 44.48\% African-Americans and a second district close to 36.20\% African-Americans. 
We write  the score function as 
\begin{align}
J_m(\xi) = \sqrt{H(44.48\%-m_1)}+\sqrt{H(36.20\%-m_2)}, 
\end{align}
where $m_1$ and $m_2$ represent the percentage of African-Americans in 
the districts with the  highest and second highest percent 
of African-Americans, respectively. $H$ is the function 
defined by $H(x)=0$ for $x \leq 0$ and $H(x)=x$ for $x > 0$. The use of the square root function steepens the score function as districts near the desired population percentage; when used in conjunction with the Monte Carlo algorithm presented below, the steepening will have the effect that districts close to the set desired minority populations are more likely to move toward achieving these populations.  
Notice that whenever $m_1\geq 44.84\%$ and $m_2 \geq 
36.20\%$ we have that $J_m=0$; this feature allows the possibility for high minority populations, but allows such instances to arise naturally and does not target such an outcome. 


\subsection*{A Family of Probability Distribution on Redistricting Plans}
We now use the score function $J(\xi)$ to assign a probability 
to each redistricting $\xi \in  \dist$ that makes redistricting plans with 
lower scores more likely. Fixing a $\beta >0$, we define the 
probability of $\xi$, denoted by $\mathcal{P}_\beta(\xi)$, by 
\begin{align}
  \label{eq:Prob}
  \Pr(\xi) = \frac{e^{- \beta \JJ(\xi)}}{\mathcal{Z}_\beta}
\end{align} 
 where $\mathcal{Z}_{\beta}$ is the normalization constant defined so that $\mathcal{P}_\beta(\dist)=1$. Specifically, 
\begin{align*}
    \mathcal{Z}_{\beta} = \sum_{\xi \in \dist} e^{- \beta J(\xi)}\,. 
\end{align*}
The positive constant $\beta$ is often called the ``inverse temperature'' in analogy with statistical mechanics and gas dynamics. 
When $\beta$ is very small (the high temperature regime), different elements of $\dist$ have close to equal probability. As $\beta$ increases (``the temperature decreases''), the measure concentrates the probability around the redistricting plans $\xi \in \dist$ which minimize $J(\xi)$. This idea has been previously used in \cite{MattinglyVaughn2014} and \cite{fifield2015}. 

\subsection*{Sampling the distribution}
We use a standard Metropolis-Hastings algorithm.  We define the proposal chain $Q$ used for
proposing new redistricting plans in the following way:
\begin{enumerate}
\item Uniformly pick a conflicted edge at random. An edge, $e=(u,v)$
  is a conflicted edge if $\xi (u) \neq \xi(v)$, $\xi(u) \neq 0$,
  $\xi(v) \neq 0$.
\item For the chosen edge $e=(u,v)$, with probability $\frac12$, either:
\begin{equation*}
   \xi '(w) = 
     \begin{cases}
       \xi(w) & w \neq u\\
       \xi(v)& u
     \end{cases}
\qquad\text{or}\qquad
\xi '(w) =
\begin{cases}
   \xi(w) & w \neq v\\
       \xi(u)& v 
\end{cases}
\end{equation*} 
\end{enumerate}
Let $\con(\xi )$ be the number of conflicted edges for redistricting
$\xi$.  Then we have $Q(\xi,\xi ')=\frac{1}{2}\frac1{\con(\xi)}$. The
acceptance probability is given by:

$$p=\min \Big(1, \frac{\con(\xi)}{\con(\xi ')} e^{-\beta (\J(\xi ')-\J(\xi))} \Big)$$
If a redistricting $\xi '$ is not connected, then we refuse the step, which is equivalent to setting
$\J(\xi ')=\infty$.

We utilize simulated annealing to sample the space: $\beta$ starts starts and remains at zero until $40,\!000$ steps are accepted, which allows the MCMC algorithm to freely explore the space of connected redistricting plans; next $\beta$ grows linearly to one over the course of $60,\!000$ accepted steps, which allows the algorithm to search for a redistricting with a low score without getting caught in local minimum; finally, $\beta$ is fixed at one for $20,\!000$ accepted steps a sufficient number of steps so that the algorithm locally samples the measure $\Pr$.
This process is repeated for each sampled redistricting.  For other Monte Carlo algorithms see \cite{macmillan2001redistricting,MattinglyVaughn2014,QuantifyingGerrymandering,Wu15,fifield2015}; for other redistricting algorithms see \cite{cirincione2000assessing,ChenRodden13,Chen15,mehrotra1998optimization,Liu16}.  

\subsection*{Determining the weight parameters}
As we have mentioned above, we have four independent weights
$(w_p,w_I,w_c,w_m)$ used in balancing the effect of the different
scores in the total score $J(\xi)$. In addition to these parameters, we
also have the low and high temperatures corresponding respectively to
the maximum and minimum $\beta$ values used in simulated
annealing. We have set the minimum value of $\beta$ to be zero which corresponds to infinite temperature.  In this regime, no district is favored over any other, which allows the redistricting plan freedom to explore the space of possible redistrictings. To consider a high $\beta$ value, we note that $\beta$ multiplies the four weights in the probability distribution function: This means that one of the five remaining degrees (the four weights and high $\beta$) is redundant and can be set arbitrarily.  We therefore chose to fix the low temperature (high value of $\beta$) to be one. 

To select appropriate weights, we employ the following tuning method:
\begin{enumerate}
\item{Set all weights to zero.}
\item{Find the smallest $w_p$ such that a fraction of the results are within a desired threshold (for the current work we ensured that at least 25\% of the redistrictings were below 0.5\% population deviation, however we typically did much better than this).}
\item{Using the $w_p$ from the previous step,  find the smallest $w_I$ such that a fraction of the redistrictings have all districts below a given isoperimetric ratio (we ensured that at least 10\% of the results were below this threshold; we chose a threshold of 60, as we have done above).}
\item{If above criteria for population is no longer met, repeat steps 2 through 4 until both population and compactness conditions are satisfied.}
\item{Using the $w_p$ and $w_I$ from the previous steps, find the smallest $w_m$ such that at least 50\% of all redistrictings have at least one district with more than 40\% African-Americans and a second district has at least 33.5\% African-Americans.}
\item{If the thresholds for population were overwhelmed by increasing $w_m$, repeat steps 2 through 6.  If the thresholds for compactness were overwhelmed, repeat steps 3 through 6.}
\item{Using the $w_p$, $w_I$, and $w_m$ from the previous steps, find the smallest $w_c$ such that we nearly always only have two county splits, and the number of two county splits are, on average, below 25 two county splits.}
\item{If the thresholds for population are no longer satisfied, repeat steps 2 through 8.  If the criteria for the compactness is no longer met, repeat steps 3 through 8.  If the criteria for the minority populations is not satisfied, repeat steps 5 through 8.  Otherwise, finish with a good set of parameters. }
\end{enumerate}
With this process, we settle on parameters $w_{p}=3000$, $w_{I}=2.5$,
$w_{c}=0.4$,  and $w_{m}=800$ and have used these parameters for all
of the results presented in the main text.  We remark that this choice of parameters allows us to sample the space more quickly.  Both the primary and local redistricting plans are available for download\footnote{git@git.math.duke.edu:gjh/districtingDataRepository.git}.  Other choices in parameters lead to similar results as is demonstrated Section~C of the appendix.

\subsection*{Thresholding the sampled redistrictings}
It is possible for the simulated annealing algorithm to draw a redistricting with a bad score when using the MCMC algorithm combined with the probability distribution given in
\eqref{eq:Prob}. 
 Additionally, using simulated annealing in the MCMC algorithm increases the chance 
of becoming trapped in a district that is a local minimum of the score function, but has a less-than-desirable score; this is because it may take more steps to escape a local minimum than we take with the high value of $\beta$. Lastly, our score functions focus on average values rather than individual districts: For example, the isoperimetric score function is the
sum of the individual isoperimetric scores of each district, and it is therefore possible to have one bad district if the rest have exceptionally small isoperimetric scores.

In maximizing the degree of compliance with HB92, we only
use samples which pass an additional set of thresholds, one for each of
the selection criteria. This additional layer of rejection sampling was
also used in reference \cite{fifield2015}, though the authors of
this work chose to reweigh the
samples to produce the uniform distribution over the set redistrictings
that satisfy the thresholds. We prefer to continue to bias our
sampling according to the score function so better redistrictings
are given higher weights; we note that the idea of preferring some redistrictings to others is consistent with the provisions HB92.

From our experience from the Beyond Gerrymandering project,
redistrictings which use VTDs as their building blocks and have less that
1\% population deviation can readily be driven to 0.1\% population
deviation by breaking the VTDs into census tracts and performing 
minimal alterations to the overall redistricting plan; we also demonstrate in Section~D.I of the appendix that splitting VTDs to achieve zero population deviation has a negligible effect on our results.  We thus only
accept redistrictings that have no districts above 1\% population
deviation. Many of our samples have deviations considerably below this
value. It is important to emphasize that we require this of every district
in the redistricting. In addition to showing that splitting VTDs to zero population deviation has no effect on our results (Section~D.I of the appendix), we also show
that when the population threshold is decreased from
1\% to 0.75\% and then to 0.5\%, the results are quantitatively extremely similar, and
qualitatively identical (Section~C.I of the appendix).

We have found that districts with isoperimetric scores under 60 are
almost always reasonably compact. Thus, we choose to accept a redistricting only if each district in the plan has an isoperimetric ratio less than 60. The Judges redistricting plan would be accepted under this threshold as its least compact district has an isoperimetric score of 53.5. Neither NC2012 nor NC2016 would be accepted with this thresholding as the least compact districts of each plan have isoperimetric scores of 434.65 and 80.1, respectively. We also note that only two of the thirteen districts for the NC2012 plan meet our isoperimetric score threshold, whereas eight of the thirteen districts of NC2016 fall below the threshold. Although we examine our principle results over a space of highly compact redistricting plans, we also demonstrate that our results are insensitive to lifting this restriction in the Appendix (see Section~C.III).

Although redistrictings which split a single county in three are
infrequent, they do occur among our samples. Since these are undesirable, we only accept
redistrictings for which no counties are split across three or more
districts. In order to satisfy population requirements,
some counties must be split into two districts; an example making this clear is that Wake and Mecklenburg Counties each contain a
population larger than a single Congressional district's ideal population. We do not explicitly
threshold based on number of split counties, though redistrictings
with more split counties have a higher scores, and hence are less favored.  We remark that none of our generated redistrictings have more county splits than the NC2012 redistricting plan, and that the NC2012 plan was never critiqued or challenged based on the number of county splits.

To build a threshold based on minority requirements of the VRA, we
note that the NC2016 redistricting was deemed by the courts to
satisfy the VRA. The districts in this plan with the two highest proportion of African-Americans to total population are composed of 44.5\%  and 36.2\% African-Americans. With this in mind, we only accept redistrictings if the districts with the two highest percentages of African-American population have at least 40\% and 33.5\% African-American voters, respectively.

Thresholding in this way sub-samples to roughly 16\% of the
samples initially produced by our MCMC runs.  Although this leads to many unused samples, it ensures that all of the utilized redistrictings meet certain minimal standards. This better adheres to the spirit of
HB92. The reported 24,518  
samples used in our study refer to those left after thresholding. The  
full data set of samples was in excess of 150,000.  That being said, we show in Section~C.I of the appendix
that results without thresholding were quantitatively very close and
qualitatively identical.

\section*{Details of the Indices}
\subsection*{Details on symmetry summary statistics}
Partisan bias is defined to be the symmetric difference between the expected number of seats won, as a function of the votes cast.  If we define $E(s|v)$ as the expected number of seats, $s$, given a statewide Democratic vote fraction, $v$, then the partisan bias is defined to be (for a 13 district state) $(E(s|0.5+v_s)-E(s|0.5-v_s))/13$, where $v_s$ is a uniform shift in the overall votes (e.g. \cite{Gelman94}).  Historically vote shifts are computed by uniformly shifting the vote fractions in the marginal distributions, and we adopt this convention in the current work; we have set $v_s=0.05$ (5\%) in all of the presented results.

The Efficiency Gap is an index that was used in the 
decision Whitford Op. and Order, Dkt. 166, Nov. 21, 2016 (see also \cite{McGhee2015}).  It 
quantifies the difference of how many ``wasted votes'' each party 
cast; a larger number means that one party wasted more votes 
than another. 
More precisely, the Efficiency Gap measures the difference of the relative 
  efficiency for the Democrats and Republicans. The efficiency for 
  each party is the sum of the fraction of votes in districts the party 
  loses plus the sum over the percentage points above 50\% in the 
  districts won. The relative efficiency is the efficiency of a given 
  party divided by the sum of percentage of votes obtained by the 
  party in each district. The efficiency gap is the difference between 
  the relative efficiencies of the two parties.\footnote{The original 
    used actual votes, but when the population of each district is 
    equal then the two measures are exactly equivalent. If the actual 
    votes is close to equal then they are almost the same.}

\subsection*{Details of  Gerrymandering Index} 
To compute the Gerrymandering Index, we begin by extracting the mean percentage of
Democratic votes in each
of the thirteen districts when the districts are ordered from most to least Republican (see
Figure~2.  For any given redistricting plan, we take the Democratic votes
for each district when the districts are again ordered from most to
least Republican and consider the differences between the mean and the observed
democratic percentage.  This gives us a set of thirteen numbers on which we consider a two-norm (each number is squared, and these squares are then summed; the sum is then square rooted).

The Gerrymandering Index is 
smallest when all of the ordered Democratic vote percentages are
precisely the mean values. However, this is likely not possible as the
percentages in the different districts are highly correlated. To
understand the range of possible values, we have plotted the complementary cumulative
distribution function of the Gerrymandering Index of our ensemble of
randomly generated reasonable redistricting plans in the main text (see Figure~3). This gives a context in
which to interpret any one score.

To provide an example, we note that for the 2012 votes, the mean percentages for the collection of redistricting plans we generate is 
{
\begin{align*}
  (0.37,0.39,0.41,0.44,0.46,0.48,0.50,0.52,0.55,0.57,0.60,0.63,0.67)\,.
\end{align*}}
If a given redistricting is associated with the sorted winning Democratic 
percentages 
{
\begin{align*}
  (0.36,0.38,0.39,0.40,0.41,0.42,0.43,0.44,0.49,0.52,0.64,0.66,0.7)\,.
\end{align*}}
then the Gerrymandering Index for the redistricting is the square root
of 
{\begin{align*}
  (0.37&-0.36)^2 + (0.39-0.38)^2+ (0.41-0.39)^2 \\&+ (0.44-0.40)^2
                                                                   +(0.46-0.41)^2+(0.48-0.42)^2+(0.50-0.43)^2\\&+(0.52-0.44)^2
+(0.55-0.49)^2+(0.57-0.52)^2+(0.60-0.64)^2\\&+(0.63-0.66)^2+(0.67-0.7)^2=0.0291
\end{align*}}
In summary, in this example the Gerrymandering Index is $\sqrt{0.0291}=0.17$.

\subsection{Details of Representativeness Index} 
 To calculate the Representativeness Index, we first construct a modified histogram of election results that captures how close an election was to flipping a congressional seat.  To do this for a given redistricting plan, we examine the least Republican district in which a Republican won, and the least Democratic district in which a Democrat won.  We then linearly interpolate between these districts and find where the interpolated line intersects with the 50\% line.  For example, in the 2012 election (NC2012 map with 2012 votes), the 9th most Republican district elected a Republican with 53.3\% of the vote, and the fourth most Democratic district won their district with 50.1\% of the vote.  We would then calculate where these two vote counts cross the 50\% line, which will be 
\begin{align}
\frac{50-(100-50.1)}{53.3-(100-50.1)}\approx0.03,
\end{align}
and add this to the number of Democratic seats won to arrive at the
continuous value of 4.03.  This index allows us to construct a
continuous variable that contains information on the number of
Democrats elected, and also demonstrates how much safety there is in the
victory. 

\begin{figure}[ht]
  \centering 
\includegraphics[width=\linewidth]{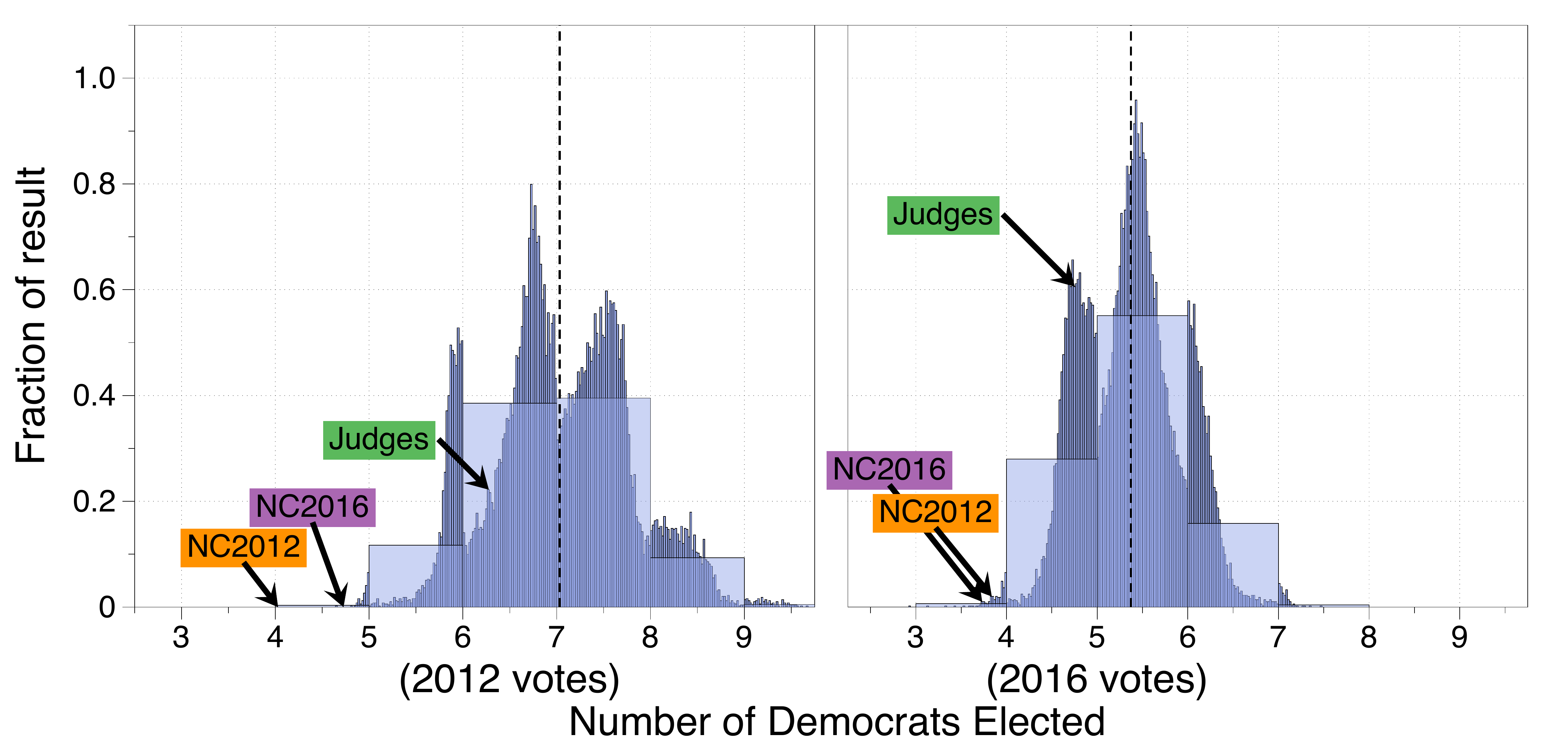}
\caption{For the 2012 votes (left) and the 2016 votes 
  (right), we plot the interpolated winning margins, which give the number of seats won by the Democrats in finer detail.  We determine the mean of this new histogram and display it with the dashed line. The Representativeness Index is defined to be the distance from this mean value.  The histograms presented in Figure~1 are overlaid on this plot for reference. }
  \label{fig:finehist2012}
\end{figure}

Fractional parts close to zero suggest that the most competitive
Democratic race is less likely to go Democratic than the most
competitive Republican race is to go Republican.  On the other hand, fractional
parts close to one suggest that the most competitive Republican race
is less likely to go Republican than the most competitive Democratic
race is to go Democratic.  Instead of simply creating a histogram of
the number of seats won by the Democrats, we construct a histogram of our new
interpolated value in Figure~\ref{fig:finehist2012}.  We define the representativeness as the distance
from the interpolated value to the mean value of this histogram (shown
in the dashed line).  These are the values we report in
Figure~3.  For the 2012 vote data, we
find that the mean interpolated Democratic seats won is 7.01, and the
Judges plan yields a value of 6.28, giving a Representativeness Index of
$|7.01-6.28|=0.73$.  The NC2012 and NC2016 plans both have
Representativeness Indices greater than two.

\section*{Discussion}

We have found evidence that that the NC2012 and NC2016 are heavily gerrymandered: they employ packing and cracking to generate a dramatic jump of partisan vote counts in the ordered district structures.  This jump is significantly larger than those found in the ensemble of redistricting plans which is a signature of gerrymandering.
When summarizing these observations in single number metrics, we consistently find that the NC2012 and NC2016
redistricting plans are outliers, suggesting that (i)
these districts are heavily gerrymandered and (ii) do not represent
the geo-political landscape expressed in a number of elections occurring between 2012 and 2016.
Further analysis reveals the NC2012 and NC2016 plans are locally engineered for partisan benefit.  All of these results are consistent with the findings of \cite{Liu16}.

On the
contrary, the districting plan produced by a bipartisan redistricting
commission of retired judges from the Beyond Gerrymandering project
produced results which are highly typical. The Judges plan is not
gerrymandered, was typically representative of the people's will,
and displayed consistent statistics with nearby redistricting plans.

The ideas presented in this report are generalizable at both the state 
and federal level.  We note that
each state may have different requirements when drawing district
boundaries and so care must be taken when considering the criteria to
be included in the generative procedure.  We hope that the analysis in
this report is utilized across different states and levels of
government to test the viability of districting plans.

\section*{Acknowledgments}
A preliminary version of the results presented in this paper were first reported to the arXiv repository \cite{Bangia17}.
The Duke Math Department, the Information
Initiative at Duke (iID), the PRUV and Data+ undergraduate research 
programs provided financial and material support. Bridget Dou and Sophie 
Guo were integral members of the Quantifying Gerrymandering
Team and have significantly influenced this work.
Mark Thomas, and the Duke Library GIS staff
helped with data extraction. Robert Calderbank, Galen Reeves, Henry
Pfister, Scott de Marchi, and Sayan Mukherjee have provided help throughout this
project. John O'Hale procured the 2016 election Data. Tom Ross, Fritz
Mayer, Land Douglas Elliott, and  B.J. Rudell invited us observe
the Beyond Gerrymandering Project and this work relies on what we learned there.

\bibliography{pnas}
{\bibliographystyle{pnas-new}

\newpage

\appendix

The appendix is organized as follows: (1) we begin by detailing the utilized data, including the HB92 criteria, voting data, the enacted NC2012 and NC2016 districting plans, and the Judges districting plan; (2) we then provide evidence that our MCMC algorithm is properly sampling the space of redistricting plans on the distribution described in the main text; (3) we demonstrate that our main results are insensitive to changing the probability distribution function in a variety of ways; (4) we provide error analyses for how our results might change due to resampling and due to splitting VTDs to achieve zero population; (5) we give details on the properties of our sampled redistricting plans.

\section{Details on Data (Examined Districting Plans of Interest; Redistricting Criteria; Voting Data)}

\subsection{Information on the three considered redistricting plans}
Maps for the NC2012, NC2016, and Judges districting plans are shown in Figure~\ref{fig:DistrictingsConsidered}.  The Democratic vote fractions for each district are given in Table~\ref{tab:table1}; the vote fractions are given for the U.S. congressional elections in 2012 and 2016.  In the table, vote fractions are adjacent to numbers in
parentheses that give the numerical label of the individual districts as
identified in the maps in Figure~\ref{fig:DistrictingsConsidered}. The last two columns contain the mean values of percentage of Democratic votes for the ensemble of redistricting plans.

\begin{figure}[ht]\centering 
\includegraphics[scale=0.25]{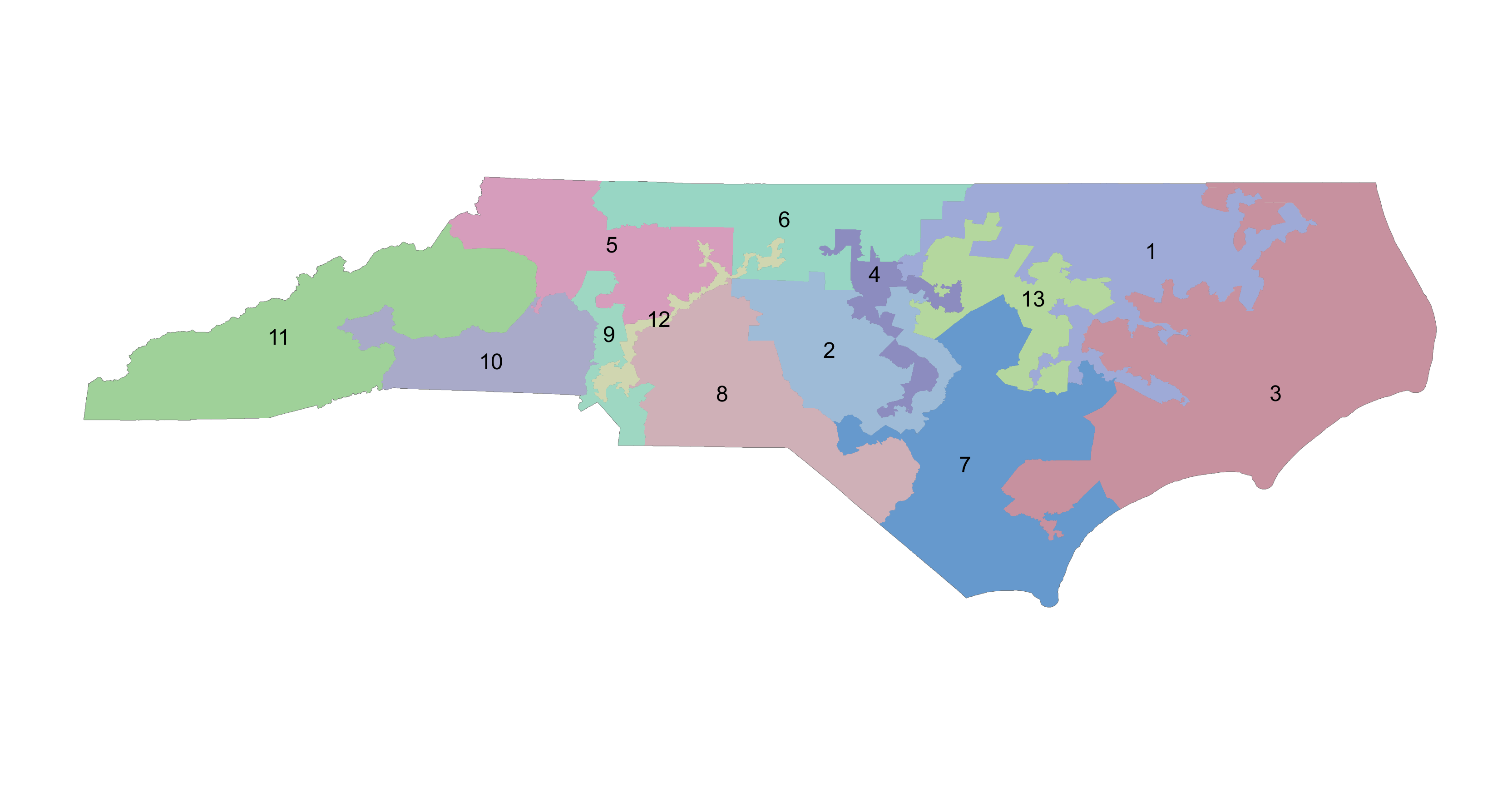} 
\includegraphics[scale=0.25]{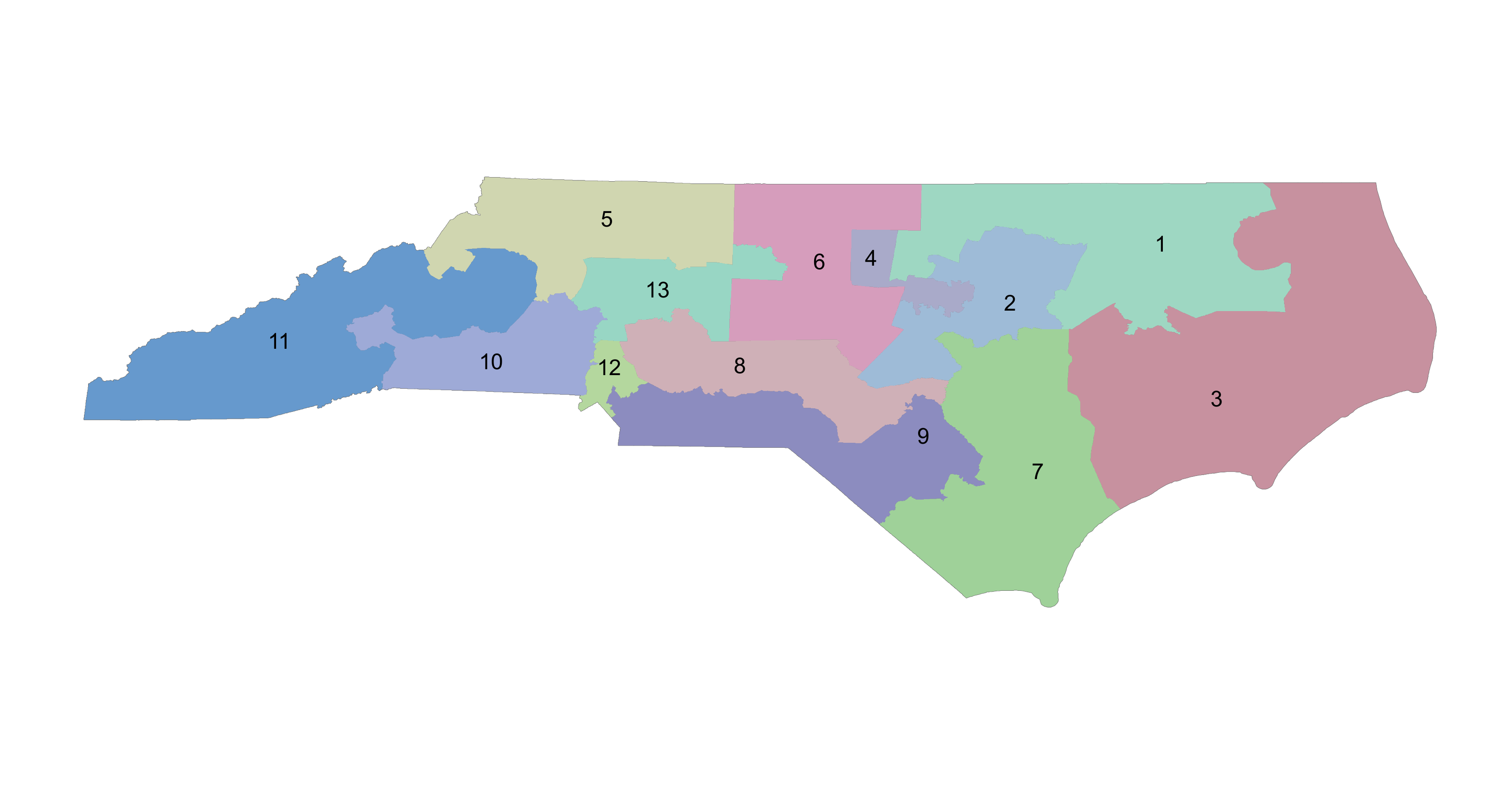} 
\includegraphics[scale=0.25]{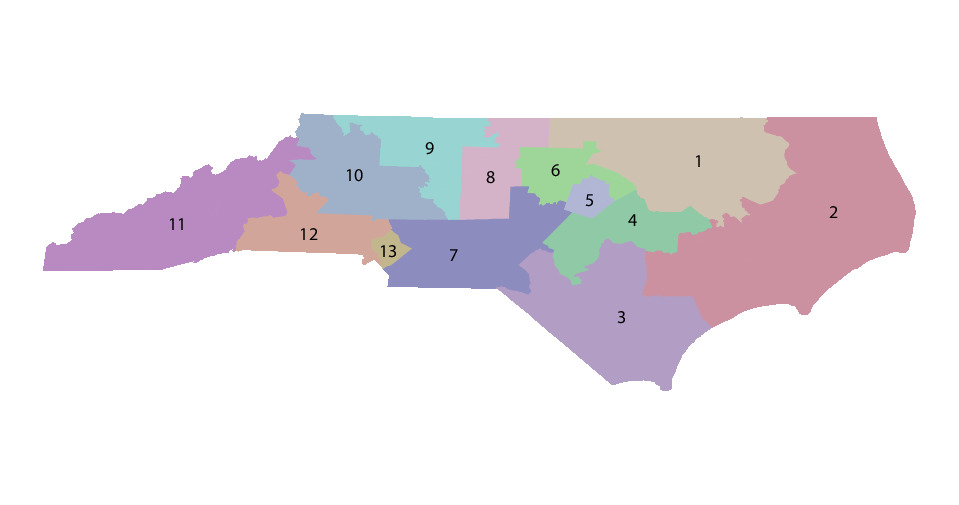} 
\caption{Map for NC2012 (top left), NC2016 (top right) and Judges (bottom). Numbers correspond to labels in Table~\ref{tab:table1}.} 
\label{fig:DistrictingsConsidered}
\end{figure}

The three most Democratic districts (labeled 1, 4 and 12
in both the NC2012 and NC2016 plans) have significantly more Democratic
votes than the expected average for these districts.  Districts 9 and 13 both show evidence of having less Democrats
than one would expect from their rankings. These conclusions are consistent across the 2012 and 2016 votes.

The raw data used to produce Figure~1 of the main text is given
in Table~\ref{table:winnerCount}. It underscores how atypical the
results produced by the NC2012 and NC2016 redistricting plans are. If one
is ready to accept four seats for Democrats in the  2012  vote then one should
equally accept nine. Similarly in the 2016 votes, if one accepts three seats for Democrats
as a legitimate outcome then one should also be willing to accept seven
seats. None of these results are particularly representative of the
votes cast.

\begin{table}[h!]
  \centering
  \begin{tabular}{r||l|l||l|l||l|l||l|l||}
    \multicolumn{1}{c||}{ }&\multicolumn{2}{|c||}{NC2012}
    &\multicolumn{2}{|c||}{NC2016}&\multicolumn{2}{|c||}{Judges}&\multicolumn{2}{|c||}{Mean}\\ 
    \cline{2-9}
    \multicolumn{1}{c||}{Rank}& 2012 & 2016  & 2012 & 2016  & 2012 & 2016  & 2012 & 2016  \\
    \hline
    1 & 37.5\pp (3)  & 34.2\pp (3)  & 38.7\pp (3) & 32.8\pp (3)  &
                                                                   35.5\pp
                                                                   (10)
                                     & 28.9\pp (10)
                                        &  37.0\pp   & 30.6\pp\\
    2  & 39.0\pp (6) & 34.6\pp (11)  & 42.5\pp (10) & 35.8\pp (11)  &40.0\pp (2)
                                     & 33.6\pp (2) & 39.1\pp &33.0\pp\\
    3 & 42.4\pp (5) & 36.2\pp (7)  & 43.7\pp (6) & 36.8\pp (10)
                              &42.6\pp (12)  & 36.3\pp (7)
                                             &41.0\pp &35.3\pp  \\
    4 & 42.5\pp (11)  & 36.6\pp (8)  & 43.9\pp (11) &  39.0\pp (7)
                              &42.7\pp (7)  & 37.6\pp (12)  &43.7\pp &38.5\pp \\
    5 & 42.6\pp (2)  & 37.4\pp (10) & 44.0\pp (2)  &  40.7\pp (6)
                              &44.5\pp (9)  &40.0\pp (9) &46.4\pp &40.6\pp  \\
    6 & 43.1\pp (10) & 38.9\pp (5) & 45.1\pp (5)   &  41.2\pp (8) &
                                                                    48.5\pp (8) &41.9\pp (3) &48.4\pp &42.2\pp   \\
    7 & 43.5\pp (13) & 40.8\pp (6)  &46.3\pp (13)  &  41.6\pp(5) &
                                                                   48.8\pp
                                                                   (11)
                                     & 42.7\pp (11) &50.2\pp & 44.3\pp\\
    8 & 46.2\pp (8) & 41.2\pp (2)  & 47.3\pp (8) &41.8\pp (9) &
                                                                50.5\pp
                                                                (4) &
                                                                      45.7\pp (4) & 52.3\pp&47.7\pp\\
    9 & 46.7\pp (9) & 44.0\pp (9)  & 49.4\pp (9) & 43.3\pp (2)  &
                                                                  57.0\pp
                                                                  (3)
                                     &48.1\pp (8) & 55.1\pp&51.2\pp  \\
    10 & 50.1\pp (7)  & 45.8\pp (13)   &  51.6\pp (7) &  43.9\pp (13)
                              & 57.5\pp (5) & 55.9\pp (1) & 57.2\pp&54.6\pp  \\
    11 & 74.4\pp (4)   & 71.5\pp (1)  & 66.1\pp (4) & 66.6\pp (12)  &
                                                                      59.2\pp
                                                                      (1)
                                     & 59.7\pp (5) &59.5\pp & 57.5\pp \\
    12 & 76.0\pp (1) & 73.0\pp (4)  & 69.8\pp (12) & 68.2\pp (4)  &
                                                                    64.6\pp
                                                                    (6)
                                     &63.3\pp (13) & 62.6\pp&61.4\pp  \\
    13 &79.3\pp (12)  &75.3\pp (12)   & 70.9\pp (1)  & 70.3\pp (1)  &
                                                                      66.0\pp
                                                                      (13)
                                     &65.3\pp (6) &67.5\pp &65.1\pp
    \\
  \end{tabular}
\vspace{1em}
  \caption{Percentage of Democratic votes in each district when
    districts are ranked from most Republican to most
    Democratic. Numbers in parentheses give label of actual district
    using the numbering convention from maps in Figure~\ref{fig:DistrictingsConsidered}. 
        }
  \label{tab:table1}
\end{table}

\begin{table}[h!]
  \centering
  \begin{tabular}{r|r|r|r|r|r|r|r|r|r|r|}
 &  \multicolumn{10}{|c|}{\# of Democratic Winners}  \\
\cline{2-11}
 & 1 & 2 &  3& 4 & 5 & 6 & 7& 8 & 9 & 10\\ 
\hline
2012 Votes& 0 & 0 & 0& 89& 2875 & 9455 &9690& 2288 & 121 & 0\\
2016 Votes&  0 &1 & 162 & 6861& 13510 & 3881 & 103 & 0  & 0 & 0\\
  \end{tabular}
\vspace{1em}
  \caption{Among the 24,518 random redistricting plans generated, the
    number which produced the indicated number of Democratic seats in
    the Congressional Delegation is presented. }
\label{table:winnerCount}
\end{table}

\subsection{Redistricting Criteria}
The Judges districting plan was established in the Beyond
Gerrymandering project\footnote{ For more information see
  \texttt{https://sites.duke.edu/polis/projects/beyond-gerrymandering/}}
   was a collaboration between UNC system
President Emeritus and Davidson College President Emeritus Thomas
W. Ross, Common Cause, and the POLIS center at the Sanford School at
Duke University. The project's goal was to educate the public on how
an independent, impartial redistricting process would work. The
project  formed
an independent redistricting commission made up of ten retired jurists -- five Democrat and five Republican. 

The
commission used
strong and clear criteria to create a new North Carolina congressional
map based on NC House Bill 92 (HB92) from the 2015 legislative season, which states
\begin{itemize}
\item \S120-4.52(f): Districts must be contiguous; areas that meet only at points are not considered to be contiguous.
\item \S120-4.52(c): Districts should have close to equal populations, with deviations from the ideal population division within 0.1\%.
\item  \S120-4.52(g): Districts should be reasonably compact, with (1) the maximum length and width of any given district being as close to equal as possible and (2) the total perimeter of all districts being as small as possible.
\item \S120-4.52(e): Counties will be split infrequently and into as few districts as possible.  The division of Voting Tabulation Districts (VTDs) will also be minimized.
\item \S120-4.52(d): Redistricting plans should comply with pre-existing federal and North Carolina state law, such as the Voting Rights Act (VRA) of 1965.
\item \S120-4.52(h): Districts shall not be drawn with the use of (1) political affiliations of registered voters, (2) previous election results, or (3) demographic information other than population.  An exception may be made only when adhering to federal law (such as the Voting Rights Act (VRA)).
\end{itemize}

 All federal rules related to the Voting Rights Act were followed but no
political data, election results or incumbents addresses were
considered when creating new districts. The commission met twice over
the summer of 2016 to deliberate and draw maps. The maps resulting from this
simulated redistricting commission were released in August 2016. The
Judges agreed on a redistricting at the level of Voting Tabulation
Districts (VTD). This coarser redistricting was refined at the level of
census blocks to achieve districts with less than 0.1\% population
deviation. The original VTD based maps are used in our study and are denoted Judges; our results are insensitive to splitting VTDs to zero population (for a careful study, see below in Section~D.\ref{subsec:zeropopdiv}).

Although unratified, HB92 was passed in the House, which provides some credence to using it as a guide to draw fair redistricting plans.  In contrast, the criteria adopted in the drawing of the 2016 congressional districts contains a ``Partisan Advantage" clause which seeks to predetermine the number of elected officials from each party\footnote{The 2016 Contingent Congressional Plan Committee Adopted Criteria may be found here: http://www.ncleg.net/GIS/Download/ReferenceDocs/2016/CCP16\_Adopted\_Criteria.pdf}; with the balance of power pre-selected by this criteria, it would be impossible determine any sense of the statistical range in the balance of power.  Because we do not wish to use partisan data to draw maps, we utilize the criteria from the non-partisan HB92.

\subsection{Data sources and extraction}
VTD geographic data was taken from the NCGA website \cite{NCGeo}
and the United States Census Bureau website \cite{shapeFilesCongress} , which
provide for each VTD its area, population count of the 2010 census,
the county in which the VTD lies, its shape and location.
Perimeter lengths shared by VTDs were extracted in ArcMap from this
data.  Minority voting age population was found on the NCGA website
using 2010 census data \cite{NCMino}.

Data for the vote counts in each VTD in each election was taken from the NCSBE Public data \cite{VotingData}.
For the 2016 election,
VTD data was reported for precincts rather than VTDs, but rather for each precinct;
2447 of the precincts are VTDs, meaning that we have data for the
majority of the 2692 VTDs.  However 172 precincts contain multiple
VTDs, 66 VTDs were reported with split data, and 7 VTDs were reported
with complex relationships.  To extrapolate VTD data on those
contained in the 172 precincts containing multiple VTDs, we split the
votes for a precinct among the VTDs it contained proportional to the
population of each VTD.  For the 66 split VTDs, VTDs were comprised of multiple precincts all contained with in a certain VTD, so we simply added up the votes among the precincts that were contained in each VTD -- there was no extrapolation for these VTDs and these results are precise. For the VTDs with complex relationships, we divided up the votes using estimates
based on the geography and population of the VTDs.  We note that
roughly 10\% of the population lies in the VTDs with imperfect data, and that we
do not expect significant deviation in our results based on the above
approximations.

In using 2012 and 2016 data we  have only used presidential election year data. Unfortunately, the 
2014 U.S. congressional election in North Carolina contained an 
unopposed race which prevents the support for both parties being 
expressed in the VTDs contained in that district. In reference 
\cite{QuantifyingGerrymandering}, the missing votes were replaced with 
votes from the Senate race. However, since we had two full elections, 
namely 2012 and 2016, 
which needed little to no alterations, we chose not to include the 
2014 U.S. congressional votes in our study.

\section{Evidence of robust sampling}
There is a possible pitfall of using simulated annealing: as the districting plan moves along the random walk it may
become trapped in a local minimum from which it is unlikely to escape.  This may be because the system has cooled too quickly, keeping it trapped in a local region, or it
may be because the likelihood of finding a path out of one local
region of redistricting plans and into another is small.
To test that the MCMC algorithm has sufficiently sampled the space, we consider three tests:
\begin{enumerate}
\item we consider three different initial conditions,
\item we double the simulated annealing time, and
\item we increase the length of the MCMC chain for a longer run that includes more samples.
\end{enumerate}
When comparing the main results to the results from the above tests, we will examine both the histogram and marginal distributions presented in Figures~1~and~3 in the main text.  We will have evidence for a robust sampling if the histograms and box plots do not significantly change.  

In addition to the above tests, we
have animated our algorithm and have found that districts may travel
from one end of the state to another; such motion suggests that our sampler is not trapped in a local well, and it is reasonable to hypothesize
that as districts exchange locations, they lose information on past
configurations.

\subsection{Independence of initial condition}

The samples from the random walks presented in the main text begin with the Judges districting plan.  To test for the effect on initial conditions, we also examine using initial conditions with the NC2012 and NC2016 districting plans.  Altering the initial condition has only a small effect on the ensemble of election results; the effects are displayed in Figure~\ref{fig:deltaInitialConditions}.   The initial
condition for the NC2012 redistricting has a 15\% chance of electing five
Democrats rather than the 12\% chance we have seen before.  We note
that these are exploratory runs, with 785 and 835 redistricting plans for the NC2012 and NC2016 initial conditions.  These sample sizes are robust enough to provide a general
trend but are subject to statistical variations. Hence the small sample sizes are a possible and likely culprit of these variations. 

\begin{figure}[ht]
  \centering 
\includegraphics[width=.4\linewidth]{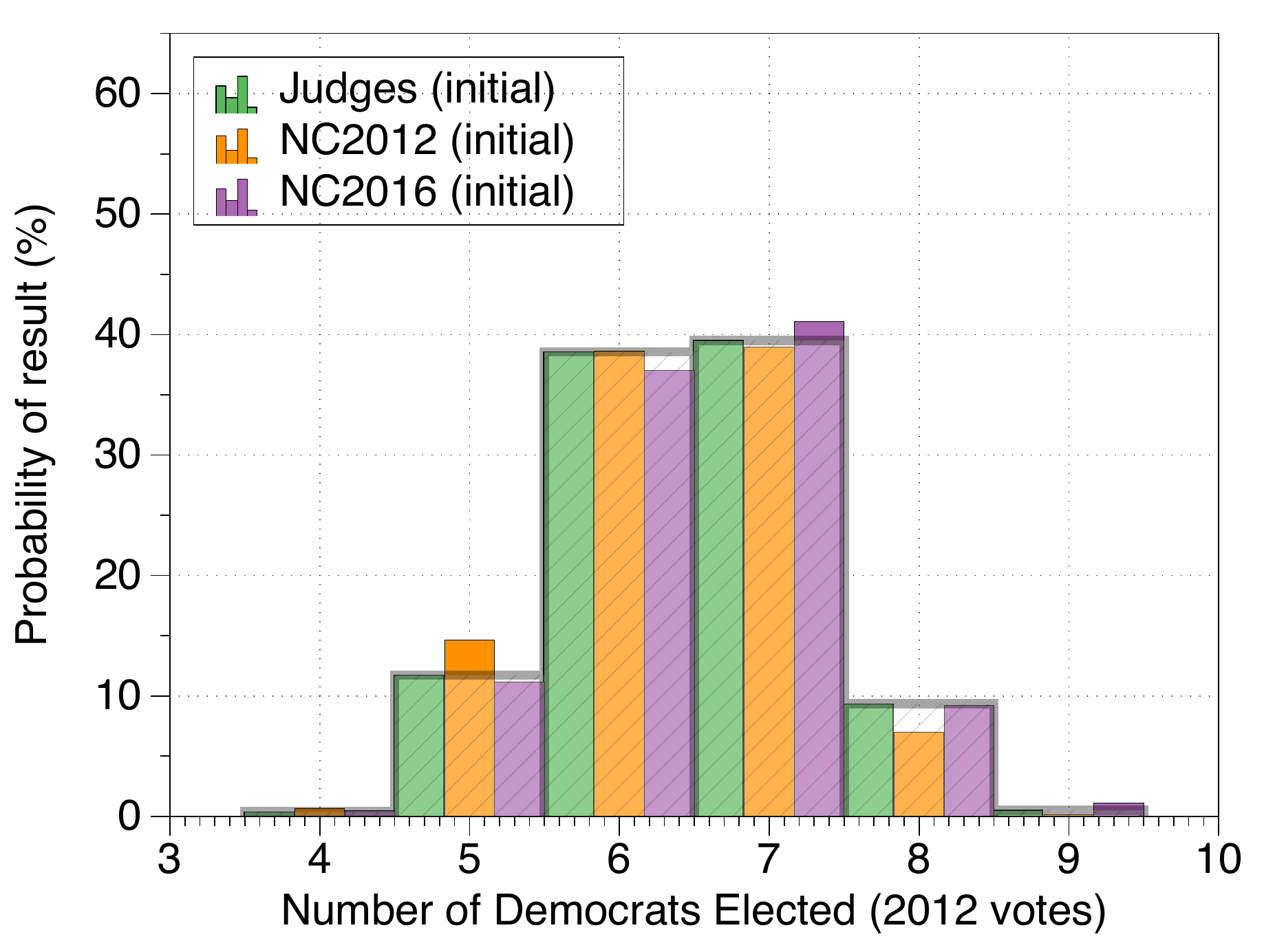}
\includegraphics[width=.4\linewidth]{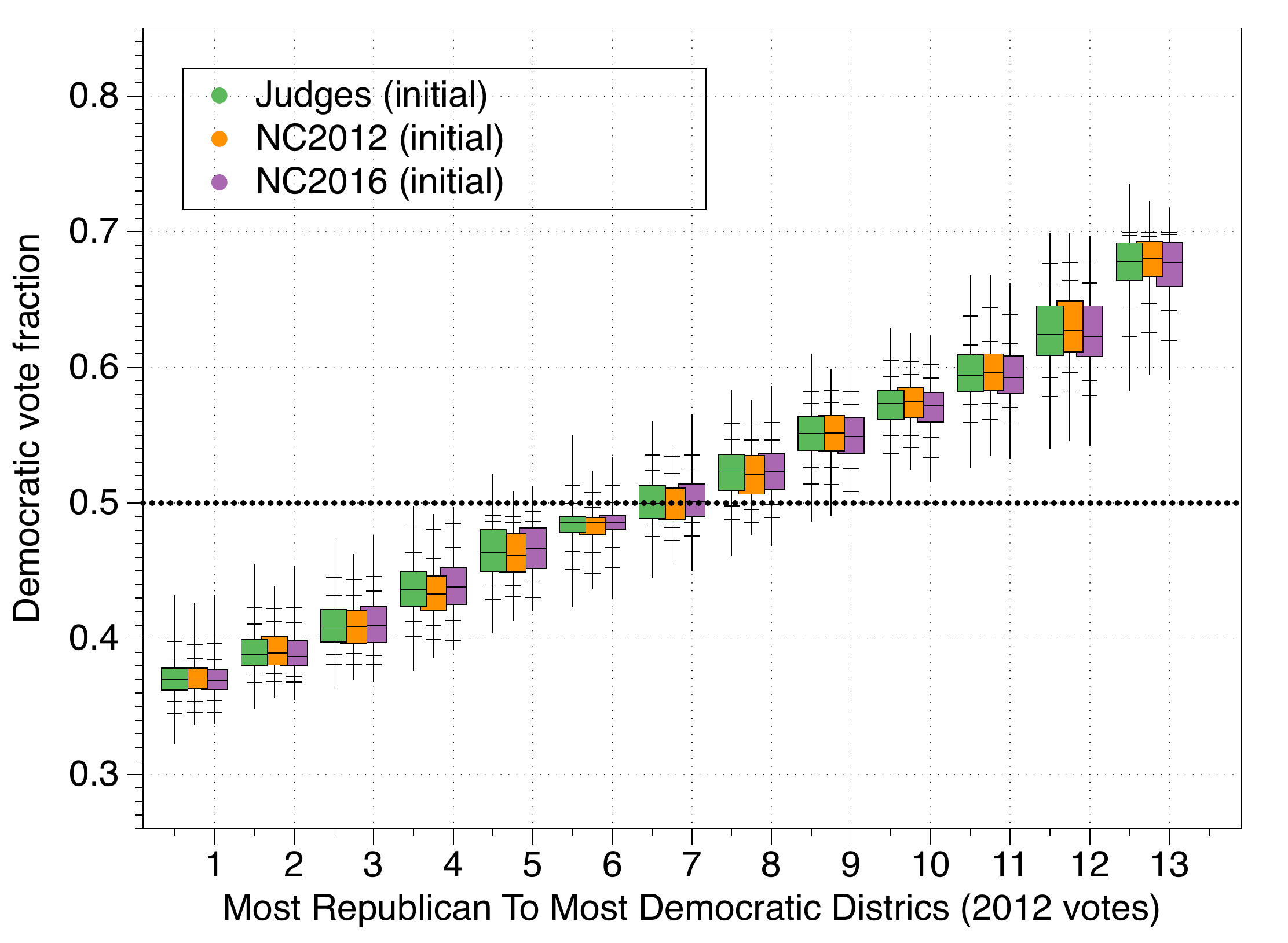}\\
   \caption{We display the probability distribution of 
     elected Democrats with respect to initial conditions (left) 
     We display our standard box-plots for the three initial conditions (right; vertical line length is the full range of possibilities, 5\% of data is outside the outer lines, 20\% of the data is outside the inner lines, and 50\% of the data is outside of the boxes).}
  \label{fig:deltaInitialConditions}
\end{figure}

\subsection{Independence of simulated annealing parameters}
We double the relaxation time of the simulated annealing process to see if it has an effect on the simulated election statistics: Instead of remaining hot ($\beta=0$) for $40,000$ steps, cooling linearly for $60,\!000$ steps, and remaining cold ($\beta=1$) for $20,\!000$ steps, we instead remain hot for $80,\!000$ steps, cool linearly for $120,\!000$ steps, and remain cold for $40,\!000$ steps.  2,411 redistricting plans for the increased cooling times are presented in Figure~\ref{fig:SAdouble}

\begin{figure}[ht]
  \centering 
\includegraphics[width=.4\linewidth]{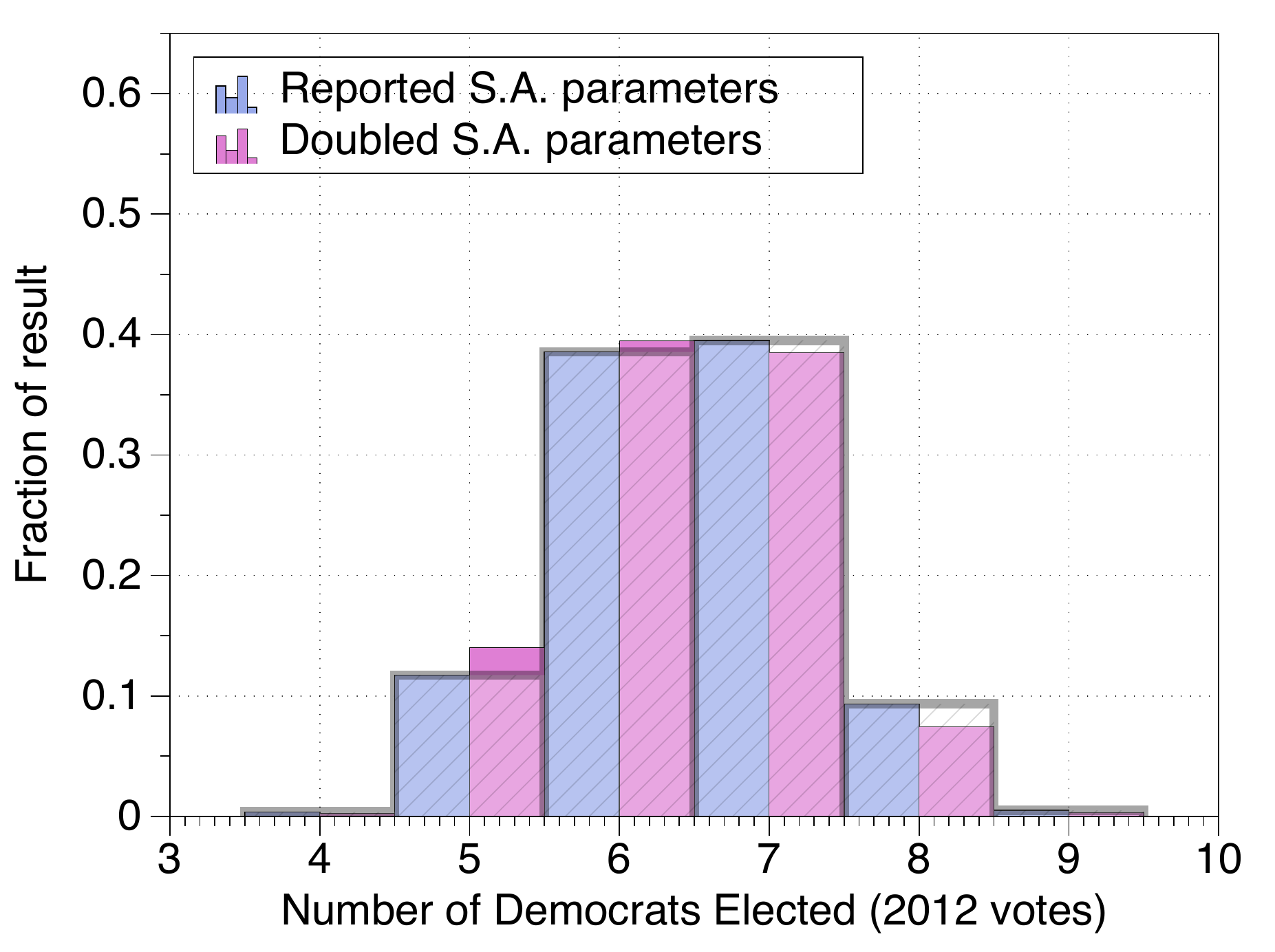}
\includegraphics[width=.4\linewidth]{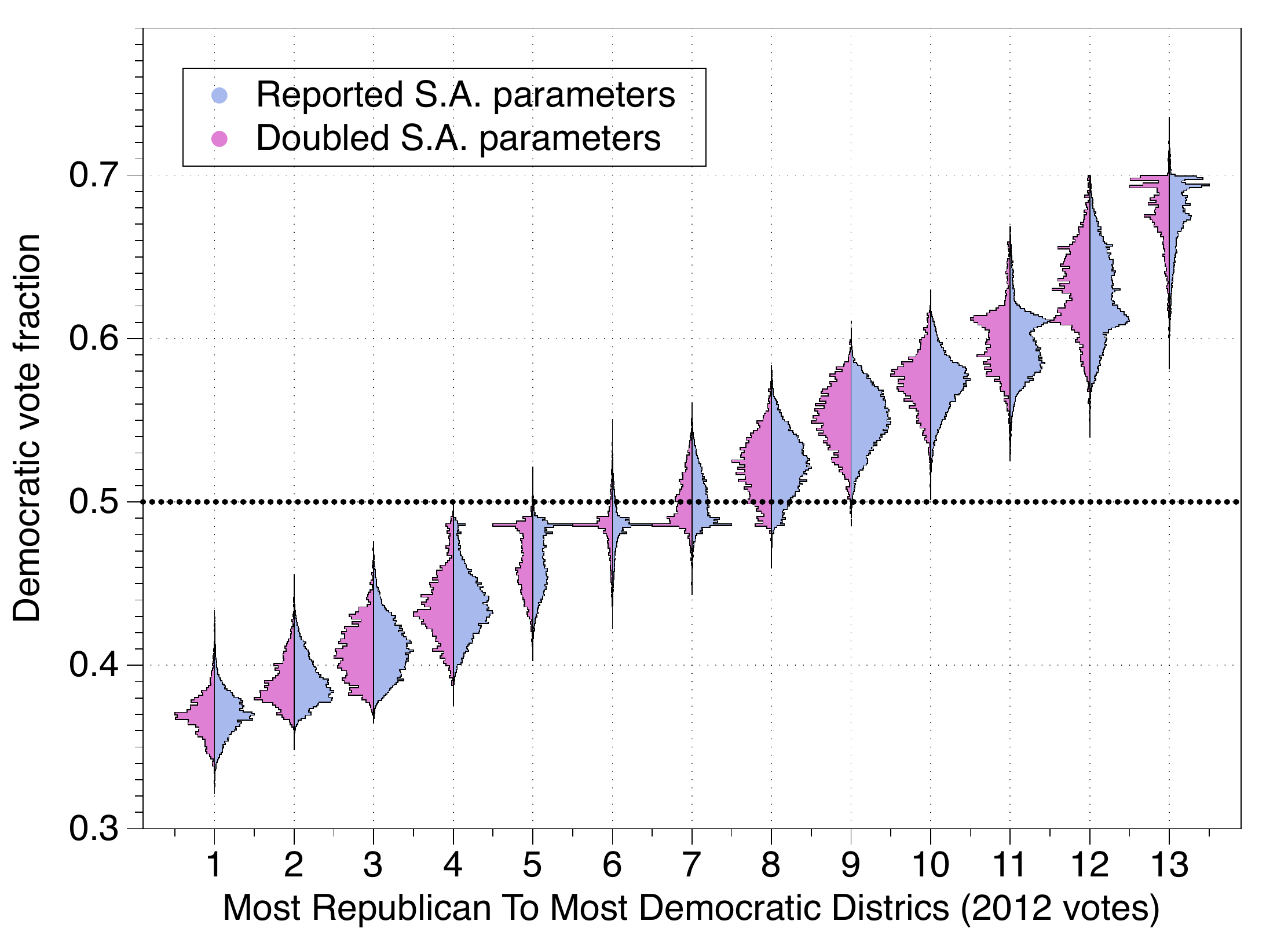}\\
   \caption{We display the probability distribution of 
     elected Democrats with respect the original versus doubled simulated annealing parameters 
     (left).  The histogram presented in the primary text overlays this image with the gray shaded histogram for comparison. 
     We display marginal distributions, rather than box plots, to further display the similarity of the sampled structure (right).}
  \label{fig:SAdouble}
\end{figure}

\subsection{Considering more samples}
The above tests on initial conditions and simulated annealing parameters provide evidence that we have properly sampled the probability distribution of redistricting plans.  To strengthen this claim,
we also continue to allow the algorithm to sample the space until we
have sampled roughly 120 thousand acceptable redistricting plans as defined
by the original thresholding criteria -- this means that each chain in the MCMC algorithm samples nearly 5 times as many districts as the 24,518 plans used in the primary results.  We then compare the results of
the elections along with the box plots and histogram plots.  We find
that there is negligible change in the distribution of outcome both
for the overall number of elected representatives and for each ordered
district (from most to least republican).  We display our results in
Figure~\ref{fig:moreSamples}.  The stability of these results together
with those presented in the previous two tests provides
robust evidence that we have correctly recovered the underlying probability distribution of redistricting plans.

\begin{figure}[ht]
  \centering 
\includegraphics[width=.4\linewidth]{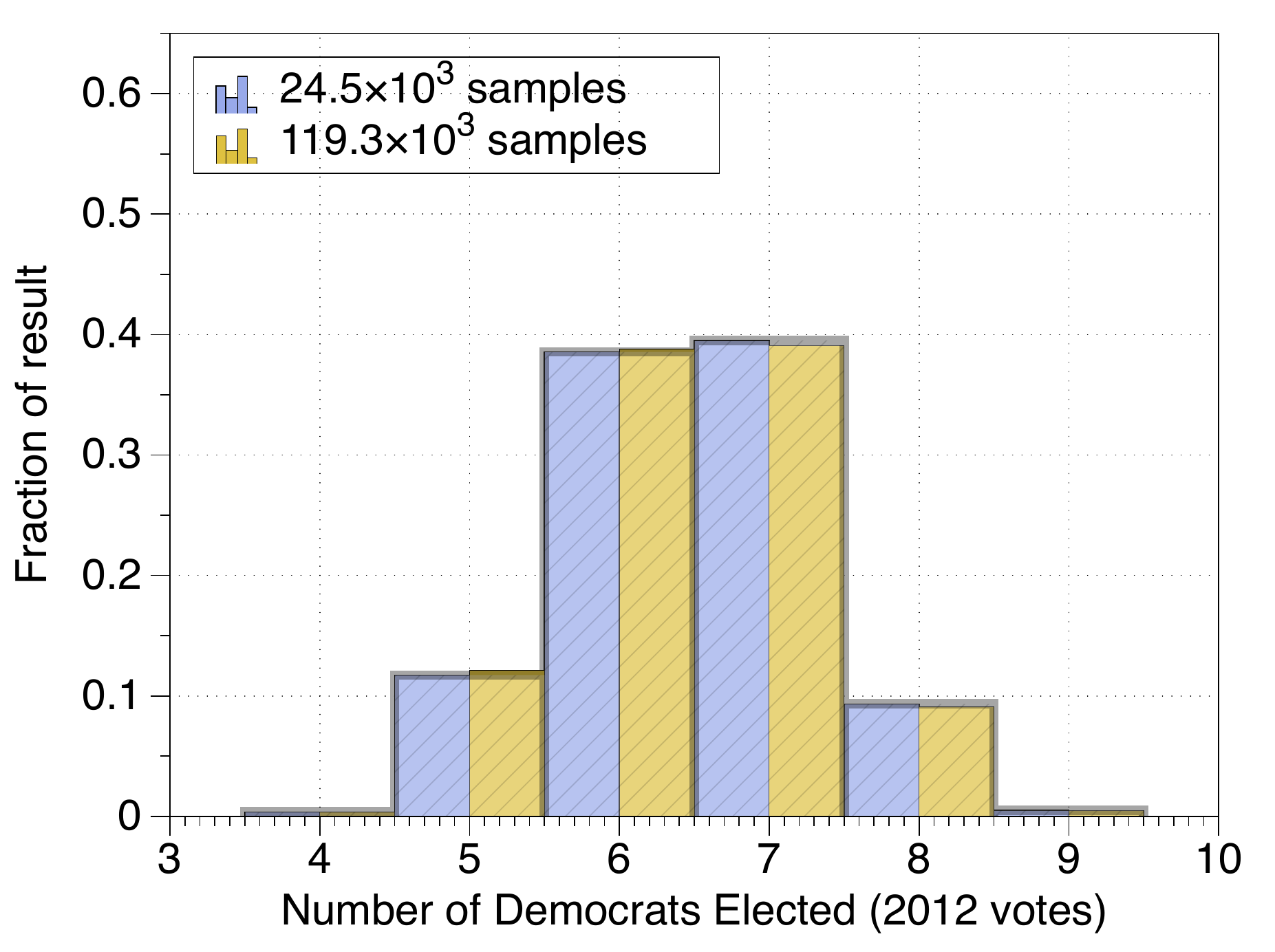}
\includegraphics[width=.4\linewidth]{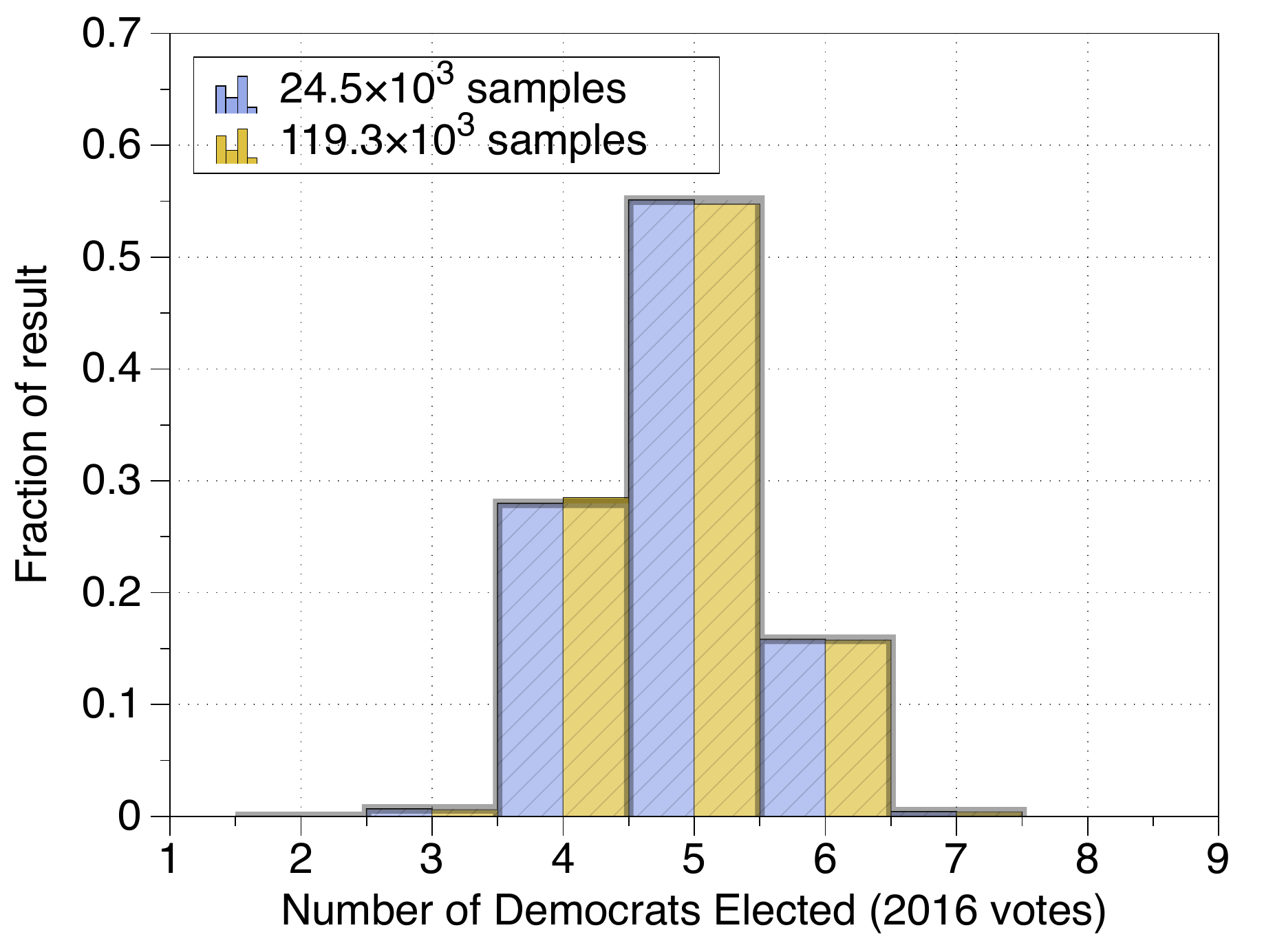}\\
\includegraphics[width=.4\linewidth]{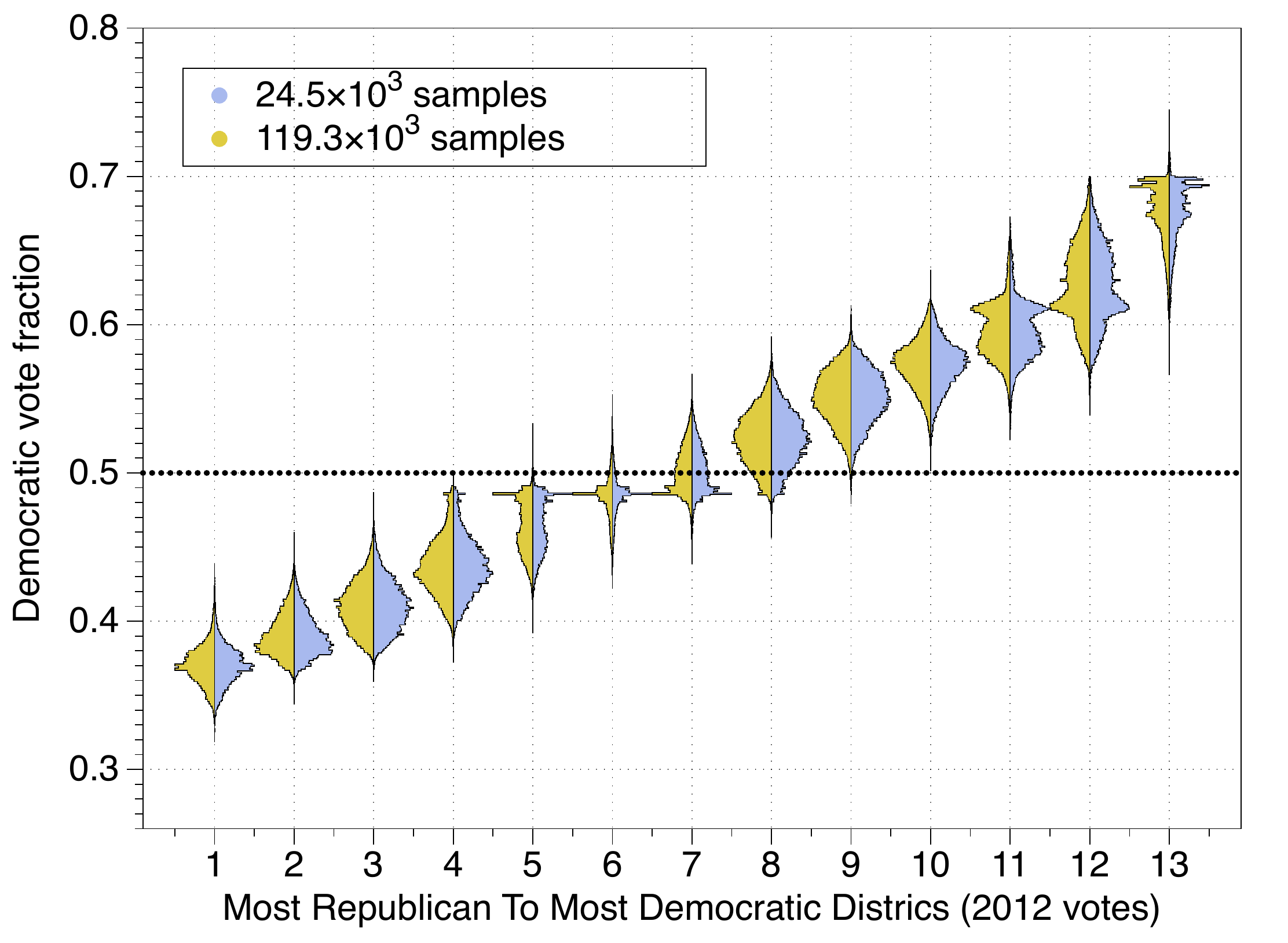}
\includegraphics[width=.4\linewidth]{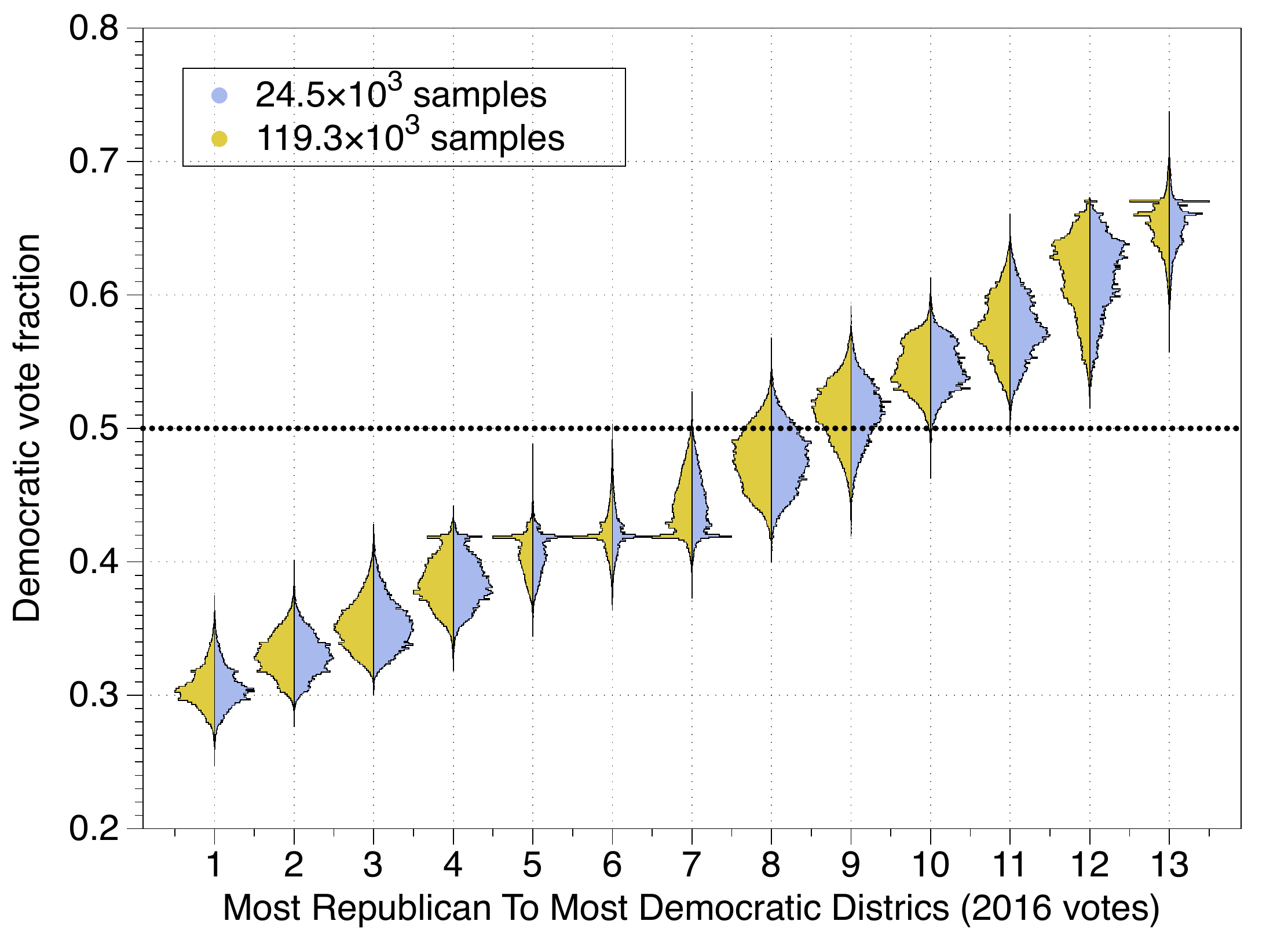}
   \caption{ We extend the samples from the main text by allowing the sampling algorithm to continue until we have sampled roughly 120 thousand districts that fall below the threshold.  We find almost no difference between the distributions in the original and extended samples.}
  \label{fig:moreSamples}
\end{figure}

\section{Insensitivity to the choice of distribution}
We have sampled over a fixed a probability distribution function.  As mentioned in the main text it is unclear that there is a single `correct' distribution from which to sample.  However, if the results are robust across a range of probability distributions, we may conclude that the precise choice of distributions is irrelevant.  To investigate this possibility, we perform the following tests:
\begin{enumerate}
\item change the threshold values,
\item change the weights in the score function, 
\item prioritize the number of counties split over district compactness, and
\item consider a different compactness energy.
\end{enumerate}
In all of these tests we find either negligible changes to the histograms and box plots, or qualitatively similar results.

\subsection{Varying thresholds}
\label{subsec:vary-thresholds}
Achieving a 0.1\% population deviation is the only statute of HB92 that we violate -- we instead consider all plans below a 1\% population deviation.  The Judges original redistricting plans in the `Beyond Gerrymandering' project each had districts that were slightly over 1\% population deviation;  VTDs were split to achieve a 0.1\% population deviation, and this splitting did not impact the election results.  In the current section we demonstrate that changing the threshold at lower than 1\% population deviations does not effect our results (in Section~D.\ref{subsec:zeropopdiv} below, we demonstrate bounds to election result changes in our ensemble when zeroing population).    To test that our modified threshold will not impact our results, we change the population threshold to 0.75\% and 0.5\%.  The results are shown in Figure~\ref{fig:deltaPopThresh}, for which we have used the 2012 US congressional voting data.  We find that tightening the population threshold has negligible impact on the number of Democrats elected, and that the variation in the histogram box-plots is barely perceptible.  In the 0.5\% population deviation threshold plots, we have discarded over half of our utilized samples and we still do not see any significant changes.

\begin{figure}[ht]
  \centering 
\includegraphics[width=.4\linewidth]{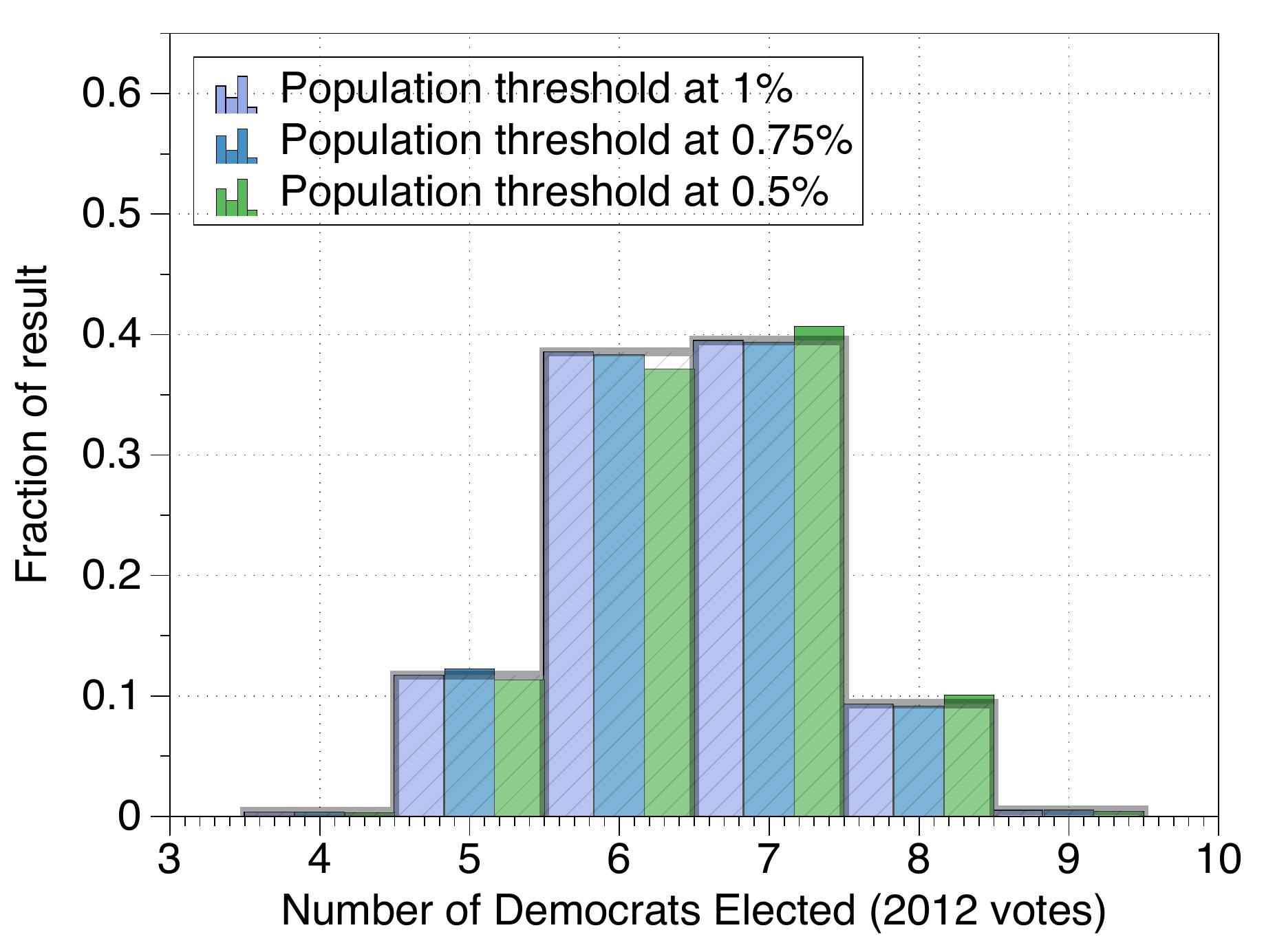}
\includegraphics[width=.4\linewidth]{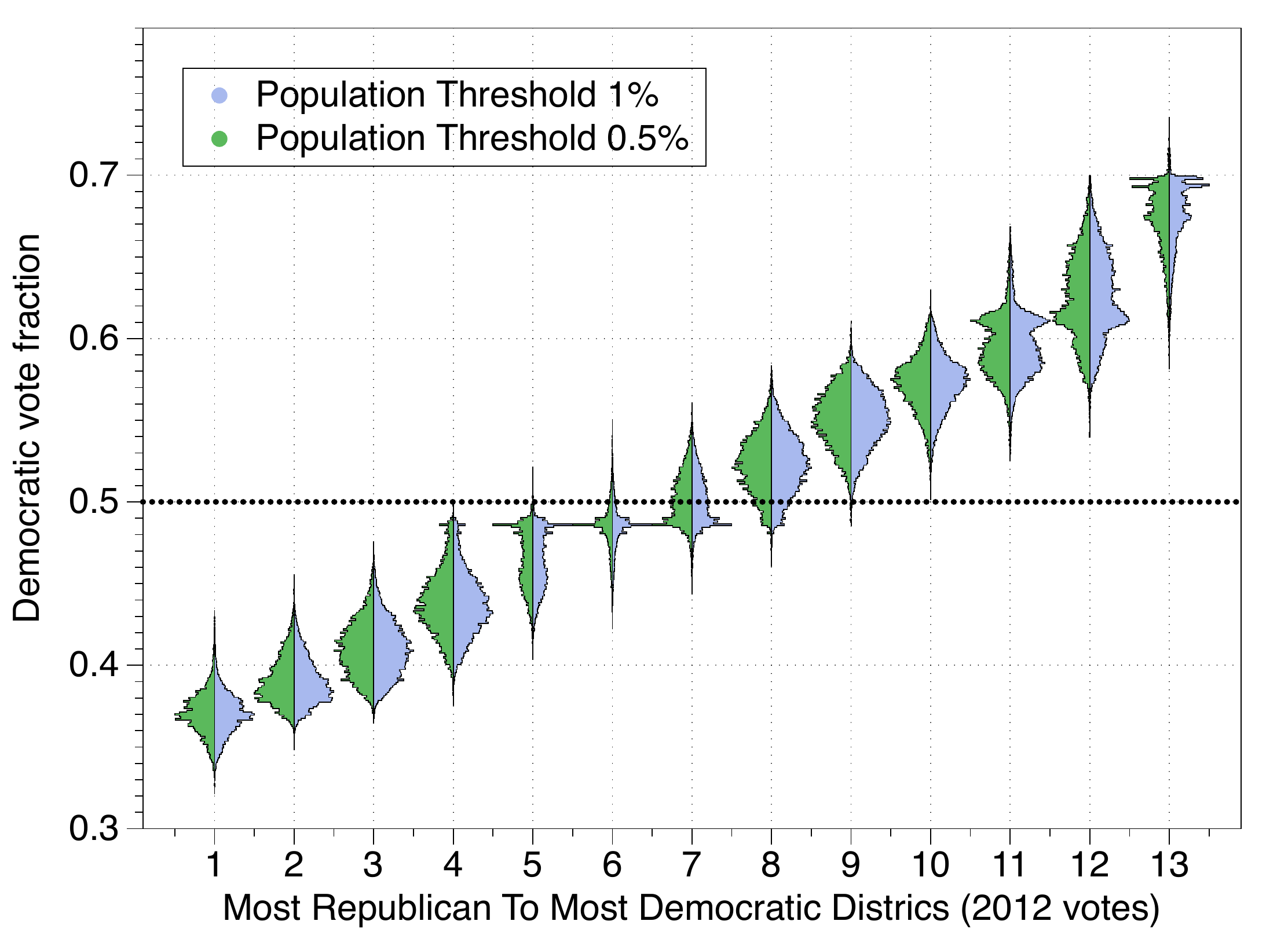}
   \caption{ We display changes of the distribution of election results with changes to the population threshold (left).  The histogram formed with 1\% population deviation overlays this image with the gray shaded histogram.  We display changes to the histogram of the box-plot when comparing 1\% population deviation threshold with 0.5\% (right).}
  \label{fig:deltaPopThresh}
\end{figure}

\begin{figure}[ht]
  \centering 
\includegraphics[width=.4\linewidth]{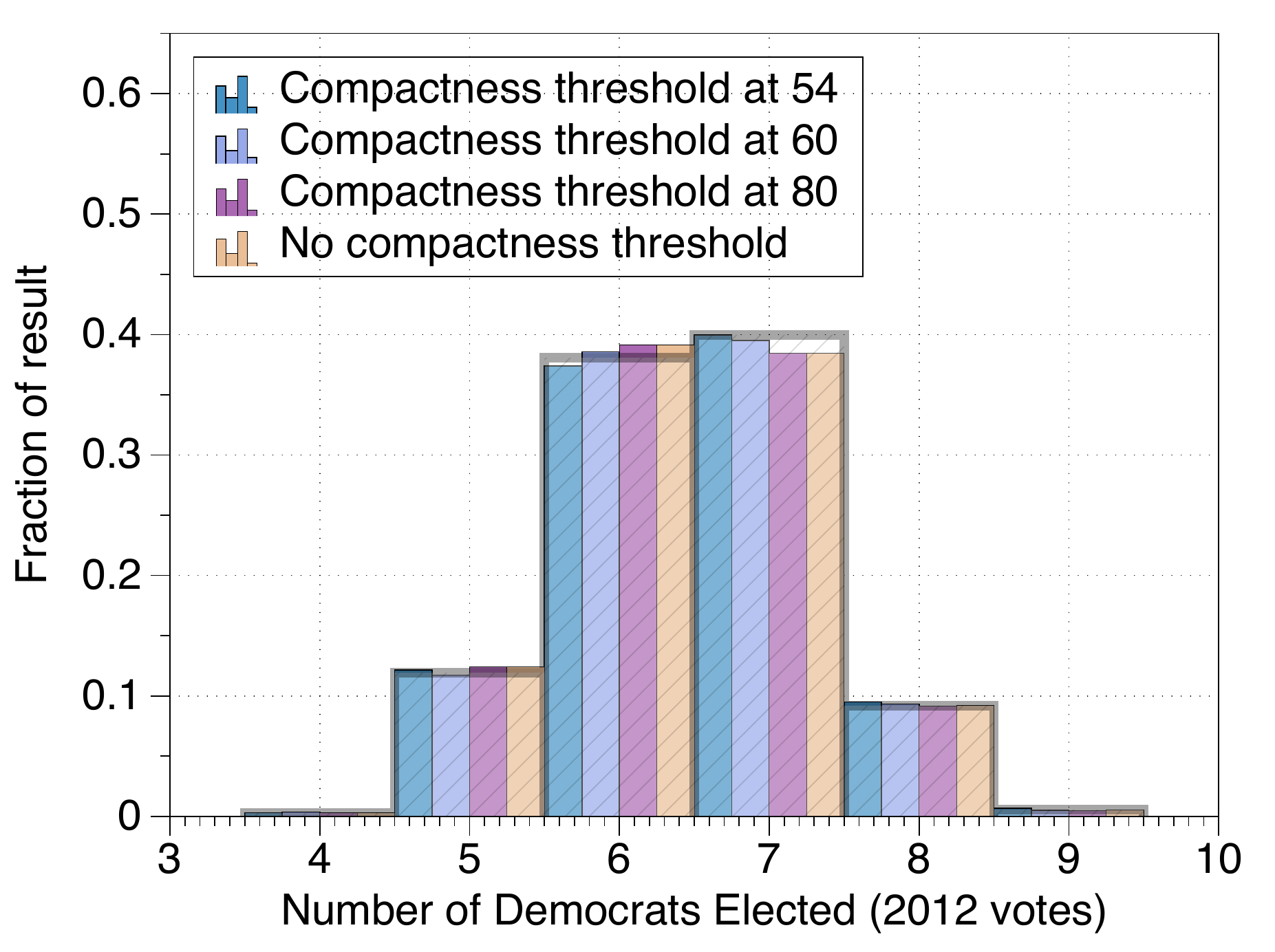}
\includegraphics[width=.4\linewidth]{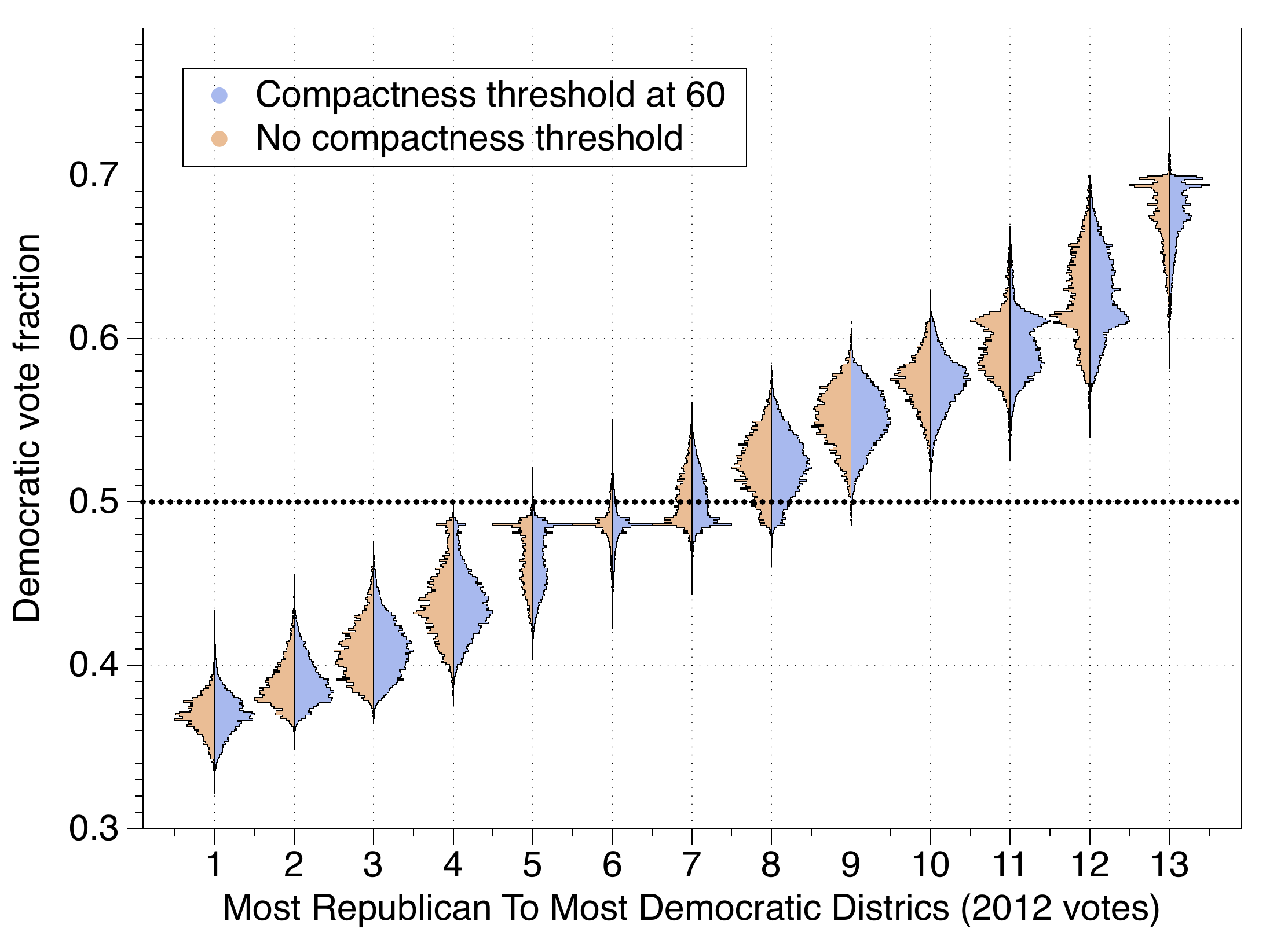}
   \caption{ We display changes of the distribution of election results with changes to the compactness threshold (left).  The histogram formed with a maximum of 60 for the isoparametric ratio overlays this image with the gray shaded histogram.  We display changes to the histogram of the box-plot when comparing a maximum of 60 in the isoparametric ratio without any thresholding on compactness (right).}
  \label{fig:deltaCompThresh}
\end{figure}

In terms of the compactness threshold, there is no corresponding law to dictate a definite value in the choice of threshold.  The NC2016 districts have a maximum isoparametric ratio of around 80, and the NC2012 districts have a maximum of over 400.  HB92 mentions that when two districting plans are compared, the one with more compact districts should be preferred.  The Judges redistricting has a district with maximum isoparametric ratio of around 54.  To test the effect of setting different compactness thresholds, we repeat our analysis with multiple compactness thresholds of 54, 80 and no threshold for all districts within a redistricting.  We note that having no threshold does not mean that we have arbitrarily large compactness values. This is because of the cooling process in the simulated annealing algorithm and the fact that we continue to penalize large compactness scores.  When considering no thresholding, we have an average maximum isoparametric ratio of around 75 and rarely see redistricting plans with maximal ratio larger than 120.  Relaxing or tightening the compactness threshold minimally changes the election results as demonstrated in Figure~\ref{fig:deltaCompThresh}.

\subsection{Varying distribution weights}
As described in the Methods section, we have proposed a methodology for determining the weights in the
score function that is primarily concerned with obtaining a high
percent of redistricting plans below our chosen threshold values.  Other parameters
may be chosen, and here we test whether making a different choice will
affect the statistics on the election outcomes.  There are four
dimensions to explore -- this means that exploring this space exhaustively would come at an large
computational cost.  To reduce this cost, we perform a sensitivity test about the reported weights used in the primary analysis:  We first significantly increase and decrease
$w_p$, $w_I$, and $w_m$.  For the fourth direction, we
could simply increase or decrease $w_c$; we could also increase and decrease the maximum value of $\beta$ instead, and choose this path instead. Because
changing $\beta$ is equivalent to changing all parameters, this forms
a fourth linearly independent search direction, and provides us with
information equivalent to changing $w_c$. This leads to eight different parameter sets, which still require a large number of runs.  To cut down on the computational cost, we take advantage of the result presented above, where we conclude that ignoring the compactness threshold has a minimal effect on our results. The compactness threshold is by far the most restrictive, so omitting it will allow us to sample more redistricting plans with fewer runs. We examine between 125 and 1100 samples for each search direction (we examine less samples with more restrictive parameters)

\begin{figure}[ht]
  \centering 
\includegraphics[width=.75\linewidth]{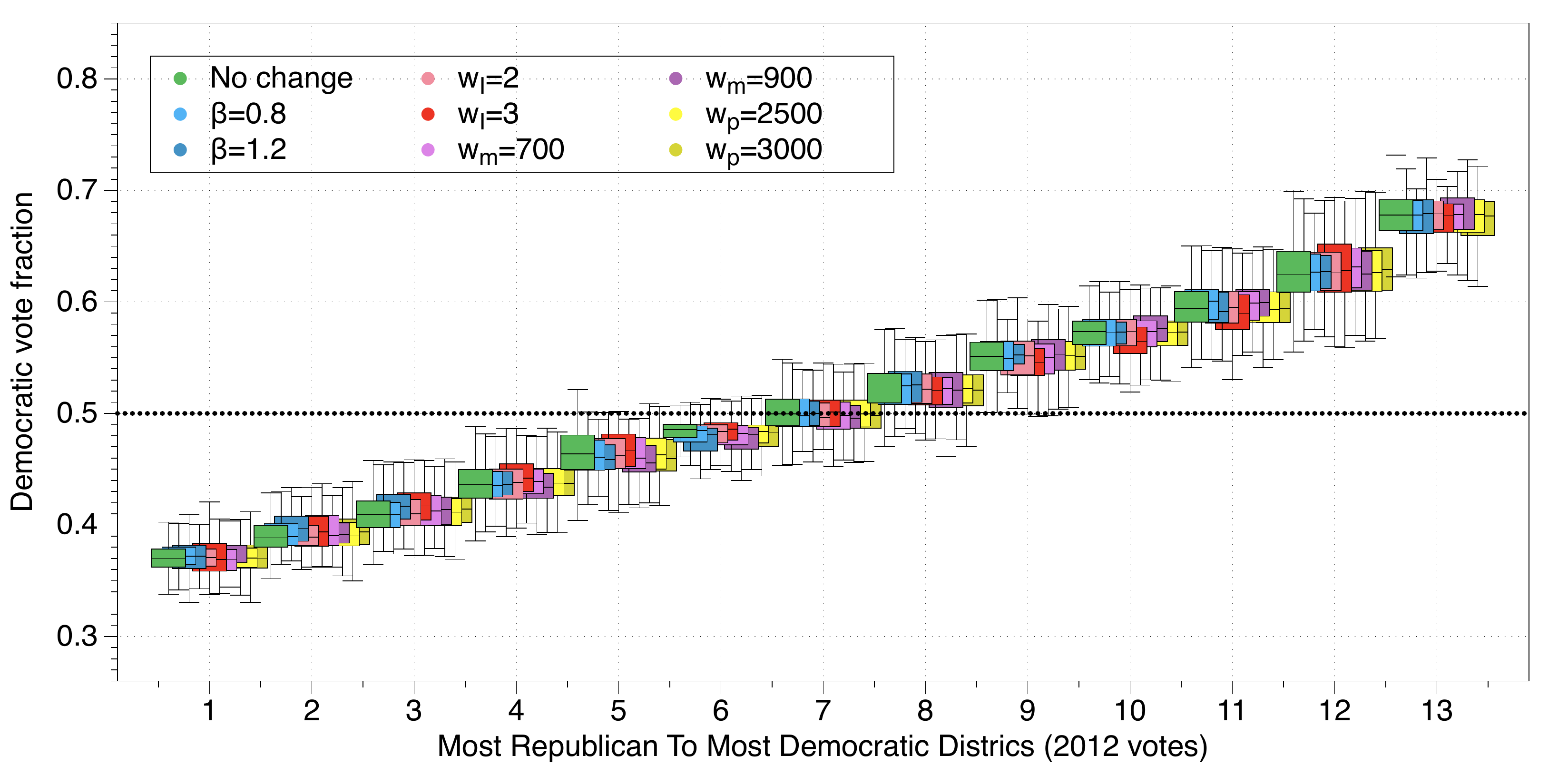}
   \caption{
   We display box-plots with outer bounds marked by 1.5$\times$ the interquartile range.   Election results change little with respect to changing the values of the weights. 
   }
  \label{fig:varyWeights}
\end{figure}

All eight search directions maintain the overall structure of the
election results.  The results are visualized in
Figure~\ref{fig:varyWeights}.  We conclude that significant changes in
the parameters will have little effect on the statistical results of
the election data.

\subsection{Different weights for lower county splits}
In the above analysis compact districts are prioritized over those with low county splits.  
In this section we determine the sensitivity of our results when we prioritize keeping a low number of county splits.  To make this examination, we double the county weight ($w_c=0.8$) and reduce the compactness weight ($w_I=2$).  By resetting the compactness threshold to be 80, we obtain just under 15 thousand redistricting plans and note that all of them have a worst district better than the worst NC2016 district.  If we had kept the threshold at 60, there would only be a couple thousand samples, thus in order to obtain more samples and because we have found that compactness does not have a large effect on the results, we select the higher threshold.  We find that despite changing the weights in this severe way, the over all election results, in addition to the distribution of results per ordered district remains remarkably stable.  The results are presented in Figure~\ref{fig:lowcs}.  We also remark that we now have a median of 16 two county splits with a mean of 16.5, in contrast with a median of 21 and mean of 21.6 from the main results (see Section~\ref{sec:characteristics}).

\begin{figure}[ht]
  \centering 
\includegraphics[width=.4\linewidth]{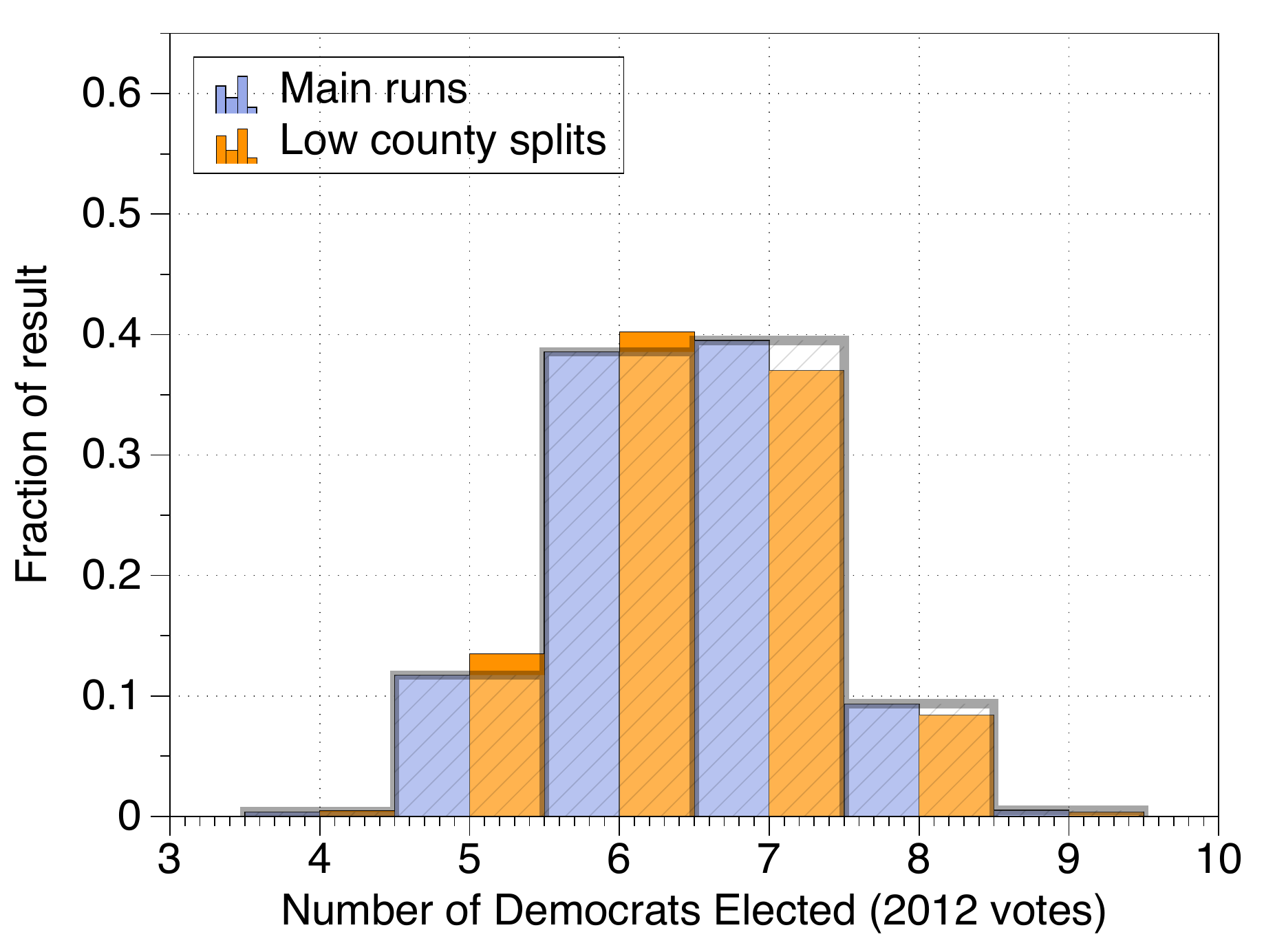}
\includegraphics[width=.4\linewidth]{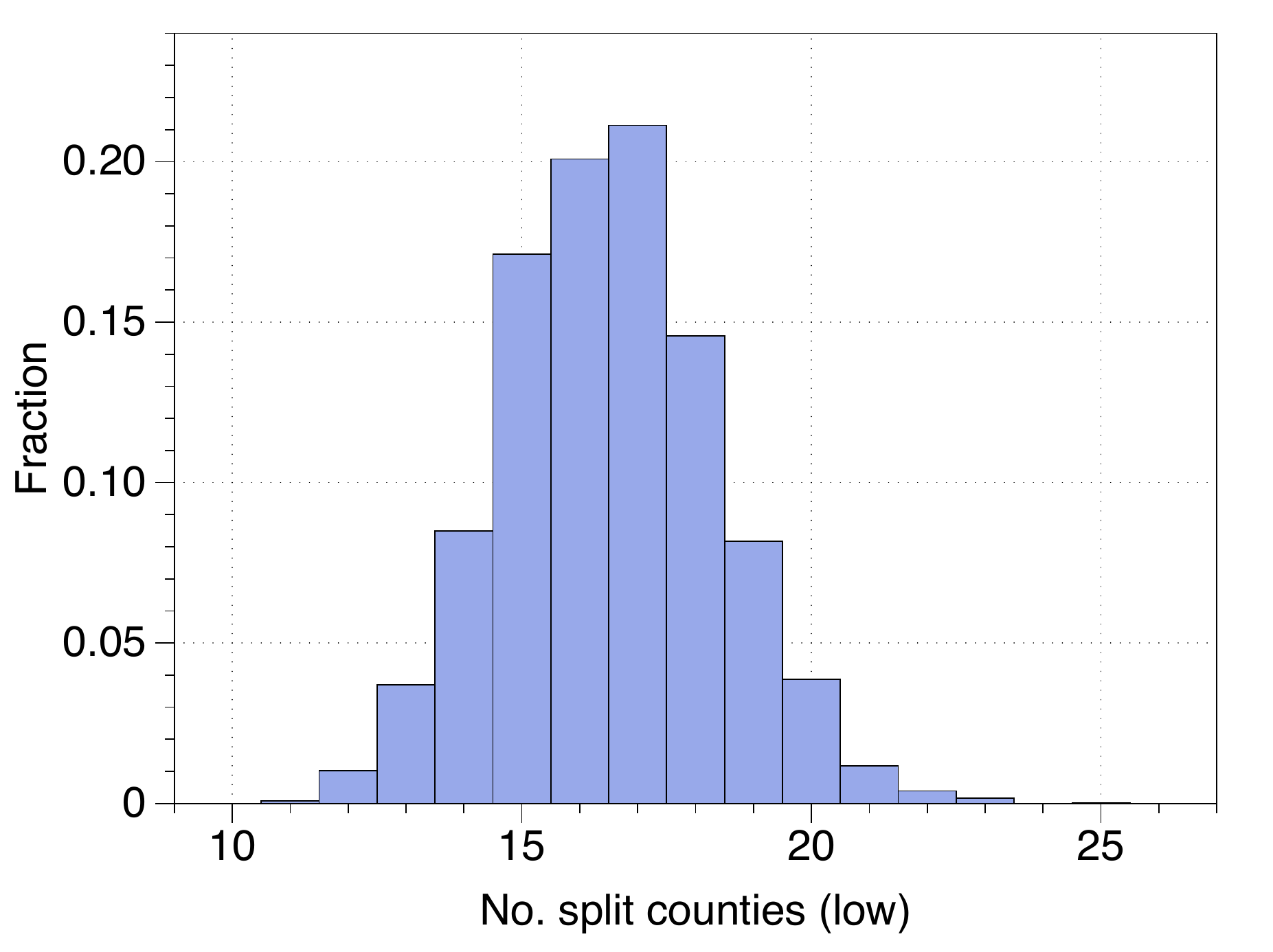}\\
\includegraphics[width=.4\linewidth]{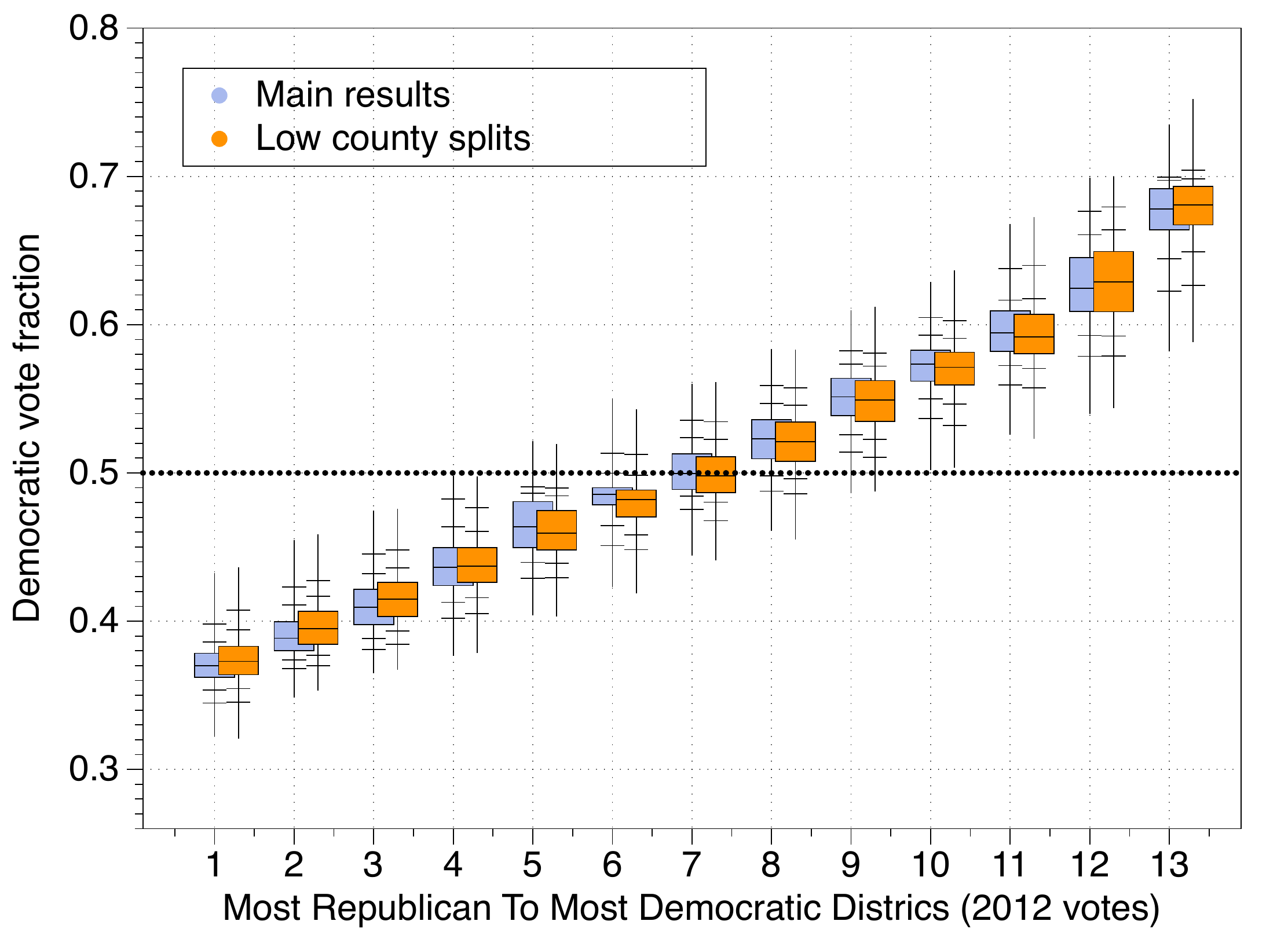}
\includegraphics[width=.4\linewidth]{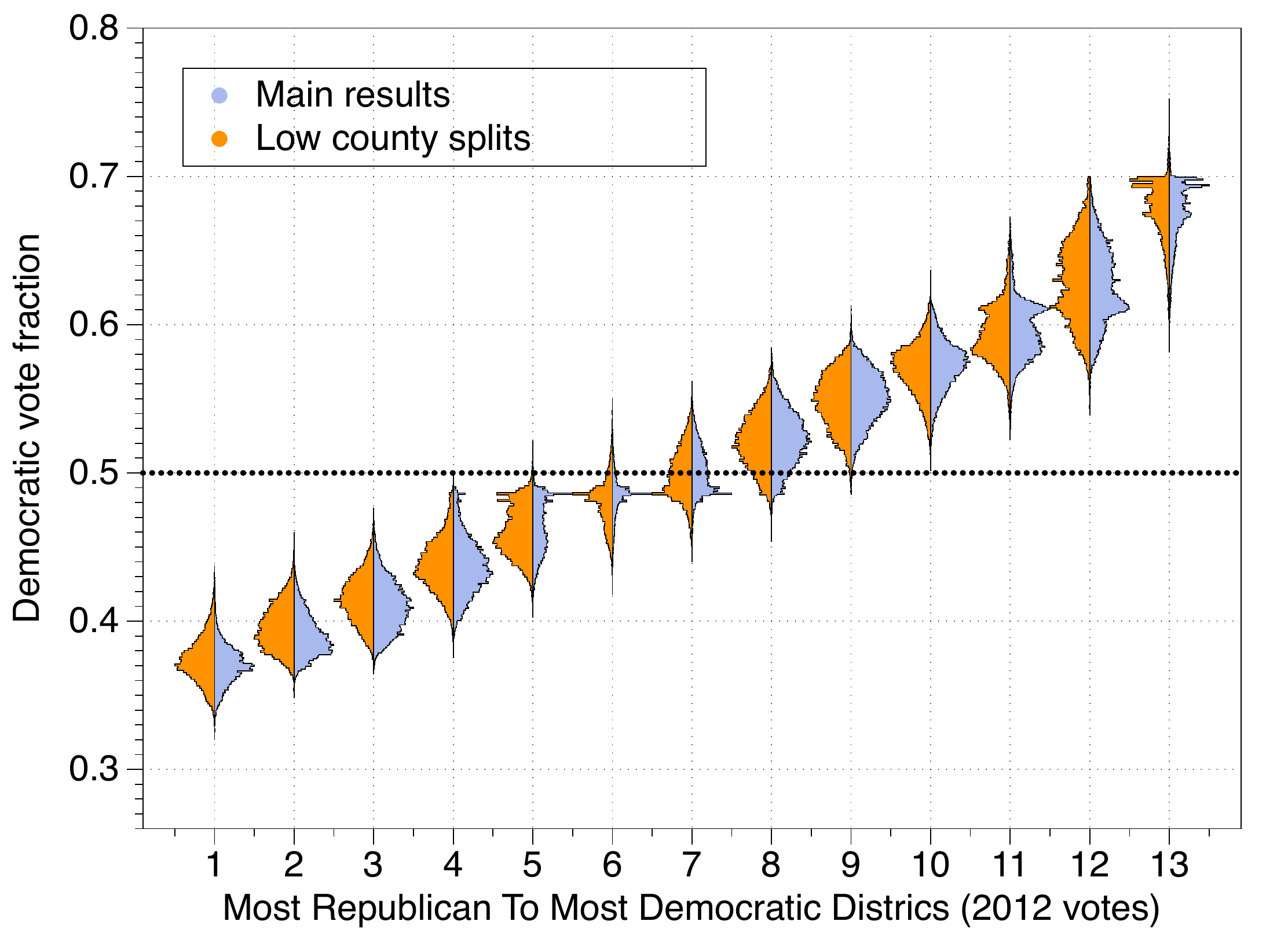}
   \caption{ By changing the weights on the energy function we alter the distribution of two county splits (top right).  Despite these changes, the over all election results (top left) and box and box histogram plots by district (bottom) remain stable.}
  \label{fig:lowcs}
\end{figure}

\subsection{Using a different compactness energy}
We have used the isoparametric ratio for the compactness energy however there are other possible choices.  Mentioned above when introducing the isoparametric score function, the dispersion score measures the spread of a district: Typically it is thought of as the ratio between the area of the minimal bounding circle and the districts area.  Although useful as a metric to compare two districting plans, the dispersion score does not minimized for jagged perimeters and cannot be used as a sufficient criteria to draw reasonable districts.  Nevertheless, we examine the space of districts in which we replace the isoparametric ratios with the dispersion ratio.  Given \S120-4.52(g)(1) of HB92, which specifies district length and width, we chose to measure dispersion as the ratio between the area of the minimal bounding rectangle and the district area.  The redistricting plans we arrive at would never be used due to the jagged perimeters, however if such a drastic change in the compactness criteria gives similar results to those which we have found before, we would have stronger evidence still for the robustness of our analysis.   

We threshold on everything but compactness as the isoparametric ratios become very high.  We keep the weights the same as in the main results.  With the 15,918 resulting redistricting plans for the new energy, the histogram of the election results and sorted district results are compared with our main results in Figure~\ref{fig:dispersionComp}. Despite this drastic change in energy definition, the results taken with different energies are remarkably similar to the main results.

\begin{figure}[ht]
  \centering 
\includegraphics[width=.33\linewidth]{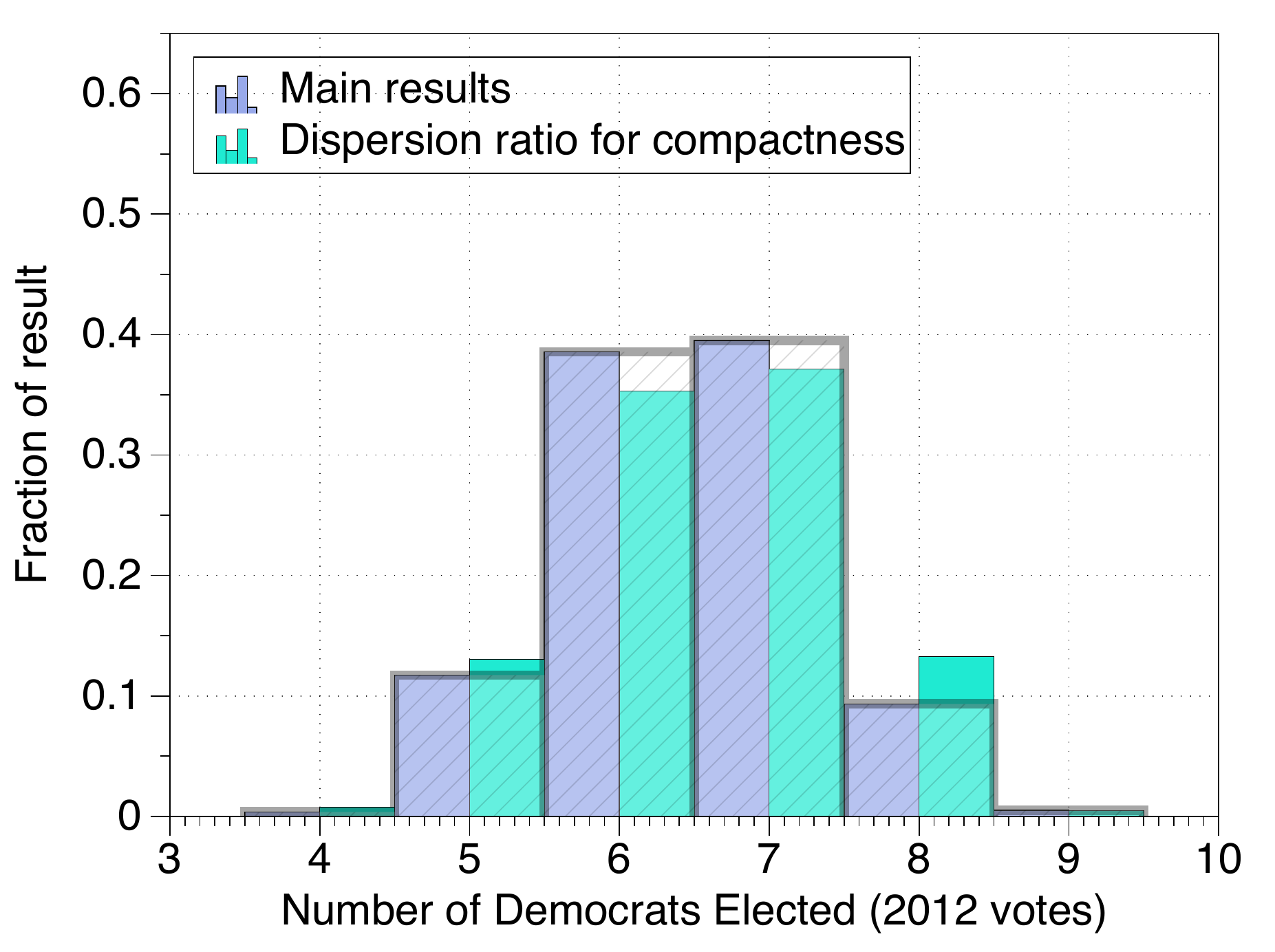}
\includegraphics[width=.33\linewidth]{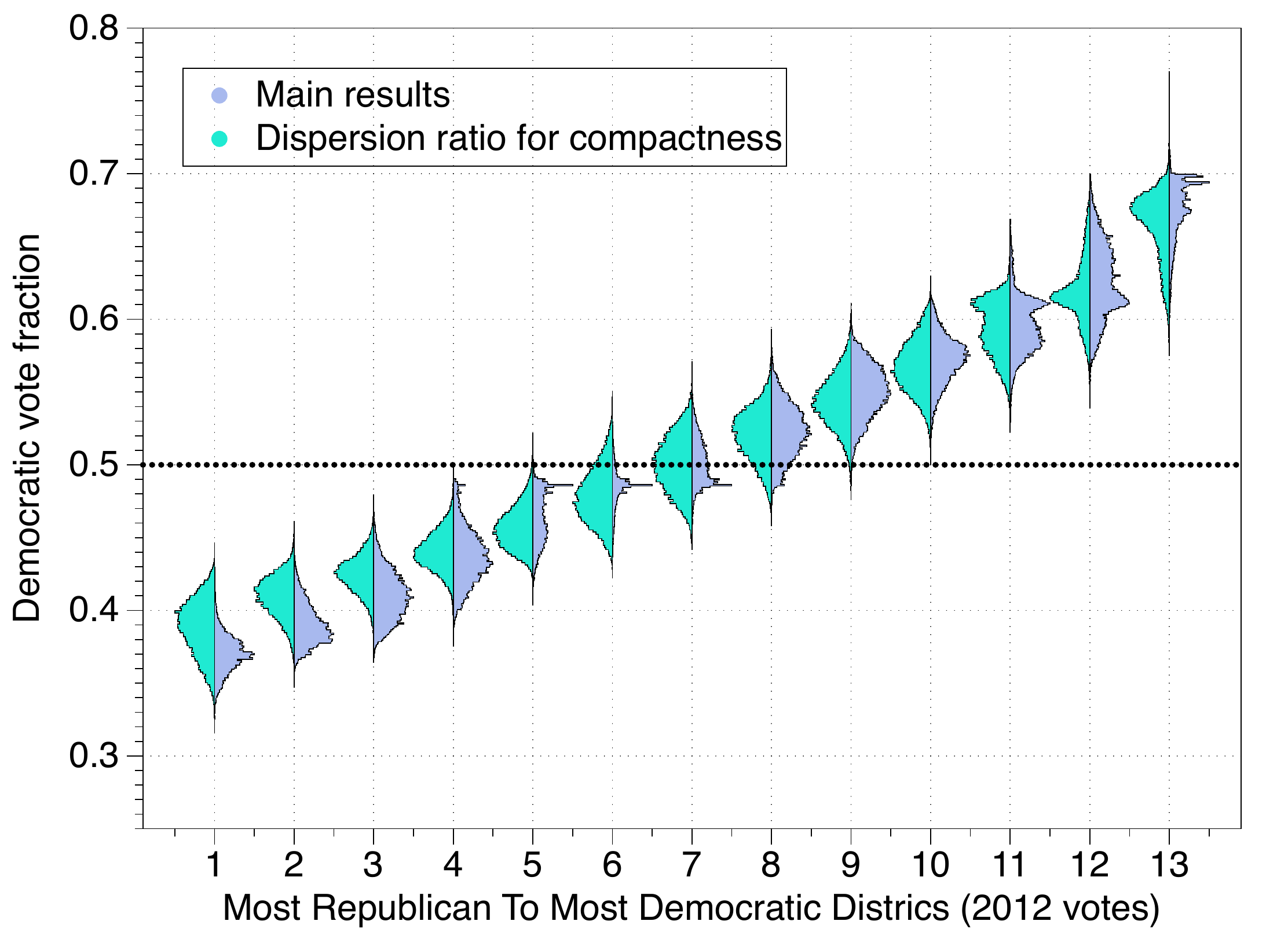}
\includegraphics[width=.33\linewidth]{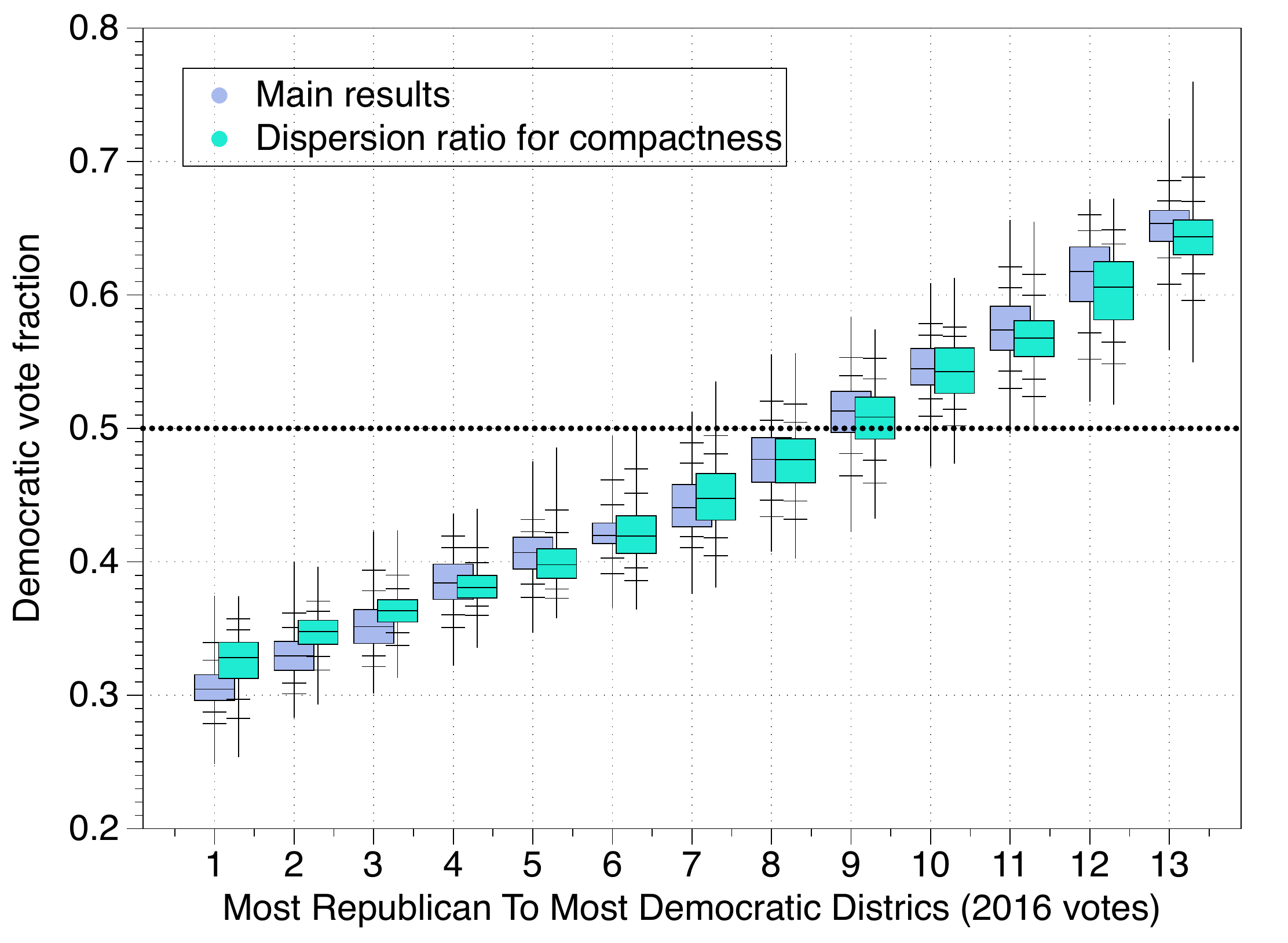}
   \caption{ We change the compactness energy from the isoparametric ratio to a type of dispersion score.  Despite this drastic change in energy definition, we find the results to be remarkably similar.}
  \label{fig:dispersionComp}
\end{figure}

\section{Error analysis}
\subsection{Error bounds for splitting VTDs to achieve zero population deviation} 
\label{subsec:zeropopdiv}
All of our redistricting plans do not split any VTDs.  In practice, VTDs must be split to achieve a population deviation below $0.1\%$.  Splitting VTDs may change the vote count in each district.  In this section, we demonstrate that splitting VTDs will have a negligible impact on the partisan vote fractions in each district.  To do this, we derive maximal and minimal error bounds on the democratic vote fractions for each redistricting plan by assuming that all of the changing votes will benefit only one party.  We repeat this process for the three redistricting plans of interest.  The results are plotted in Figure~\ref{fig:zeroPop}.  The shifts in the margins and the three plans are barely visible.  The small range of possible errors along with the test presented in Section~C.\ref{subsec:vary-thresholds} validate the idea that our results are robust when splitting VTDs to achieve a zero population deviation.

\begin{figure}[ht]
  \centering 
\includegraphics[width=.4\linewidth]{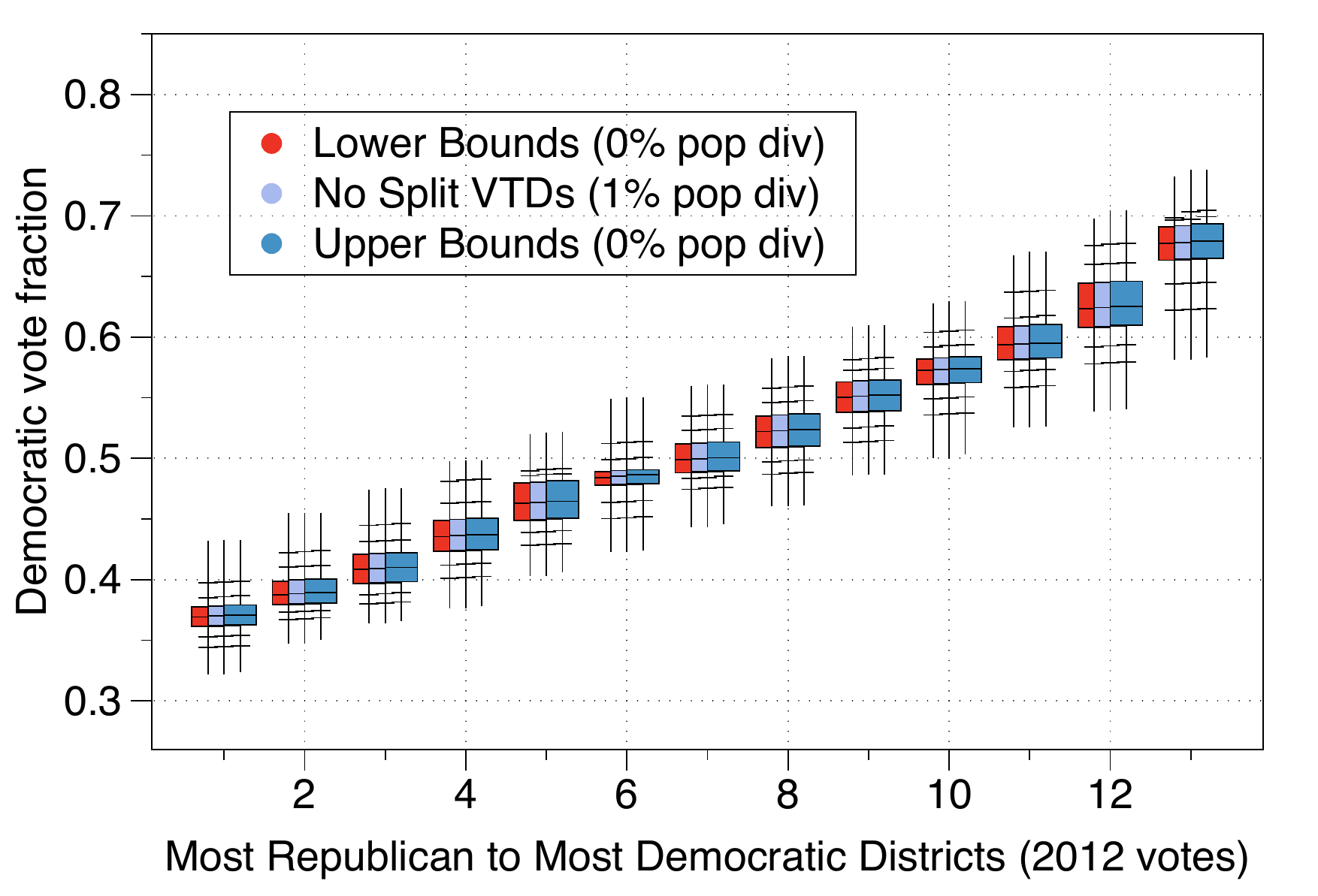}
\includegraphics[width=.4\linewidth]{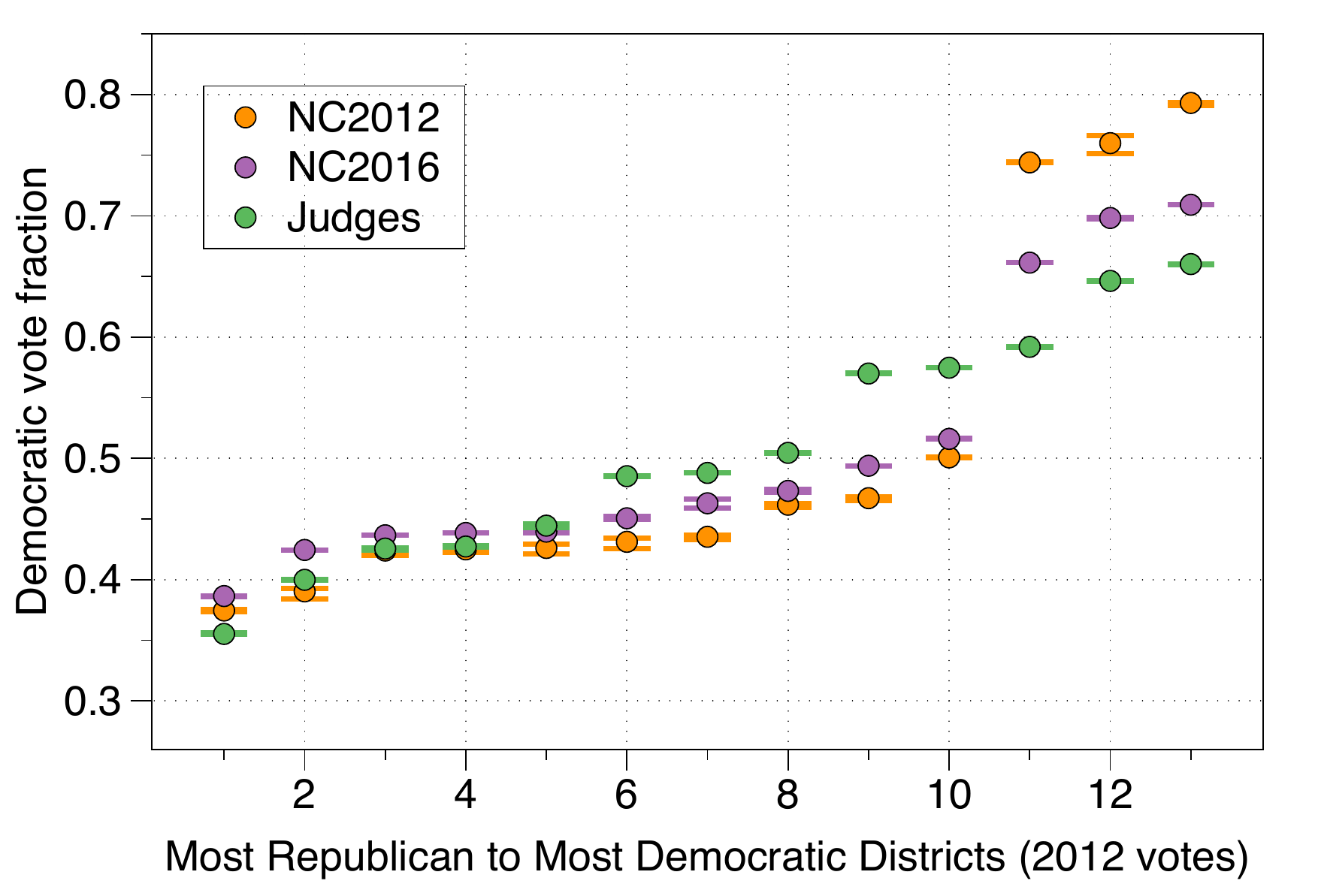}
   \caption{ We display the marginal statistics for the ensemble when zeroing population helps the democrats (dark blue) and republicans (red), and compare with the ensemble determined at the VTD level (light blue) (left).  We then display the range of possible vote fraction changes for the Judges, NC2012, and NC2016 plans (right).}
  \label{fig:zeroPop}
\end{figure}

\subsection{Bootstrapping}
We use bootstrapping to estimate the error on our samples.  On our ensemble of 24,518 redistricting plans, we sample, with replacement, 24,518, and obtain a new mean, median, and upper and lower 50\% bounds for the boxes in the box plots.  We repeat the process 10,000 times.  The maximum spread of resampled means, medians, and upper and lower 50\% bounds for the boxes is less than 0.21\% over all ordered marginal district distributions in the ensemble.

\section{Details on examining nearby redistricting plans}
To randomly sample nearby districts, we run
the same MCMC algorithm described above, but add a small modification: If a proposed step ever would increase the Hamming distance between any of the districts from the original redistricting
in question (either NC2012, NC2016, or the Judges) by more than 40 VTDs,
the step is rejected. Alternatively, one can think of $J(\xi)=\infty$ for any
$\xi \in \dist$ which has a district that differs from the original
redistricting by more than 40 VTDs.  As before, we then threshold the
results for NC2016 and the Judges on the population score, the county
score, and the minority score. We do not threshold on the
Isoperimetric Score as (i) keeping the redistricting near the original is
likely sufficient (ii) the threshold would be too severe for the NC2016 redistricting, since it already violates the threshold, and (iii) we demonstrate (see Figure~\ref{fig:threshVVnothresLocal}) that thresholding on compactness does not qualitatively affect our results. We use no thresholds for the NC2012 districting because most of the redistricting plans close to NC2012 would fail the population threshold since the compactness energy dominates here, as the districts are highly non-compact; the nearby NC2012 districts typically have a district with a few percentage points of population deviation.  Subsampling with the above thresholds leads to ensembles of 2,523 redistricting plans about the NC2012 plan, 2,334 redistricting plans about the NC2016 pland, and 2,554 redistricting plans about the Judges plan.

In order to justify the lack of thresholding on NC2012, we examine the difference in the local complementary cumulative distribution function for the Judges plan when thresholded and not.   We find that there is only a modest difference between the thresheld and non-thresheld results from the Judges which provides evidence that using the non-thresheld results for NC2012 is unimportant for obtaining a representative space of nearby districts (see Figure~\ref{fig:threshVVnothresLocal}).  
\begin{figure}[ht]
\centering 
\includegraphics[width=.4\linewidth]{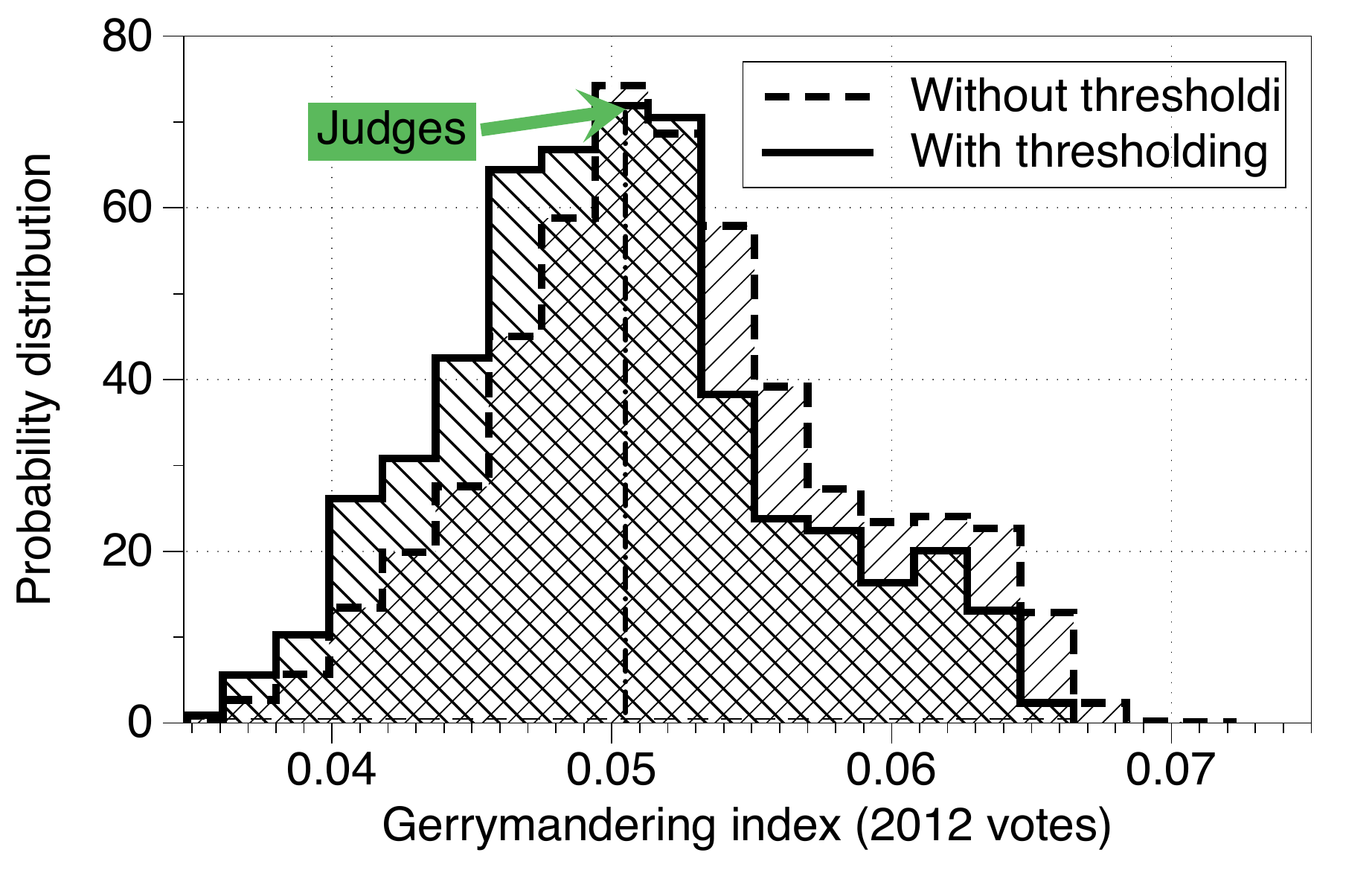}
\caption{There is not a large difference between the thresholded 
and unthreshelded results when considering local districting plans near the Judges districting plan.}
\label{fig:threshVVnothresLocal}
\end{figure}

\section{Characteristics of the redistricting plans in the ensemble}
\label{sec:characteristics}
We report the properties of the over 24,000 random restricting plans we have
generated using the algorithm described in the preceding sections. All
of the random redistricting plans passed the threshold test described
in the previous section. As such, they all have no district
with population deviation above 1\%. However, most have a deviation
much less than 1\%: the mean population deviation taken over the
more-than $13\times 24000= 312,000$ districts is 0.16\% with a standard deviation of 0.14\%.  
Figure~\ref{fig:popIsoStats} gives a finer view for the distribution of the
population deviation.  We order each redistricting by the maximum population deviation over all districts.  To simultaneously give a sense of the median population deviation of the districts with a given maximum population deviation, we examine the local statistics of the ordered districts to find the maximum and minimum values of the local median (plotted as the blue envelope) along with the standard deviation of the median (green envelop) and expected value of the median (dotted line).  With this plot we notice that over 50\% of redistricting plans have
a worst case population deviation under 0.4\% and many of these redistricting plans have a median population deviation well below 0.2\%.

\begin{figure}[ht]
  \centering 
\includegraphics[width=.4\linewidth]{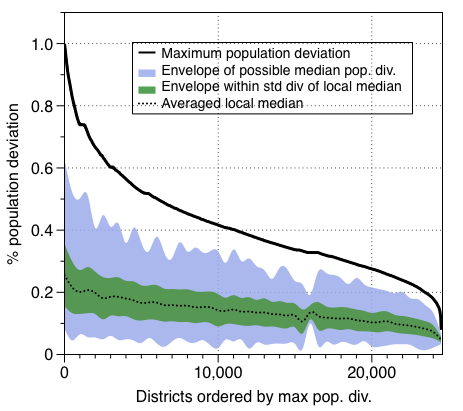}
\includegraphics[width=.4\linewidth]{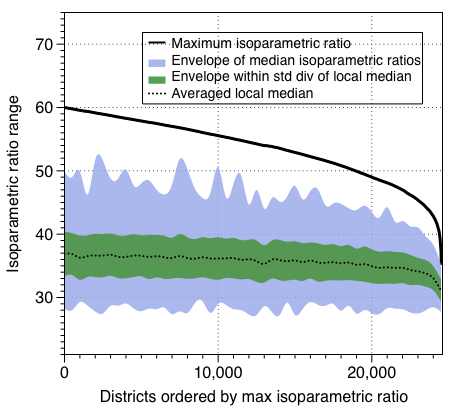}
   \caption{ The redistricting plans ordered by the worst case 
     district in terms of either population deviation (left) or 
     isoparametric ratio (right). The solid dark line give the worst 
     case districts value while the dotted line gives the average 
     across redistricting with a given max value of 
     median districts value. The outer shading gives the max and min 
     value of this median while the inner-shading covers one standard 
     deviation above and below the mean of the  medians.}
  \label{fig:popIsoStats}
\end{figure}

To compare the population deviation of our generated districts with the districting plans of NC2012, NC2016 and the Judges, we note that all of these districting plans all had to split VTDs in order to achieve a population deviation below 0.1\%.  Before splitting VTDs, NC2012, NC2016 and Judges had a district with maximum population deviation of 0.847\%, 0.683\%, and 0.313\% respectively, and had median population deviations of 0.234\%, 0.048\%, and 0.078\% respectively, meaning that the districts sampled by our algorithm are very similar to the three districts we have compared our results with in terms of population deviation.

Turing to the isoparametric ratios, recall that all of the districts have an
isoparametric constant under our threshold value of 60.  The
mean isoparametric ratio of the more-than $13\times 24000=312,000$ districts is 36.9 with a standard deviation of 9.  Examining the second part of Figure~\ref{fig:popIsoStats} gives an analogously finer view for the distribution of the isoparametric ratios of all districts.  The figure shows that most redistricting plans have a median isoparametric ratio in the mid-thirties and that roughly 50\% of our redistricting plans have a district with isoparametric ratio no worse than 55 for an isoparametric ratio.

 When comparing our generated districts, we note that the NC2012, NC2016 and Judges redistricting plans have districts with maximum isoparametric ratio of 434.6, 80.1, and 54.1 respectively, and have median isoparametric ratios of 114.4, 54.5, and 38.2 respectively.  The NC2012 and NC2016 districts would be rejected under our thresholding criteria, however we have seen above that lifting threshold conditions on compactness will note effect election results.
 
\begin{figure}[ht]
  \centering 
\includegraphics[height=5.7cm]{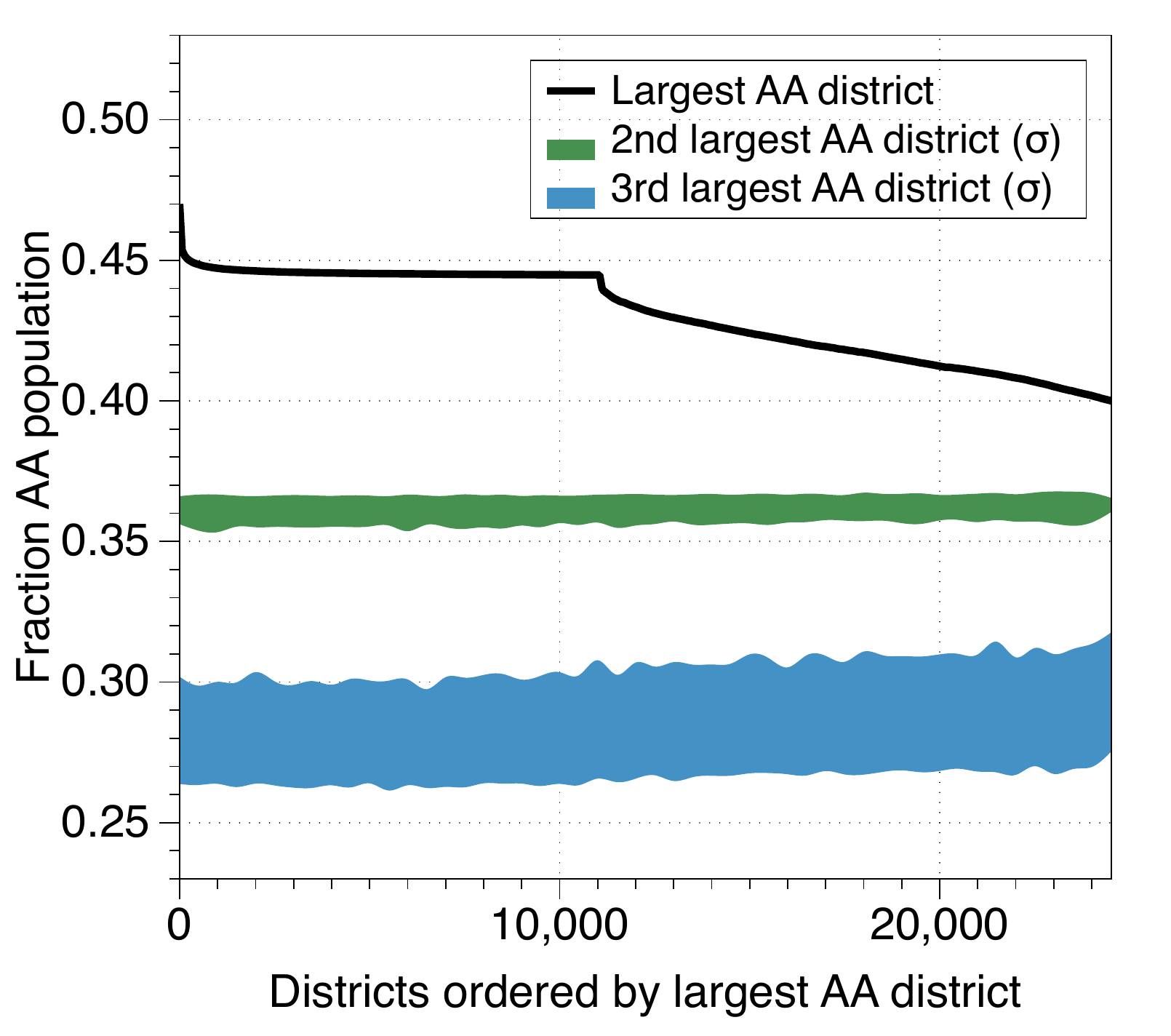}
\includegraphics[height=5.7cm]{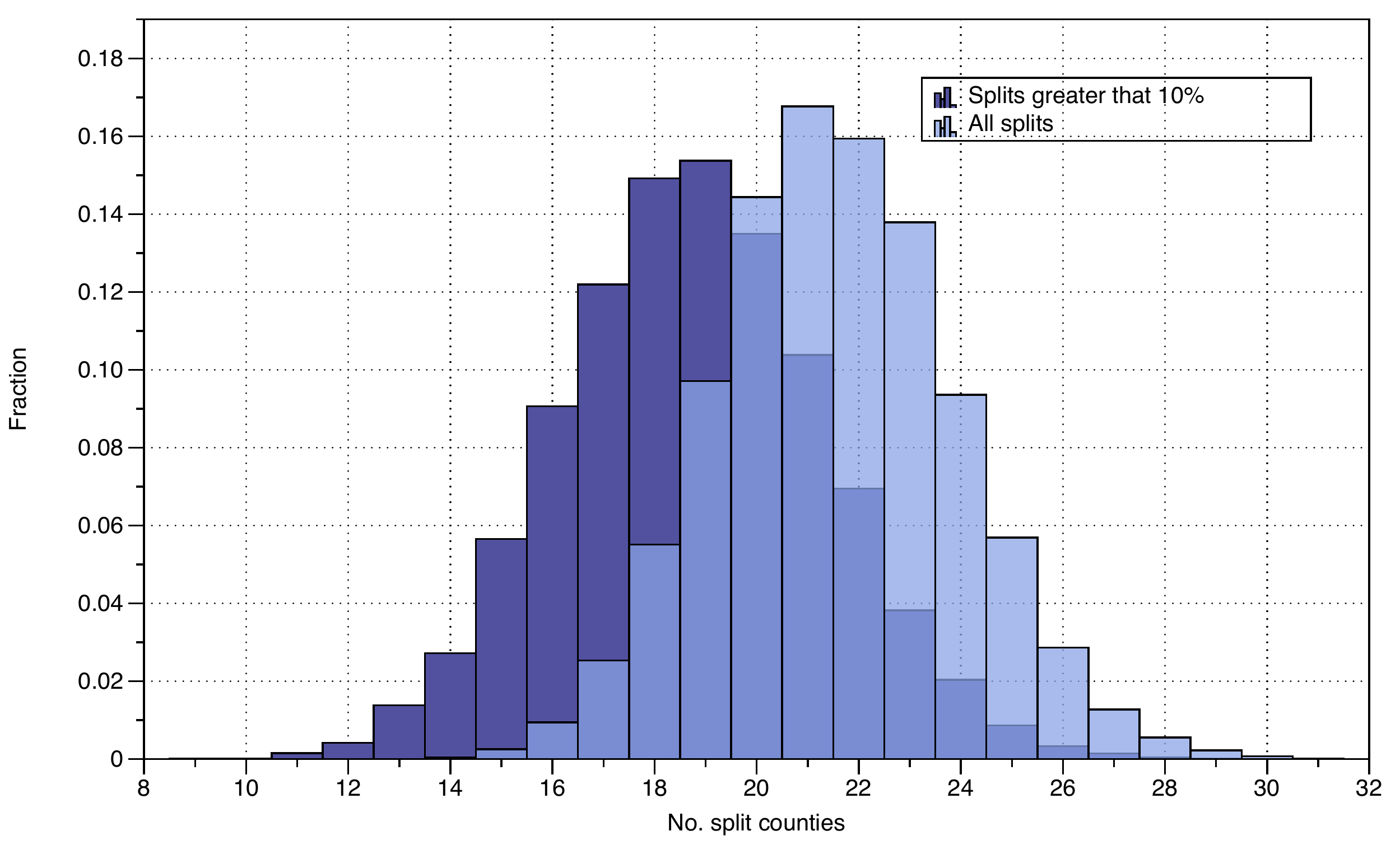}
\caption{ The redistricting plans are ordered by the district with 
  the largest African-American percentage (left).  Subsequent ranges 
  show standard deviations for districts with the second  and  third 
  largest  African-American representation.  We plot the histogram of 
  the number of county splits in each districting (right). The lighter 
  histogram gives the number total split counties while the darker 
  histogram gives only the number which splits the county into two 
  parts each containing  more than 10\% of the total VTDs. 
}
  \label{fig:minoCountyStats}
  \end{figure}

We examine the four districts with the highest minority population in each districting in Figure~\ref{fig:minoCountyStats}.  The redistricting plans are ordered on the district with the highest minority population over the over 24000 accepted redistricting plans.  The kink in this line at 44.46\% occurs due to the minority energy function which does not favor any population above this limit (this number was based on NC2016 districting which was ruled to adhere to the VRA).   Roughly half of the redistricting plans have a district with greater than 44.46\% of the population as African-Americans, whereas the other half has between 40\% and 44.46\% in the district with the largest number of African-Americans.  For the district with the second highest African-American representation, we remark that over 80\% of all redistricting plans have more than 35\% African-American representation in the second largest district; there is not a single redistricting that has the second largest African-American district with more than 40\% of the population African-American.

Finally we display the histogram of the number of split counties over our generated redistricting plans.  There is a median of 21 split counties with a mean of 21.6, and a range from 14 to 31.  We remark that NC2012, NC2016, and Judges districting plans had 40, 13, and 12 split counties respectively.

We display several of our generated redistricting plans in Figure~\ref{fig:Districting1}.  We display statistics of these selected redistricting plans in Table~\ref{tab:varyWeights}, and compare them with the NC2012, NC2016 and Judges districting plans.

\begin{figure}[ht]\centering 
\includegraphics[scale=0.25]{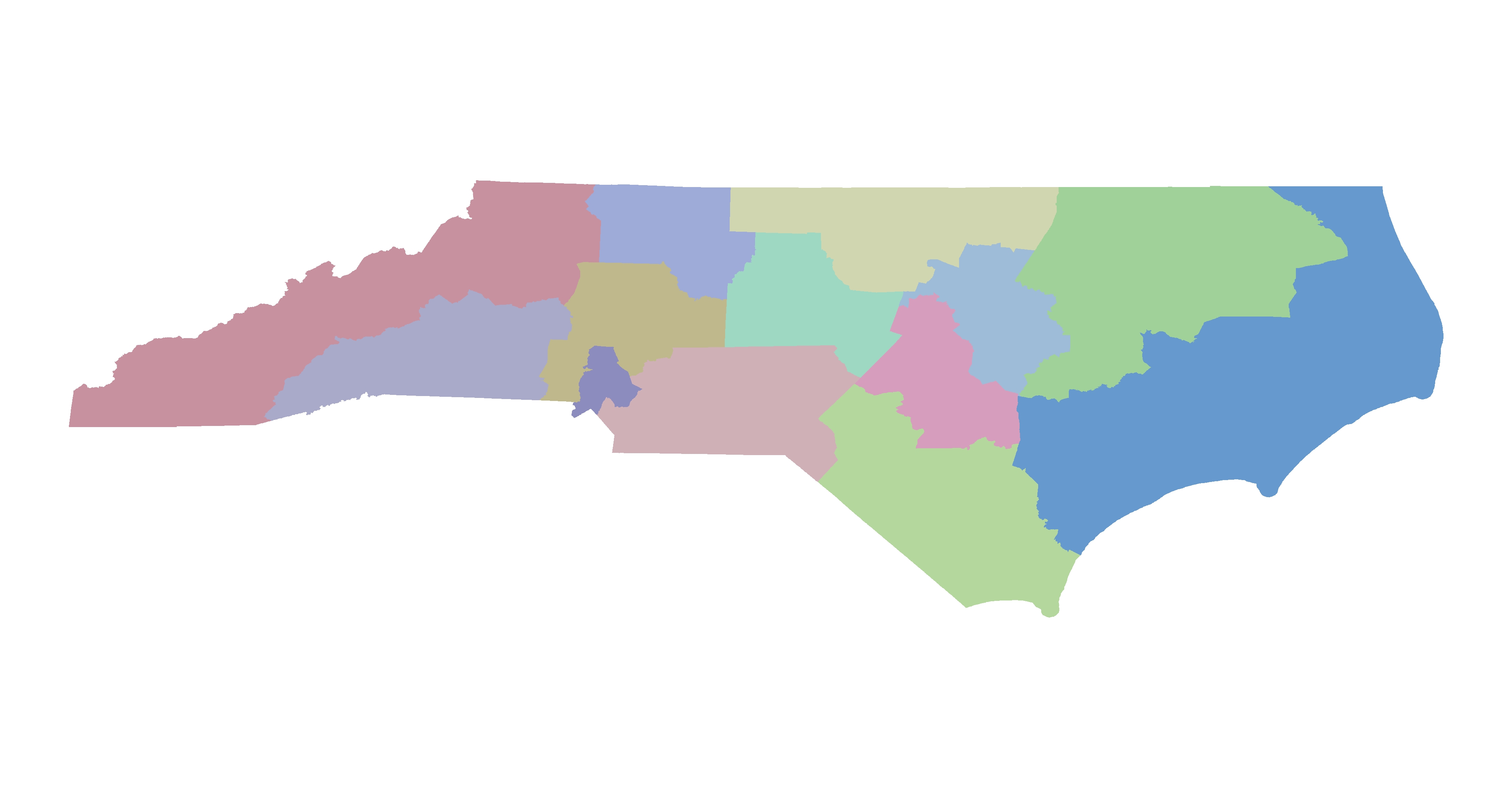} 
\includegraphics[scale=0.25]{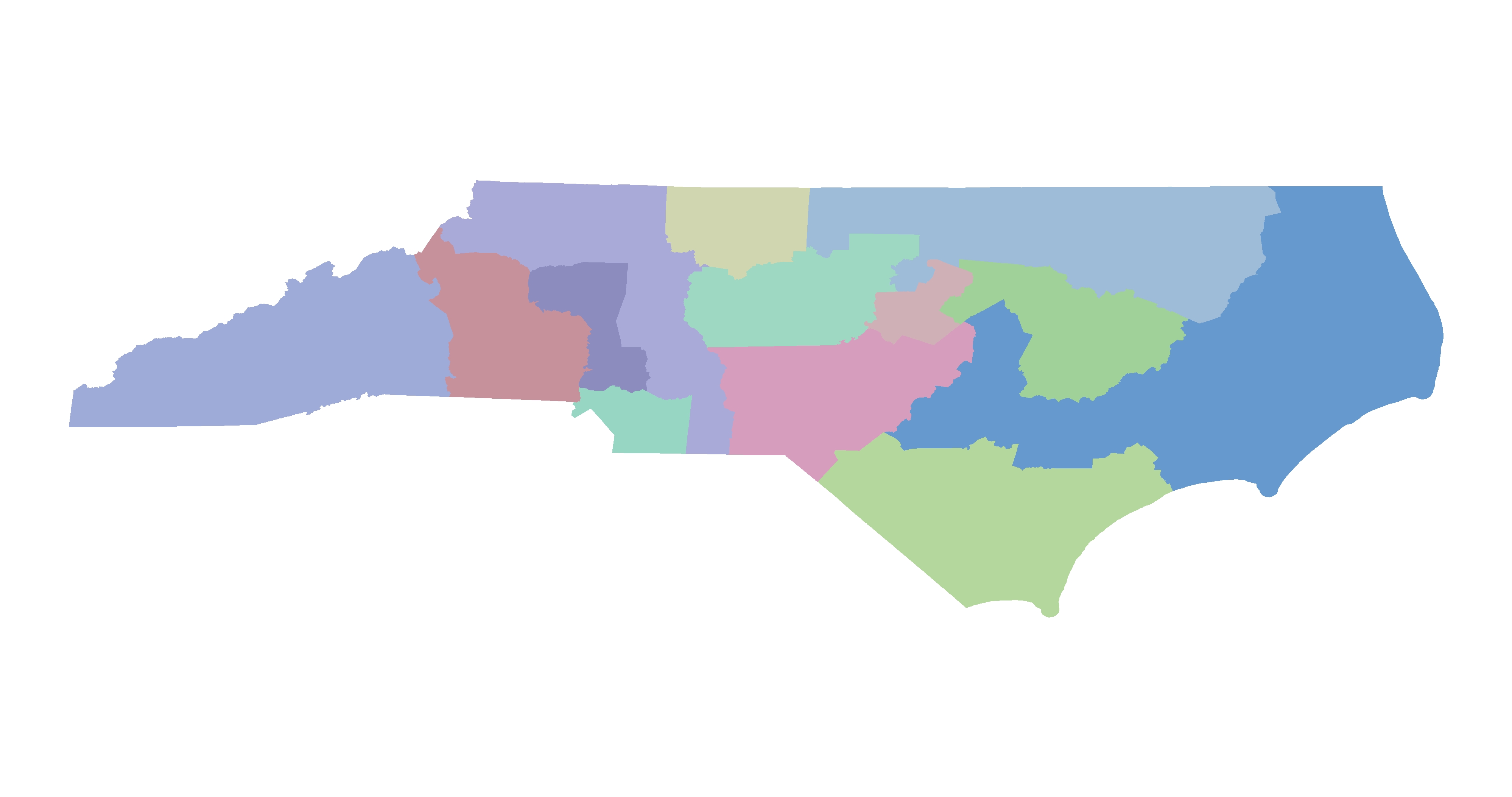} 
\includegraphics[scale=0.25]{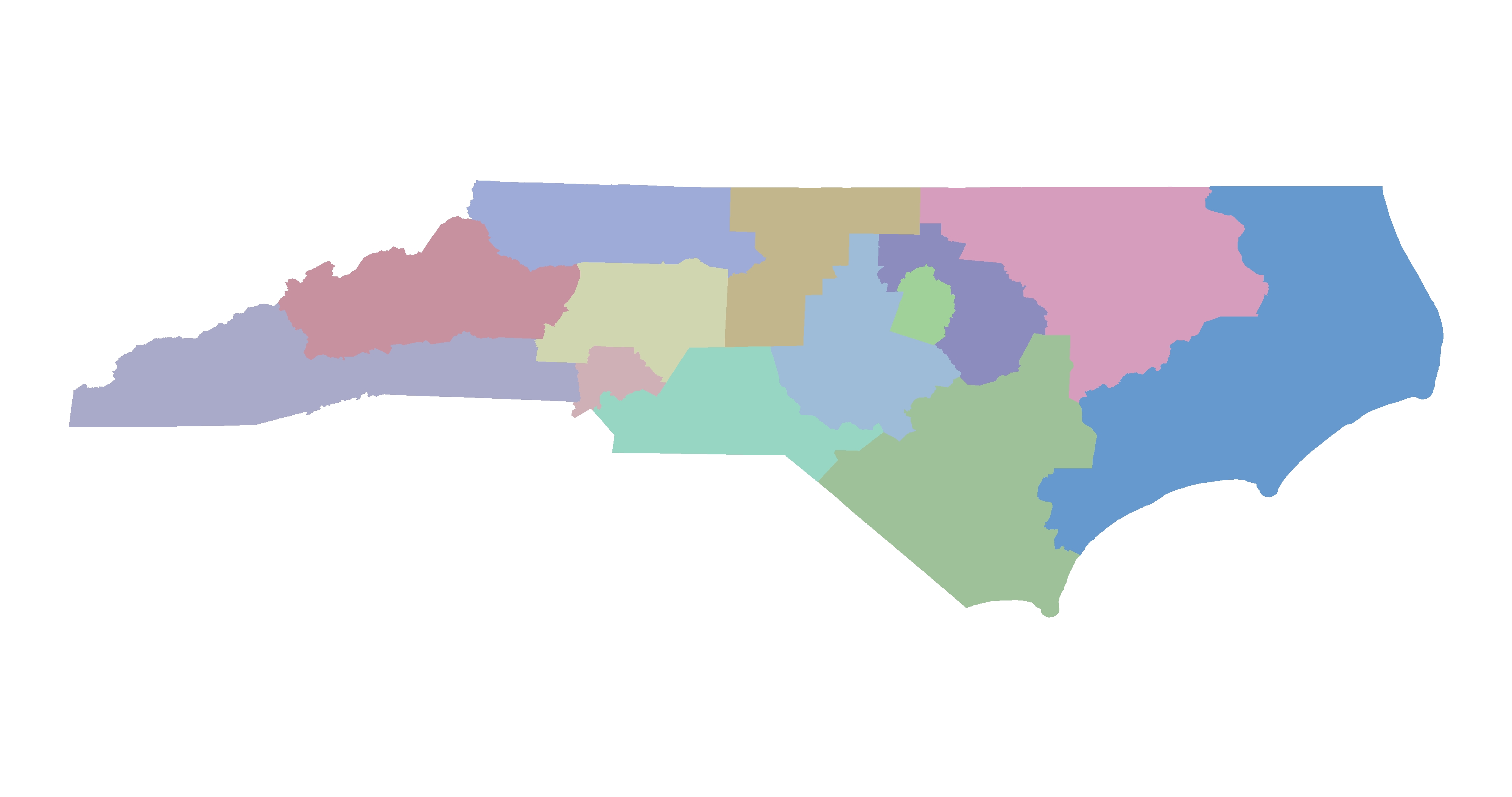} 
\includegraphics[scale=0.25]{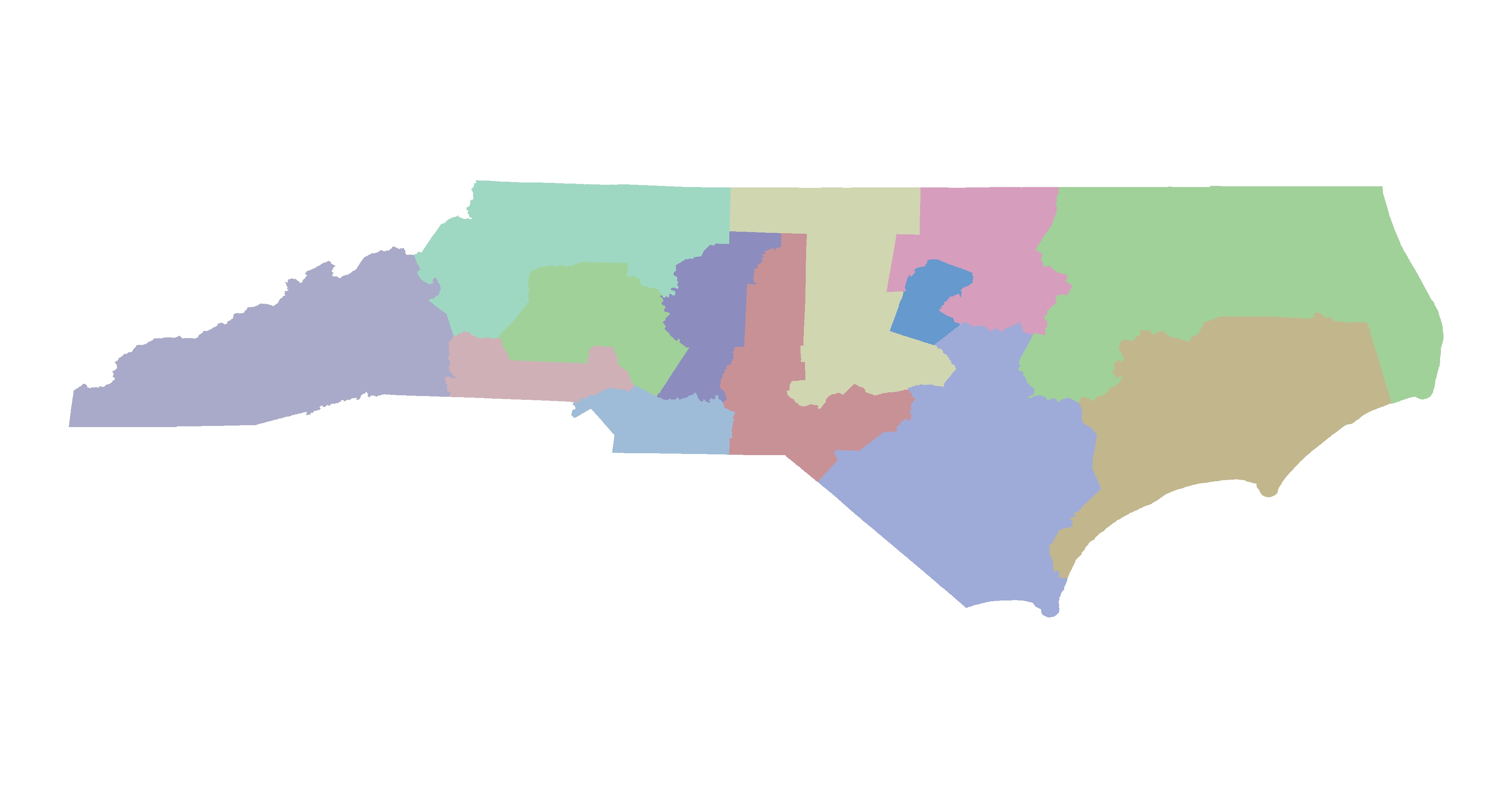} 
\includegraphics[scale=0.25]{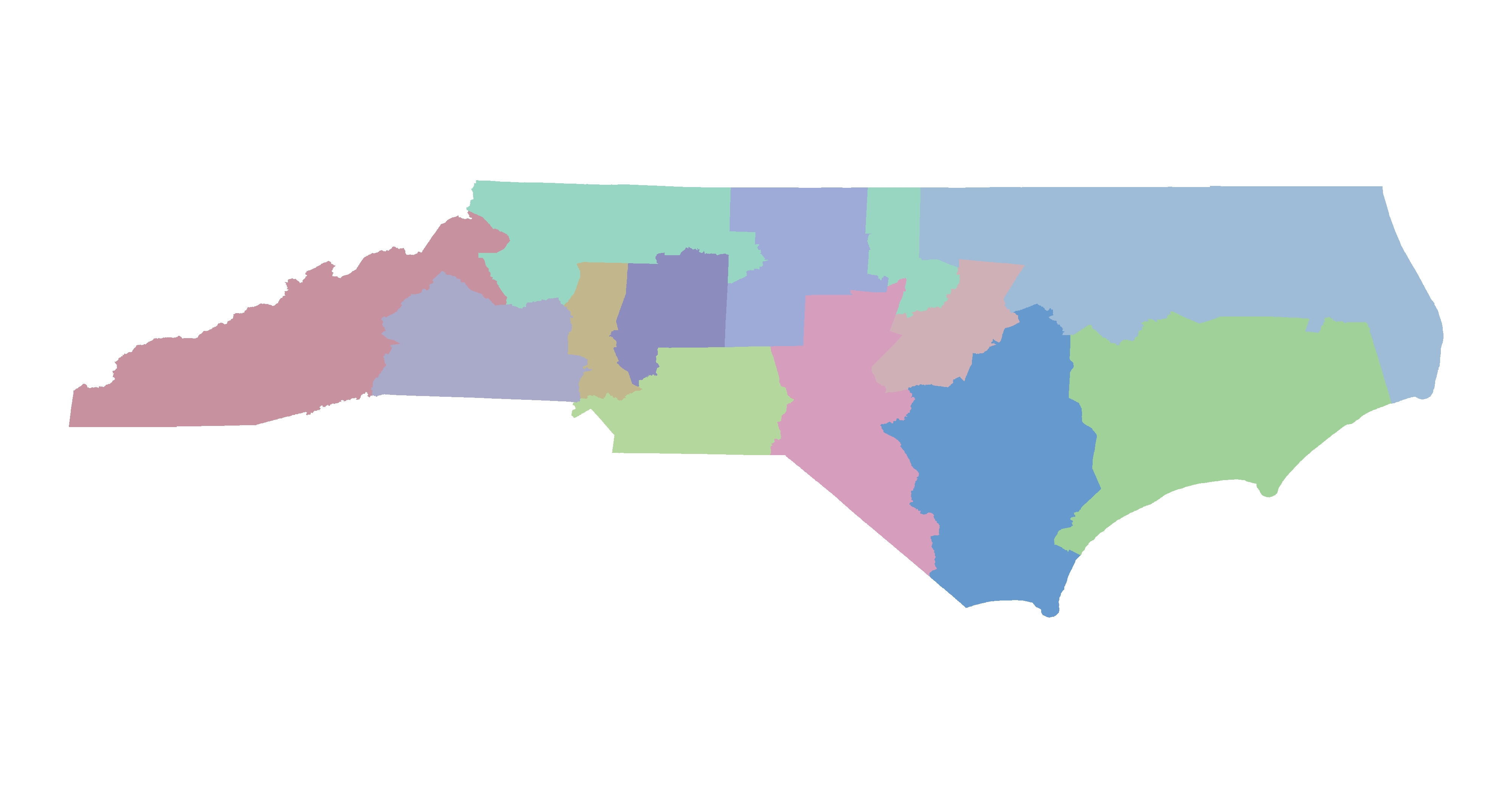}
\includegraphics[scale=0.25]{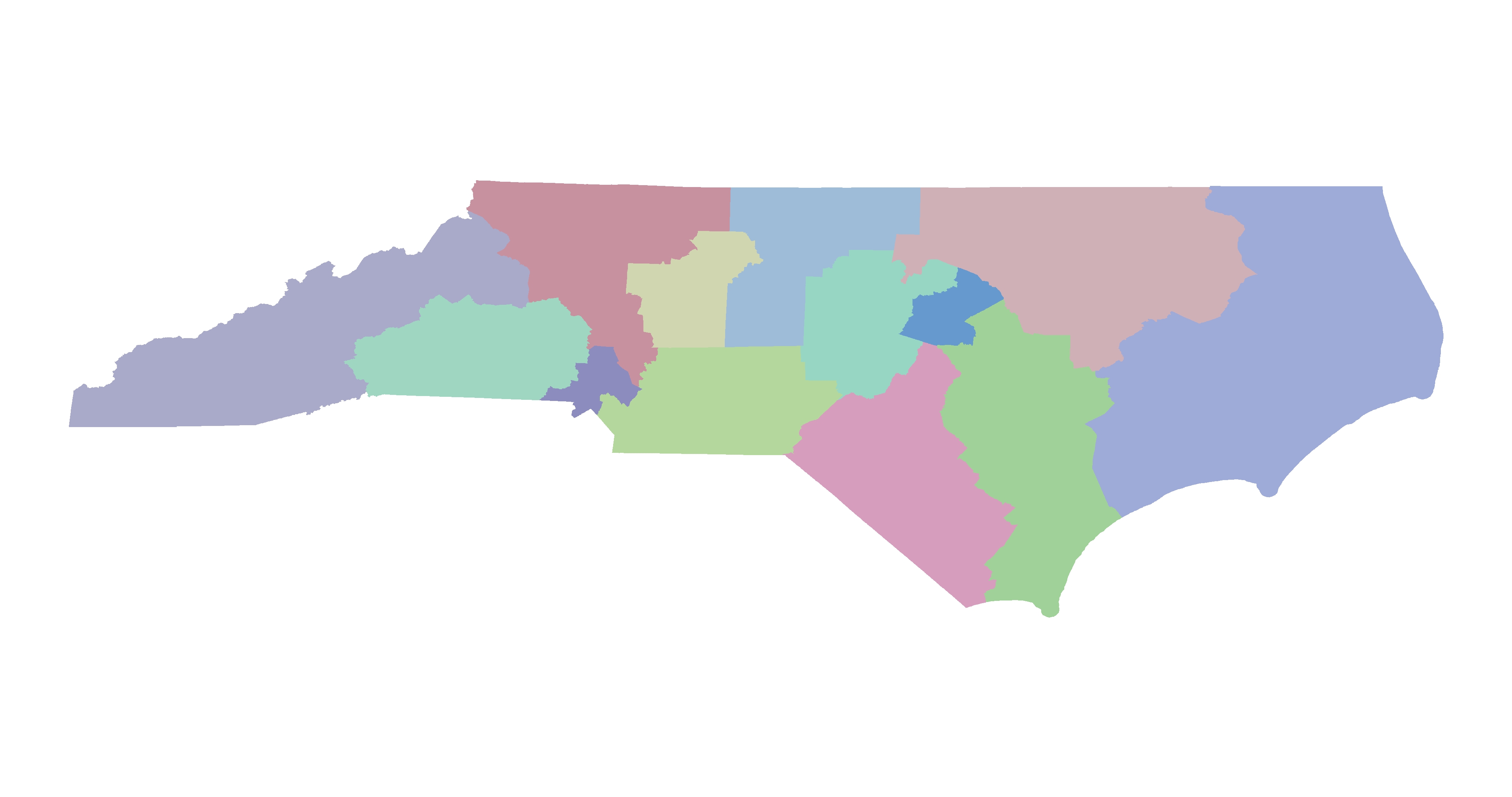} 
\caption{Six sample redistricting generated by MCMC.}
\label{fig:Districting1} 
\end{figure}

\begin{table}
\begin{tabular}{l || llllll}& &  & \multicolumn{2}{c}{\underline{Largest AA}} &
                                                                      \#
                                                                      Split
                                                                      \\
&\multicolumn{1}{c}{$J_p$} & \multicolumn{1}{c}{$J_I$} & \multicolumn{1}{c}{1st} & \multicolumn{1}{c}{2nd} & Counties\\
\hline
\hline
NC2012 & 1.59$\times 10^{-2}$ & 2068.53& 52.54 &50.01 & 40\\
NC2016& 7.93$\times 10^{-3}$ & 699.82 &  45.10& 36.28 & 13\\
Judges& 4.47$\times 10^{-3}$ & 527.14 & 41.68  & 32.96 & 12 \\
\hline
First sample& 1.38$\times 10^{-2}$ & 471.8 & 44.50 & 36.24 & 25\\
Second sample& 8.03$\times 10^{-3}$ & 489.44   & 44.48 & 36.31 &23\\
Third sample& 4.51$\times 10^{-3}$& 454.37 & 44.49 & 36.20 &23\\
Fourth sample& 8.98$\times 10^{-3}$& 465.03 &   41.65 &
                                                                  36.38 & 24\\
Fifth sample& 4.51$\times 10^{-3}$ & 463.86&  42.01 & 33,06 & 21 \\
Sixth sample&5.07$\times 10^{-3}$ & 460.12 &  44.95 & 36.26 & 21
\end{tabular}

 \centering
\vspace{1em}
\caption{
  We display the various energies for each of the districting plans that we
  present in the appendix.  Note that reported numbers for
  districting plans are before VTD splits. 
}
  \label{tab:varyWeights}
\end{table}

We store the over 24,000 redistricting plans generated by the Monte Carlo algorithm.  Each redistricting file holds the FID of the VTD, which are consistent with the FIDs in the Harvard's Election Data Archive Dataverse \cite{NC2012}.  Each district, labeled by the FID, is associated with a district labeled 1-13 in the second column.  See git@git.math.duke.edu:gjh/districtingDataRepository.git

\end{document}